\documentclass[traditabstract]{aa}
\usepackage{tabularx}
\usepackage{amssymb}
\usepackage{amsmath} 
\usepackage{txfonts}
\usepackage{balance}
\usepackage{natbib}
\usepackage{threeparttable}
\bibpunct{(}{)}{;}{a}{}{,} 
\usepackage[usenames, dvipsnames]{color}
\usepackage{xspace}
\usepackage[normalem]{ulem} 
\usepackage{multirow}
\usepackage{rotating}
\usepackage{cancel}
\usepackage{graphicx}
\usepackage{caption}
\usepackage{ifthen}
\usepackage{booktabs}
\usepackage{tabularx}
\usepackage{float}
\usepackage{gensymb}
\usepackage{subcaption} 
\usepackage{booktabs}
 
\begin{document}
\title{Galaxy populations in the most distant SPT-SZ clusters -- I.\\ Environmental quenching in massive clusters  at $1.4\lesssim z\lesssim1.7$}
\titlerunning{Environmental quenching in massive clusters at $1.4\lesssim z\lesssim1.7$}
\authorrunning{Strazzullo et al.}

\author{V.~Strazzullo\inst{\ref{LMU}}
  \and M.~Pannella\inst{\ref{LMU}}
  \and J.~J.~Mohr\inst{\ref{LMU}}$^,$\inst{\ref{MPE}}$^,$\inst{\ref{ECUniverse}}
  \and A.~Saro\inst{\ref{LMU}}$^,$\inst{\ref{INAFtrieste}}
  \and M.~L.~N.~Ashby\inst{\ref{cfa}}
  \and M.~B.~Bayliss\inst{\ref{MIT}}
\and S.~Bocquet \inst{\ref{LMU}}
  \and E.~Bulbul\inst{\ref{cfa}}
  \and G.~Khullar\inst{\ref{KICPChicago}}$^,$\inst{\ref{AAUChicago}}
  \and A.~B.~Mantz\inst{\ref{KIPAC}}$^,$\inst{\ref{Stanford}}
  \and S.~A.~Stanford\inst{\ref{UCdavis}}
\and B.~A.~Benson\inst{\ref{FNAL}}$^,$\inst{\ref{AAUChicago}}$^,$\inst{\ref{KICPChicago}}
\and L.~E.~Bleem\inst{\ref{ANL}}$^,$\inst{\ref{KICPChicago}}
\and M.~Brodwin \inst{\ref{UMKC}}
\and R.~E.~A.~Canning\inst{\ref{Stanford}}$^,$\inst{\ref{KIPAC}}
\and R.~Capasso\inst{\ref{LMU}}$^,$\inst{\ref{ECUniverse}}
\and I.~Chiu\inst{\ref{ASIAA}}
\and A.~H.~Gonzalez\inst{\ref{UFlorida}}
\and N.~Gupta\inst{\ref{LMU}}$^,$\inst{\ref{ECUniverse}}$^,$\inst{\ref{melbourne}}
\and J.~Hlavacek-Larrondo\inst{\ref{UdeM}}
\and M.~Klein\inst{\ref{LMU}}$^,$\inst{\ref{MPE}}
\and M.~McDonald\inst{\ref{MIT}}
\and E.~Noordeh\inst{\ref{KIPAC}}$^,$\inst{\ref{Stanford}}
\and D.~Rapetti\inst{\ref{CASA}}$^,$\inst{\ref{Ames}}
\and C.~L.~Reichardt\inst{\ref{melbourne}}
\and T.~Schrabback\inst{\ref{AIfA}}
\and K.~Sharon\inst{\ref{Umich}}
\and B.~Stalder\inst{\ref{LSST}}
}

\institute{Faculty of Physics, Ludwig-Maximilians-Universit\"{a}t, Scheinerstr.\ 1, 81679 Munich, Germany \email{vstrazz@usm.lmu.de}\label{LMU}
\and Max Planck Institute for Extraterrestrial Physics, Giessenbachstr. 1, 85748 Garching, Germany\label{MPE}
\and Excellence Cluster Universe, Boltzmannstr.\ 2, 85748 Garching, Germany \label{ECUniverse}
\and INAF-Osservatorio Astronomico di Trieste, via G. B. Tiepolo 11, I-34143 Trieste, Italy \label{INAFtrieste}
\and Harvard-Smithsonian Center for Astrophysics, 60 Garden Street, Cambridge, MA 02138 \label{cfa}
\and Kavli Institute for Astrophysics and Space Research, Massachusetts Institute of Technology, 77 Massachusetts Avenue, Cambridge, MA 02139 \label{MIT}
\and Kavli Institute for Cosmological Physics, University of Chicago, 5640 South Ellis Avenue, Chicago, IL 60637 \label{KICPChicago}
\and Department of Astronomy and Astrophysics, University of Chicago, 5640 South Ellis Avenue, Chicago, IL 60637 \label{AAUChicago}
\and Kavli Institute for Particle Astrophysics and Cosmology, Stanford University, 452 Lomita Mall, Stanford, CA 94305 \label{KIPAC}
\and Department of Physics, Stanford University, 382 Via Pueblo Mall, Stanford, CA 94305 \label{Stanford}
\and Physics Department, University of California, Davis, CA 95616 \label{UCdavis}
\and Fermi National Accelerator Laboratory, Batavia, IL 60510-0500 \label{FNAL}
\and Argonne National Laboratory, High-Energy Physics Division, 9700 S. Cass Avenue, Argonne, IL, USA 60439 \label{ANL}
\and Department of Physics and Astronomy, University of Missouri, 5110 Rockhill Road, Kansas City, MO 64110 \label{UMKC}
\and Academia Sinica Institute of Astronomy and Astrophysics, 11F of AS/NTU Astronomy-Mathematics Building, No.1, Sec. 4, Roosevelt Rd, Taipei 10617, Taiwan \label{ASIAA}
\and Department of Astronomy, University of Florida, Gainesville, FL 32611 \label{UFlorida}
\and{School of Physics, University of Melbourne, Parkville, VIC 3010, Australia}\label{melbourne}
\and Department of Physics, University of Montreal, Montreal, QC H3C 3J7, Canada \label{UdeM}
\and Center for Astrophysics and Space Astronomy, Department of Astrophysical and Planetary Science, University of Colorado, Boulder, C0 80309, USA \label{CASA}
\and NASA Ames Research Center, Moffett Field, CA 94035, USA \label{Ames}
\and Argelander-Institut f{\"u}r Astronomie, Auf dem H{\"u}gel 71, D-53121 Bonn, Germany \label{AIfA}
\and Department  of  Physics,  University  of  Michigan,  450  Church  Street,  Ann  Arbor,  MI,  48109 \label{Umich}
\and LSST, 950 North Cherry Avenue, Tucson, AZ 85719 \label{LSST}
\vspace{-.6cm}
}

\date{ }

\abstract {We present the first results from a galaxy population study
  in the highest redshift galaxy clusters identified in the 2500
  deg$^2$ South Pole Telescope Sunyaev Zel'dovich effect (SPT-SZ)
  survey, which is sensitive to M$_{500}\gtrsim 3\times
  10^{14}$M$_{\odot}$ clusters from $z\sim0.2$ out to the highest
  redshifts where such massive structures exist.  The cluster
  selection is to first order independent of galaxy properties, making
  the SPT-SZ sample particularly well suited for cluster galaxy
  population studies.  We carried out a four-band imaging campaign
  with the {\it Hubble} and {\it Spitzer} Space Telescopes of the five
  $z\gtrsim 1.4$, S/N$_{SZE}>$5 clusters, that are among the rarest
  most massive clusters known at this redshift.  All five clusters
  show clear overdensities of red galaxies whose colors agree with the
  initial cluster redshift estimates, although one (SPT-CLJ0607-4448)
  shows a galaxy concentration much less prominent than the others.
  The highest redshift cluster in this sample, SPT-CLJ0459-4947 at
  $z\sim1.72$, is the most distant $M_{500}>10^{14}~M_{\odot}$ cluster
  discovered thus far through its intracluster medium, and is one of
  only three known clusters in this mass range at $z\gtrsim 1.7$,
  regardless of selection.  Based on $UVJ$-like photometric
  classification of quiescent and star-forming galaxies, we find that
  the quiescent fraction in the cluster central regions
  ($r/r_{500}<0.7$) is higher than in the field at the same redshift,
  with corresponding environmental quenching efficiencies typically in
  the range $\sim 0.5-0.8$ for stellar masses
  $\log(M/M_{\odot})>10.85$.  We have explored the impact of emission
  from star formation on the selection of this sample, concluding that
  all five clusters studied here would still have been detected with
  S/N$_{SZE}>$5, even if they had the same quiescent fraction as
  measured in the field. Our results thus point towards an efficient
  suppression of star formation in the central regions of the most
  massive clusters, occurring already earlier than $z \sim 1.5$. }

 
\maketitle

\section{Introduction}
\label{sec:intro}

The long-known environmental influences on galaxy population
properties observed at low and intermediate redshifts have often
motivated the study of galaxy evolution in galaxy clusters. Commonly
observed features -- such as the color-density and morphology-density
relations -- suggest a faster evolution of galaxies towards quiescent,
bulge-dominated systems in denser environments
\citep[e.g.,][]{dressler1980,tanaka2004,postman2005,cooper2006,poggianti2008,pannella2009b,peng2010,muzzin2012,mok2013,woo2013,kovac2014}. At
the center of the most massive haloes, cluster cores turn out to be
the most extreme regions of the Universe, where the evolution of
galaxies and thus their resulting properties are most biased by a
range of environmental effects \citep[e.g.,][]{moran2007}.

Indeed, cluster cores in the nearby Universe host the most massive
early-type galaxies, containing stars nearly as old as the Universe,
and producing the tight ``red sequence'' in the color-magnitude
diagram of cluster galaxies
\citep{visvanathan1977,bower1992,kodamaearimoto} that is often
considered to be a defining signature of high-density environments at
low and intermediate redshifts.  Most studies of the evolution of the
red sequence and of the cluster galaxy luminosity function up to
$z\sim1-1.3$ largely agree on a broad-brush picture where the
high-mass end of the cluster galaxy population is largely in place
even before redshift one, with the bulk of its stars formed in a
massive star formation event in the cluster progenitor environments at
$z\sim2$ or higher
\citep[e.g.,][]{depropris1999,depropris2007,andreon2006,andreon2013,strazzullo2006,strazzullo2010b,lin2006,lidman2008,mei2009,mancone2010,wylezalek2014,foltz2015},
followed by efficient suppression of star formation in a major part of
the massive galaxy population, creating a first red sequence
\citep[e.g.,][]{kodama2007,zirm2008,strazzullo2016}. Direct
observations of the star formation suppression in high-redshift
clusters add important constraints to this broad-brush picture
\citep[e.g.,][]{muzzin2014,balogh2016,noble2016,rudnick2017b}
concerning time scales, relevance, and actual nature of environmental
effects \citep[e.g.,][and references
  therein]{wetzel2012,wetzel2013,hirschmann2014,bahe2017}. In this
respect, observations of the onset of star formation suppression in
very distant clusters clearly provide a strong leverage on the
environmental effects most relevant at early times.

Over the last decade, cluster surveys have pushed the high-redshift cluster
frontier well beyond $z\sim1$ and into the $z\sim2$ regime, bridging
the cluster and proto-cluster realms
\citep[e.g.,][]{andreon2009,henry2010,gobat2011,spitler2012,stanford2012,zeimann2012,yuan2014,newman2014,wang2016,mantz2017}.
Cluster galaxy studies have thus started approaching the expected main
epoch of star formation, and indeed have revealed significant star
formation, nuclear and merging activity in clusters at $z \gtrsim1.4$,
with star-forming galaxy fractions sometimes approaching or
even exceeding the field levels, suggesting a rapidly decreasing
impact of environmental quenching at this cosmic time
\citep[e.g.,][]{hilton2010, tran2010,tran2015,hayashi2010,
  hayashi2011,santos2011,stanford2012,zeimann2012, brodwin2013,
  bayliss2014,santos2015,wang2016,alberts2016,wagner2017,nantais2017}.

On the other hand, passively evolving galaxies with typically
early-type morphology are often found in $z\gtrsim 1.4$ clusters and
even up to $z\sim2$ \citep[e.g.,][]{kurk2009, papovich2010,
  strazzullo2010b, strazzullo2013, tanaka2012, tanaka2013,snyder2012,
  spitler2012,newman2014, cooke2016}, although their predominance,
even at high stellar masses, is not necessarily as high as at lower
redshifts. The highest redshift clusters in particular, close to
$z\sim2$, often host a mixed massive galaxy population including both
very active and quenching or quiescent systems
\citep[e.g.,][]{kurk2009,tanaka2013,strazzullo2016,hatch2017},
although in some cases quiescent galaxies already heavily dominate the
massive population, forming a tight, well defined red sequence even at
very early times \citep[e.g.,][]{andreon2011,andreon2014,newman2014}.
Even considering only the most massive among the very distant
clusters, which may be expected to also host the most evolved galaxy
populations, a complex picture has emerged, where star formation is
already efficiently suppressed in the central regions of some clusters
\citep{strazzullo2010b,newman2014}, while it is still ongoing at
significant rates in others \citep[][see more detailed discussion in
  Sect.~\ref{sec:conclusions}]{santos2015}.

The variety of results described above occur at a redshift where
cluster-to-cluster variations likely start to become significant,
cluster samples are usually of very small size and are selected with a
variety of different methods, and the study of cluster galaxies is
complicated by observational difficulties and selection effects.
Depending on galaxy sample selection and observations, one may
highlight different characteristics of galaxy populations
\citep[e.g.,][]{tran2010, tran2015, smail2014}. Furthermore, galaxy
population properties might possibly exhibit a dependence on cluster
mass or assembly history. Therefore, galaxy- vs.\ intracluster medium
(ICM)-selected cluster samples might, for instance, suggest seemingly
inconsistent results, which are in fact due to specific aspects of
galaxy evolution in dense environments, to first order related to the
different cluster masses typical of the differently selected samples
\citep[e.g.,][]{culverhouse2010}. Poor statistics and concerns about
possible biases associated with cluster selection thus still challenge
our understanding of cluster galaxy populations at these redshifts.
  
Historically, X-ray cluster searches have provided the optimal
selection of cluster samples with a well-understood cluster selection
function, no direct bias with respect to the properties of cluster
galaxies, and the availability of cluster mass estimates. Such mass
estimates also crucially provide cluster scale radii for proper
comparison of properties with a radial dependence, such as galaxy
population properties.  However, $z\sim1.5$ is close to the limit
where current X-ray satellites are able to detect clusters. Only a few
of the known $z \gtrsim 1.4$ clusters are X-ray selected
\citep{mullis2005,stanford2006, henry2010,
  fassbender2011,santos2011,mantz2017}; most of the very distant
clusters have been identified instead through their galaxies. However,
cluster selection based on galaxies is by definition biased with
respect to galaxy population studies, to a greater
\citep[e.g.,][]{andreon2009,spitler2012} or lesser extent
\citep[e.g.,][]{eisenhardt2004, eisenhardt2008, papovich2008,
  papovich2010, gobat2011, stanford2012, zeimann2012, muzzin2013a}
depending on the actual selection criteria adopted. This is especially
true at a redshift where galaxy population properties may more
significantly depend on the dynamical state and/or mass of the host
halo.

In this work, we carry out an investigation of early environmental
effects on galaxy populations in a sample of the most massive clusters
at $z\sim1.5$, selected through the Sunyaev Zel'dovich effect
\citep[SZE;][]{sunyaevzeldovich1972} in the South Pole Telescope
\citep[SPT;][]{carlstrom2011} mm-wave survey over 2500 deg$^{2}$
\citep[SPT-SZ;][]{bleem2015}.  The cluster SZE signature or
signal-to-noise (S/N) is related to the total thermal energy in the
ICM, resulting in a relatively low scatter ($\sim$20\%) in cluster
mass at fixed SZE signature and redshift
\citep{andersson2011,bocquet2015}.  Moreover, the mapping from SZE
signature to mass has only a weak redshift dependence
\citep[e.g.,][]{deHaan16}.  Therefore, the SPT-SZ cluster sample can
be considered to a first approximation to be a mass-selected sample,
whose selection is independent of both the redshift and the properties
of the cluster galaxy population.

For this analysis, we start with the optically confirmed cluster
sample from the SPT-SZ survey associated with SZE detections having
S/N$>$4.5. These clusters have measured photometric redshifts accurate
to $\delta z/(1+z)\lesssim 0.02-0.04$ up to $z\sim1.5$
\citep{bleem2015}.  The purity of the original SZE--only candidate
catalog, as estimated through simulations and confirmed by optical/NIR
follow-up observations, is 95\% for the S/N$>$5 sample, and 75\% for
the S/N$>$4.5 sample \citep{song2012b}.  Confirmation through
optical/NIR follow-up effectively removes the noise fluctuations
responsible for the contamination in the candidate cluster catalog.
About 40 optically confirmed SPT-SZ clusters lie at $z>1$, and a tail
of five S/N$>$5 systems are at $z>1.4$.  We focus here on this highest
redshift tail, a representative sample of the most massive, collapsed
structures at $z\gtrsim1.4$ selected over 2500~deg$^{2}$.

The (negative) SZE signatures of our cluster sample are contaminated
at some level by mm-wave emission from galaxies and AGN.  Because we
empirically calibrate this observed signature directly to halo mass,
the effects of this contamination are already reflected in the
resulting mass--observable relation, which is characterized by an
amplitude, power law trends in mass and redshift, and the amplitude of
the intrinsic scatter in the observable at fixed mass and redshift.  This
empirically calibrated mass--observable relation indicates that the
cluster mass threshold of the SPT-SZ sample is M$_{500} \sim 3
\times10^{14}$M$_{\odot}$ from $z\sim0.2$ out to the highest redshifts
where such massive structures exist
\citep{bleem2015,bocquet2015,deHaan16}.  We estimate the completeness
of the $z\gtrsim1.4$, S/N$>$5 cluster sample studied here to be 70\%
above the mass of our least massive cluster
($M_{500}=2.74\times10^{14}M_\odot$).  The completeness above the mass
corresponding to S/N$=$5 at $z=1.4$
($M_{500}=2.65\times10^{14}M_\odot$) is 63\%.

Individual clusters with higher contamination from galaxy and/or AGN
emission and that lie close to the selection threshold could drop out
of the sample.  However, high frequency cluster radio AGN are rare
\citep{lin07}.  Although studies of the cluster radio AGN population
out to redshifts $z\sim1$ are ongoing, \citet{gupta2017} have already
characterized the high frequency cluster radio AGN population in an
X-ray selected local cluster sample.  Assuming a relatively strong
redshift evolution scenario, they estimate that no more than
$\sim$10\% of SZE selected clusters at $z\sim 1.5$ would fall out of a
pristine S/N$>$4.5 sample due to cluster radio AGN
contamination. Similarly, strong emission from star formation could
impact the SZE detection, making it more difficult to select clusters
characterized by higher star formation rates.  We explore this effect
in Section~\ref{sec:samplebias}, and we conclude that the cluster
sample studied here would not be significantly impacted even if the
star-forming galaxy fraction and star formation rates of cluster
galaxies were the same as in the field at the cluster redshift (i.e.,
even if the environmental quenching efficiency were negligible).

We present the main properties of this cluster sample in
Section~\ref{sec:data} (see Table~\ref{tab:sample}).  These five
M$_{500} \sim3 \times10^{14}$M$_{\odot}$ (M$_{200} \sim 5
\times10^{14}$M$_{\odot}$) SPT-SZ clusters that we study here are
among the few known examples of the rarest, first massive clusters to
have formed
\citep{mullis2005,rosati2009,andreon2009,stanford2012,brodwin2012,bayliss2014,tozzi2015}.
They are thus the likely progenitors of the most massive clusters in
the nearby Universe.

\begin{table*}[htbp!]
\caption{The cluster sample studied in this work. Columns 1, 2, 3, 4
  list the cluster name, coordinates, S/N of the SZE detection, and
  redshift used for the selection
  of this cluster sample, all from \citet{bleem2015}. Column 5 lists
  the cluster mass estimated as described in
  Sect.~\ref{sec:data}. Columns 6 and 7 list the cluster photometric
  redshifts as derived in Sect.~\ref{sec:redshifts}, and the adopted
  values in the analysis presented here.
  Column 8 lists the
  estimated $r_{500}$.
  \label{tab:sample}}
\begin{threeparttable}
  \vspace{0.2cm}
   \begin{tabular}{c c c c c c c c}
     \toprule
cluster & coordinates & $\xi_{\small{SPT}}$ & selection  & M$_{500,c}$ & photo-z & redshift used & r$_{500}$	\vspace{0.1cm}\\
\multicolumn{3}{c}{\footnotesize{}}  & \multicolumn{1}{c}{redshift}  & \multicolumn{1}{c}{[10$^{14} $M$_{\odot}]$ }   & \multicolumn{1}{c}{\footnotesize{(this work)}} & \multicolumn{1}{c}{\footnotesize{(in this work)}}  & \multicolumn{1}{c}{[Mpc] }   \\
\midrule
SPT-CLJ0421-4845 & $04^h21^m16.9^s,  -48\degree 45'40''$ & 5.8 & $1.42\pm 0.09$       & $2.90^{+0.65}_{-0.72}$ & $1.38^{+0.02}_{-0.02}$  & $1.38$                     & 0.60$\pm$0.04 \vspace{0.1cm}\\
SPT-CLJ0607-4448 & $06^h07^m35.6^s,  -44\degree 48'12''$ & 6.4 & $1.43\pm 0.09$       & $3.28^{+0.76}_{-0.75}$& $1.38^{+0.02}_{-0.02}$   & $1.401^b$ & 0.62$\pm$0.04   \vspace{0.1cm}\\
SPT-CLJ2040-4451 & $20^h40^m59.6^s,  -44\degree 51'37''$ & 6.7 & 1.478$^a$       & $3.44^{+0.75}_{-0.80}$ & $1.47^{+0.02}_{-0.03}$  & $1.478^a$ & 0.61$\pm$0.04 \vspace{0.1cm}\\
SPT-CLJ0446-4606 & $04^h46^m 55.8^s, -46\degree 06'04''$ & 5.7 & $\geq 1.5$           & $2.74^{+0.65}_{-0.69}$ & $1.52^{+0.13}_{-0.02}$  & $1.52$                 & 0.56$\pm$0.04  \vspace{0.1cm}\\
SPT-CLJ0459-4947 & $04^h59^m42.5^s   -49\degree 47'14''$ & 6.3 & $\geq 1.5$           & $2.85^{+0.64}_{-0.68}$ & $1.80^{+0.10}_{-0.19}$  & $1.72^c$   & 0.53$\pm$0.04  \vspace{0.1cm}\\
\bottomrule
   \end{tabular}
   \begin{tablenotes}
   \item[$a$] Spectroscopic redshift for SPT-CLJ2040  \citep[$z=1.478\pm0.003$;][]{bayliss2014}.
   \item[$b$] Spectroscopic redshift for SPT-CLJ0607   \citep[$z=1.401\pm0.003$;][]{khullar2018}.
   \item[$c$] Best redshift constraint currently available for
     SPT-CLJ0459 ($z=1.72\pm0.02$; Mantz et al. in prep, see
     Sect.~\ref{sec:redshifts}).
   \end{tablenotes}
     \end{threeparttable}
\end{table*}

In this paper we focus on cluster redshift constraints, the red galaxy
population, and quiescent galaxy fractions in the central cluster
regions within $r_{500}$\footnote{Overdensity radii $r_{500}$ and
  $r_{200}$ are the clustercentric radii within which the mean density
  is 500 and 200 times, respectively, the critical density of the
  Universe at the cluster redshift. The cluster masses M$_{500}$ and
  M$_{200}$ reported in the following refer to the mass within these
  radii.}, as determined from new observations from a dedicated,
homogeneous imaging follow-up of the full sample with the {\it Hubble}
({\it HST}) and {\it Spitzer} Space Telescopes.  This follow-up
program was designed to meaningfully constrain main galaxy population
properties with a minimum observational effort, acquiring imaging in
only four passbands chosen to enable the selection of a candidate
member sample, to allow a broad statistical separation of quiescent
and star-forming sources, to provide measurements of galaxy stellar
masses and structural properties, and to constrain the cluster
redshift.  Forthcoming papers based on the cluster sample and data set
used here will present the investigation of structural vs.\ stellar
population properties of cluster vs.\ field galaxies and their structural
evolution, mergers in massive cluster vs.\ field environments at
$z\sim1.5$, galaxy stellar mass functions (on 3.6$\mu$m-selected
samples), galaxy number density profiles, cluster stellar mass
fractions and the halo occupation distribution.

We adopt a flat $\Lambda$CDM cosmological model with $\Omega_{M}$=0.3,
and H$_{0}$=70~km~s$^{-1}$~Mpc$^{-1}$. A \citet{salpeter1955} initial
mass function (IMF) is assumed throughout.  Magnitudes are quoted in
the AB system.

\section{Data, photometry, and  measurements}
\label{sec:data}

We selected the cluster sample used in this work from the
\citet{bleem2015} SPT-SZ cluster catalog, taking all clusters with a
photometric (except for SPT-CLJ2040, see Table~\ref{tab:sample})
redshift $z>1.4$ and an SZE significance
S/N$>$5. Table~\ref{tab:sample} summarizes the main properties of the
clusters. Cluster names are shortened to SPT-CLJxxxx hereafter.

Because of the photometric redshift uncertainties, our $z>1.4$
selected cluster sample had some associated ambiguities.  Indeed, out
of the seven S/N$>$5 clusters in the \citet{bleem2015} catalog at
$1.2\le z\le1.4$
\footnote{The $z=1.2$ limit is 2-4$\sigma$ below our $z>1.4$ threshold
  given the estimated photometric redshift uncertainties.}, five are
now spectroscopically confirmed, and one turns out to be at $z\geq
1.4$ \citep{stalder2013,khullar2018}. The two remaining clusters have
photometric redshifts of 1.23 and 1.30. Therefore, the sample studied
in this work contains five of the six S/N$>$5 SPT-SZ clusters deemed
to be at $z\geq1.4$, and the possibility that we are missing any
significant number of other $z>1.4$ clusters is small, given the
results of the recent spectroscopic follow-up \citep{khullar2018}. We
thus consider the sample studied here to be representative of the
$z\geq1.4$ massive cluster population as selected from the SPT-SZ
S/N$>$5 catalog.

In Table~\ref{tab:sample} we present masses M$_{500}$ and associated
radii $r_{500}$ for each cluster, derived from the cluster SPT-SZ
observable and redshift using the latest empirical calibration of the
SPT-SZ mass--observable relation.  For this purpose, we adopt the best
fit scaling relation parameters from \citet[][see their Table
  3]{deHaan16}. Specifically, we consider the
SPT$_{\textrm{CL}}$+H$_{0}$+BBN data set, a combination of the SPT-SZ
S/N$>$5 cluster sample, X-ray $Y_\mathrm{X}$ based mass estimates for
82 of those clusters calibrated externally through weak lensing
\citep{vikhlinin09,hoekstra15}, and external priors on $H_0$
\citep{riess11} and $\Omega_\text{b} h^2$ \citep{cooke14}.  Our mass
estimates include corrections for the Eddington bias, and the mass
uncertainties that we present correspond to the sum in quadrature of
two components.  The first ``systematic'' component, corresponding to
a $\sim15$\% uncertainty, reflects the current uncertainty on the
mass-observable scaling relation parameters, which is due to
cosmological parameter uncertainties and the limitations of the
current direct mass calibration dataset.  The second ``statistical''
component, corresponding to $\sim$20\% for a S/N$=$5 cluster, is the
combination of the measurement uncertainty of the SZE S/N and the
intrinsic scatter of the underlying SZE signature at fixed mass and
redshift.  For a more complete discussion of the mass calibration of
the SPT-SZ sample and of its impact on individual cluster mass
estimates, we refer the reader to recent studies of the baryonic
components and X-ray properties of SPT-SZ clusters
\citep{chiu18,bulbul18}.  As alluded to previously, the cluster masses
M$_{500}$ ranging from 2.7 to 3.4$\times10^{14} M_\odot$ (with
corresponding M$_{200}$ ranging from 4.5 to 5.6$\times10^{14}M_\odot$)
make these clusters among the most massive systems identified to date
at $z\sim1.5$.

All clusters in this sample have been homogeneously observed in a
dedicated follow-up program with {\it HST} and {\it Spitzer}.  We
describe below these data, their reduction and the derived photometric
measurements, as well as data from part of the GOODS-S survey used as
a control field.

\subsection{{\it HST} observations, data reduction and photometry}

{\it HST} observations with the Advanced Camera for Surveys (ACS) in
the F814W band ($\sim$4800~s for each cluster), and with the {\it Wide
  Field Camera 3} (WFC3) in the F140W band ($\sim$2400~s per cluster)
were acquired in Cycle 23 (GO~14252, PI: Strazzullo). The exception is
the F140W band imaging of cluster SPT-CLJ2040, for which we used
observations ($\sim$9200~s) taken as part of program GO-14327
(hereafter {\it See Change}, PI: Perlmutter).  We used the DrizzlePac
release 2.1.0 to produce science ready images with standard procedures
from the preprocessed flat-fielded single exposure frames retrieved
from the STScI archive. More specifically we used AstroDrizzle
(v. 2.1.11) to subtract the background, perform cosmic-ray removal and
drizzle all frames to a common astrometric solution, with a square
kernel and a PIXFRAC=0.8, before combining them in a final stacked
image with a pixel scale of 0.06". The tasks Tweakreg and Tweakback
were used to register images in the different bands to the same sky
coordinates and remove some residual astrometric offsets.

Source extraction and photometry were carried out with SExtractor
\citep{sextractor} in double image mode with detection performed on
the F140W image. With the aim of removing stars, point-like sources
identified from SExtractor's MAG\_AUTO vs. FLUX\_RADIUS sequence were
removed down to a F140W band magnitude m140=22 AB mag. This selection
is purely based on a morphological criterion, thus unresolved
non-stellar sources, like very bright AGNs or very compact galaxies,
might be selected as point-like sources as well. Out of a total of 120
point-like sources removed across all five cluster fields, the colors
of 116 sources ($\sim$97\%) are not compatible with those of m140$<$22
galaxies at any redshift $z>0.6$. The stellar nature is potentially
dubious for only $\sim$3\% of the removed sources. We thus estimate
that the contamination of our point-like source sample from
non-stellar objects is at the few percent level at most.

Galactic extinction correction for each field was applied according to
\citet{schlafly2011}. We adopted SExtractor MAG\_AUTO as an estimate
of total magnitude, while m814-m140 colors were measured from 1''
(diameter) aperture magnitudes in the F814W and F140W bands, applying
an aperture correction for the different PSF between the two bands
determined using growth curves of bright unsaturated point like
sources in the image. Realistic depths and errors on the aperture
magnitudes were estimated by measuring the flux rms in 1'' apertures
placed at random locations in the image. We also empirically estimated
the completeness of the F140W band data by comparing number counts and
unmatched sources between the full-depth ($\sim$9200~s) F140W band
image of the cluster SPT-CLJ2040 from the {\it See Change} program,
and a reduced-depth image obtained by coadding exposures to the same
exposure time of the other clusters observed in our PID~14252
program. Based on this estimate, F140W band catalogs are $>95\%$
complete in the magnitude range used in this work (m140$<$24 AB mag
for all clusters).

\subsection{{\it Spitzer} observations, data reduction and photometry}
\label{sec:spitzerobs}
{\it Spitzer} observations with the Infrared Array Camera
\citep[IRAC,][]{fazio2004} were carried out in Cycle 12 (PID\,12030,
      PI: Strazzullo).  Each cluster was
      observed for 5500\,s in both the 3.6 and 4.5\,$\mu$m bands,
      except SPT-CLJ2040, which lies in a region of higher background.
      SPT-CLJ2040 was therefore observed with a total integration time
      of 7500\,s in each IRAC band.  All integrations consisted of
      overlapping full-array exposures with 100\,s frame times,
      dithered using a cycling pattern with a medium throw.  The
      resulting coverage pattern generated by the 5\farcm12-wide IRAC
      field of view combined with the dithering was sufficient to
      completely cover all the observed clusters.

The IRAC exposures were reduced using standard procedures following
steps adopted for other similar science targets
\citep{bleem2015,paternomahler2017}, adjusting for the relatively high
sensitivity of the PID\,12030 observations.  The reduction was based
on the IRAC corrected basic calibrated data (cBCD).  Median stacks for
all cBCD frames in each IRAC band and field were made after masking
bright sources.  These median stacks were subtracted from the
individual cBCD frames to eliminate residual images from prior
observations of bright sources, and to compensate for gradients in the
backgrounds of each field.  A custom column-pulldown corrector was
applied to fix the ubiquitous low-lying array columns that result from
observing sources close to saturation.  The resulting modified cBCD
frames were then combined into spatially registered mosaics using the
wrapper IRACproc \citep{schuster2006}.  With the high redundancy of
these IRAC observations (typically 55 overlapping exposures per
pixel), cosmic rays were automatically removed by outlier rejection
during mosaicking.  The final mosaics were generated with 0\farcs48
pixels and a tangent-plane projection set to match that of the {\it
  HST}/WFC3 F140W mosaics, in order to facilitate subsequent use of the
latter as priors for source extraction and to optimize coordinated
photometric measurements.

For the purpose of the work presented here, photometry was carried out
on the IRAC mosaics with {\tt T\_PHOT} \citep[see][for a detailed
  description]{merlin2015,merlin2016b}. We used priors from the WFC3
F140W band imaging down to m140=25.5\,AB\,mag, and IRAC point spread
functions generated separately for each field by stacking a few tens
of bright unsaturated stars. Given the relatively small fields studied
here, we adopted a {\it "single image mode"} fitting (fitting the
entire input catalog at once), providing the most robust results
especially in terms of uncertainty estimates and covariance between
close neighbors. We also tested a second pass {\it "dance"} run
allowing refinements of the input source centroid positions by
approximately one third of the IRAC point spread function, but the
retrieved fluxes, uncertainties and estimated covariances were
essentially unaffected with respect to the main run.  The estimated
effective sensitivity was very similar for all clusters, with an
estimated S/N$\sim$3 for point-like sources at $\sim$25.2\,AB\,mag in
both the 3.6 and 4.5\,$\mu$m mosaics. In the following we consider
that with this approach and observations we are not able to measure
reliable photometry for sources fainter than 25.2\,AB\,mag, as well as
for those with a {\tt T\_PHOT} {\it covariance index} $>0.85$
\citep[potentially contaminated by nearby sources to a significant
  level, see][] {merlin2016a}. This has a marginal effect on the
subsequent analysis, as discussed in Sects.~\ref{sec:samplesel},
\ref{sec:uvj}, \ref{sec:conclusions}. At $z\sim1.5$, a 3.6$\mu$m
magnitude of [3.6]$\sim$25.2~AB corresponds to stellar masses well
below $\sim10^{10}$M$_{\odot}$; given the mass completeness limits set
by the F140W-band selection as discussed in
Sects.~\ref{sec:masses},~\ref{sec:galpops}, our sources of interest
are expected to be generally detected at high S/N in the IRAC mosaics.

\subsection{Stellar mass estimates and control field}
\label{sec:massesandcontrolfield}

For the purpose of comparing cluster galaxy properties with field
counterparts, defining a control field for statistical background
subtraction, and estimating stellar masses for galaxies in our cluster
fields, we used photometry and derived properties of galaxies in the
CANDELS GOODS-S field \citep{grogin2011}. We used multiwavelength
photometry from \citet{guo2013} and photo-z's and stellar masses from
spectral energy distribution (SED) fitting from \citet{pannella2015}
and \citet{schreiber2015}. Because the available F140W imaging of the
GOODS-S field \citep{skelton2014} is shallower than that used in this
work, we opted for measuring a synthetic m140 magnitude for all
sources in the \citet{guo2013} catalog, by convolving the best-fit SED
of each source from \citet{schreiber2015} with the response curve of
the F140W filter \citep[see][]{strazzullo2016}. From the internal
comparison of analogously derived F160W synthetic magnitudes with the
observed ones, and the external comparison of our synthetic m140
magnitudes with the F140W photometry published in \citet{skelton2014},
we estimate the uncertainty on our synthetic m140 magnitudes, in the
magnitude range of interest for this work, to be $\lesssim$0.1 mag,
with an essentially negligible impact on the analyses presented here.

\subsubsection{Stellar masses}
\label{sec:masses}

This work is largely based on deep photometry in just four
bands. Since such photometric coverage is not ideally suited for a
full-fledged SED fitting approach, we estimated stellar masses by
converting 3.6$\mu$m flux to stellar mass with a mass-to-light ratio
(M/L) based on the m814-m140 color. We calibrated the M/L vs. color
relation on galaxies in the GOODS-S field at a redshift within
$\pm$0.15 from each cluster redshift, and in the same magnitude range
of our sample. Based on the scatter around the median M/L vs. color
relation for the selected GOODS-S galaxies, we estimate a typical
error on our stellar mass estimates of $\sim20-30\%$. This is the
internal uncertainty of the empirical mass calibration against the
stellar masses in the control field sample. It thus relies on the
assumptions adopted in estimating stellar masses for the control field
sample, and it does not represent an uncertainty on stellar mass on an
absolute scale.

In a minority of cases (see details in Sect.~\ref{sec:colselmembers}
below) we were not able to measure reliable 3.6$\mu$m fluxes (as noted
in Sect.~\ref{sec:spitzerobs} above), thus we estimated stellar masses
from the F140W band flux with a M/L calibrated on the m814-m140 color
(with a typical stellar mass uncertainty of $\sim30-40\%$ estimated as
described above). We need to resort to these F140W-scaled mass
estimates for only a very small fraction of the mass complete samples
discussed below, and we have verified that any small systematics of
these F140W-scaled mass estimates with respect to our default
3.6$\mu$m-scaled ones do not produce any appreciable effect on the
results presented in this work.

For the purpose of the analysis described in the following, the
stellar mass completeness limit adopted for each cluster is defined as
the stellar mass of a solar-metallicity, unattenuated \citet{bc03}
simple stellar population (SSP) with a formation redshift
$z_{f}\sim8$, having at the cluster redshift a F140W-band magnitude
equal to the limiting m140 adopted for the given cluster, as detailed
in Section \ref{sec:samplesel}.

\subsubsection{Control field}
\label{sec:controlfield}

The GOODS-S field catalogs described above (photometry, photometric
redshifts, SED fitting results including stellar masses and restframe
photometry) were also used for comparison with galaxy population
properties in the field at the cluster redshifts, and for the purpose
of statistical background subtraction. In fact, given the small field
probed by our observations (in particular the HST/WFC3 imaging), we
could not adopt a local control field for background estimation in the
vicinity of each cluster.  As mentioned in
Sect.~\ref{sec:massesandcontrolfield}, the GOODS-S observations at the
core of these measurements are not exactly the same as those in the
cluster fields.  However, both the cluster and field samples are used
in a regime where the adopted catalogs are complete and photometric
errors are small (typically $<0.1$~mag, or at most $0.2$~mag for F814W
magnitudes for the faintest red sources in our samples of
interest). The resulting small differences in the photometric and
stellar population parameter measurements used here are estimated to
be negligible with respect to the uncertainties involved, and thus not
relevant for the way these measurements are used in this work.

\begin{figure*}
 \includegraphics[height=0.26\textwidth,viewport= 56 428 557 721, clip]{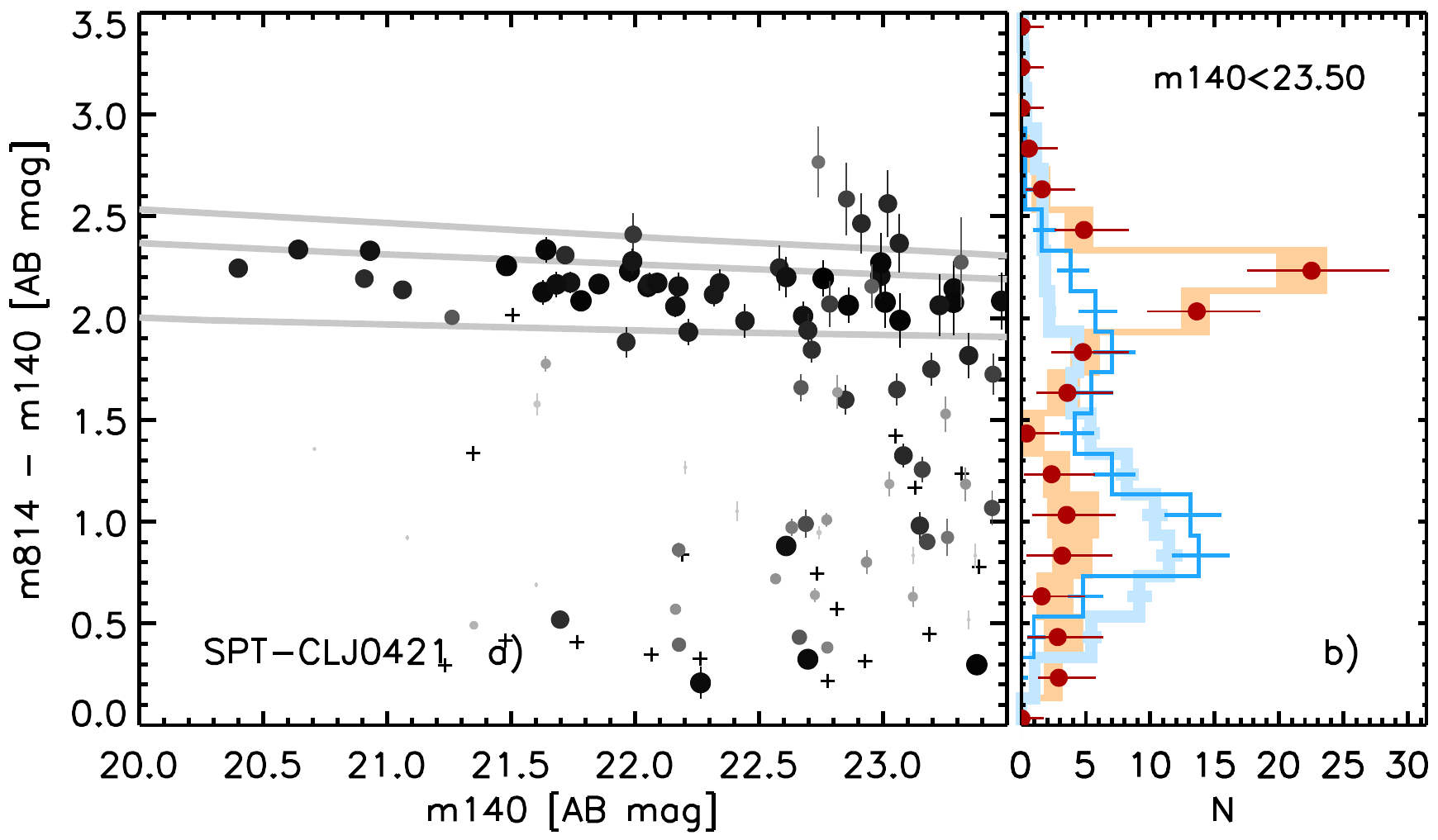}%
 \includegraphics[height=0.26\textwidth,viewport= 490 428 564 721, clip]{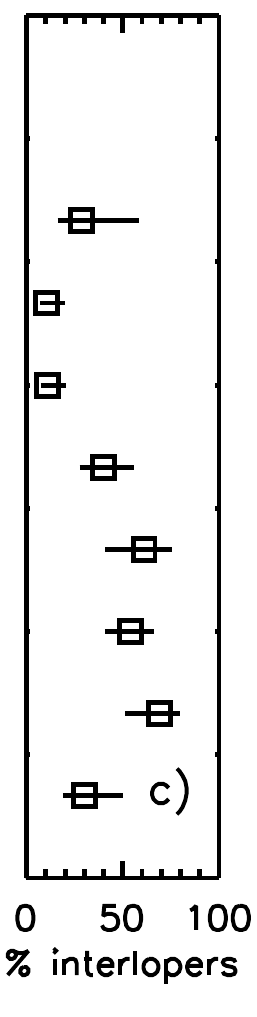}%
 \includegraphics[height=0.26\textwidth,viewport= 79 428 557 721, clip]{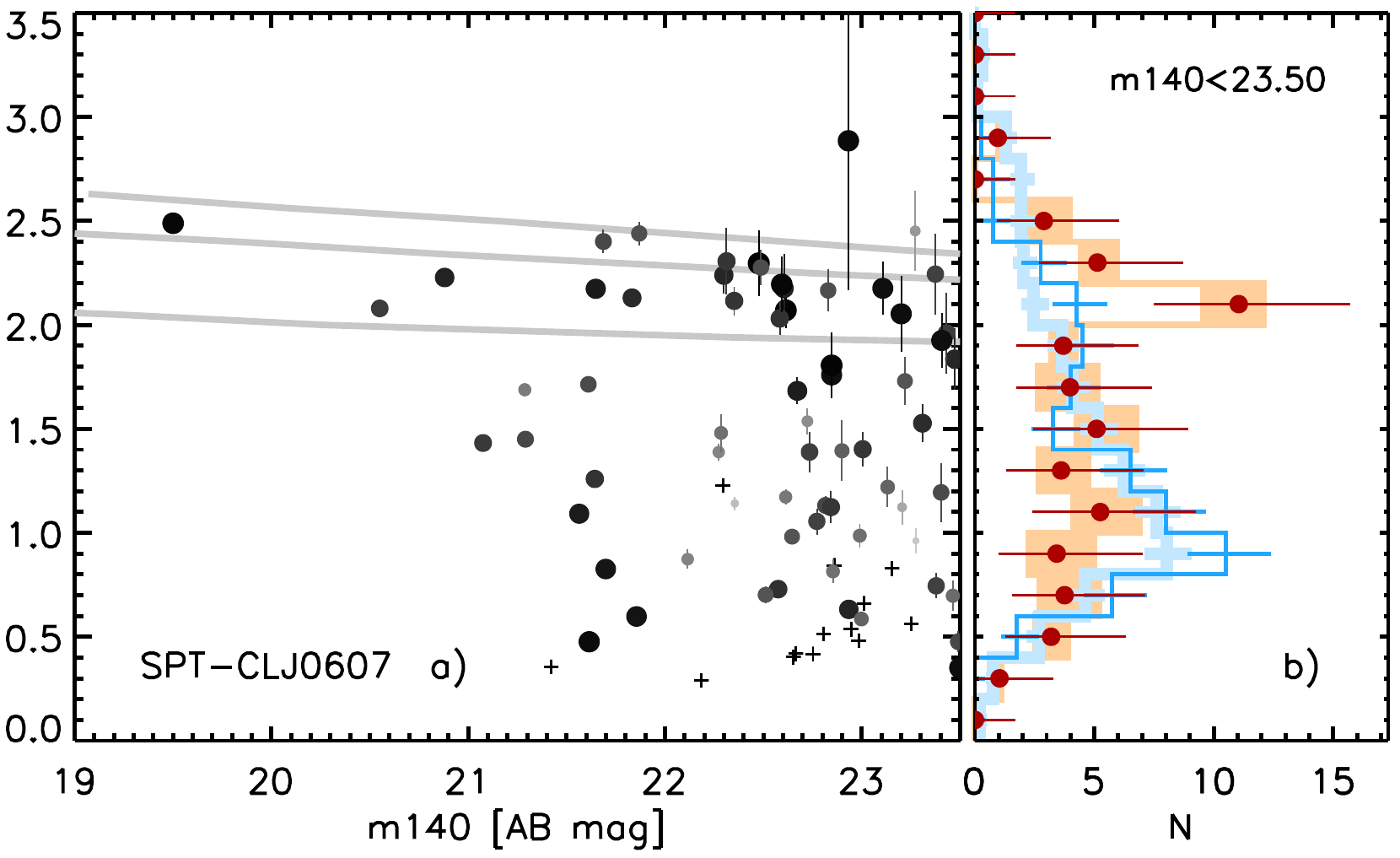}%
 \includegraphics[height=0.26\textwidth,viewport= 490 428 564 721, clip]{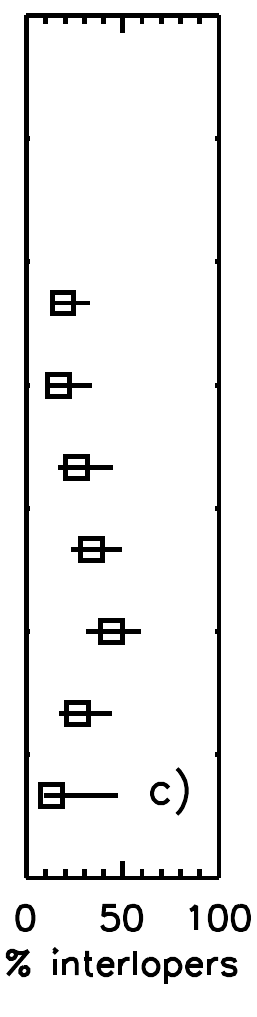}
 \includegraphics[height=0.26\textwidth,viewport= 56 428 557 721, clip]{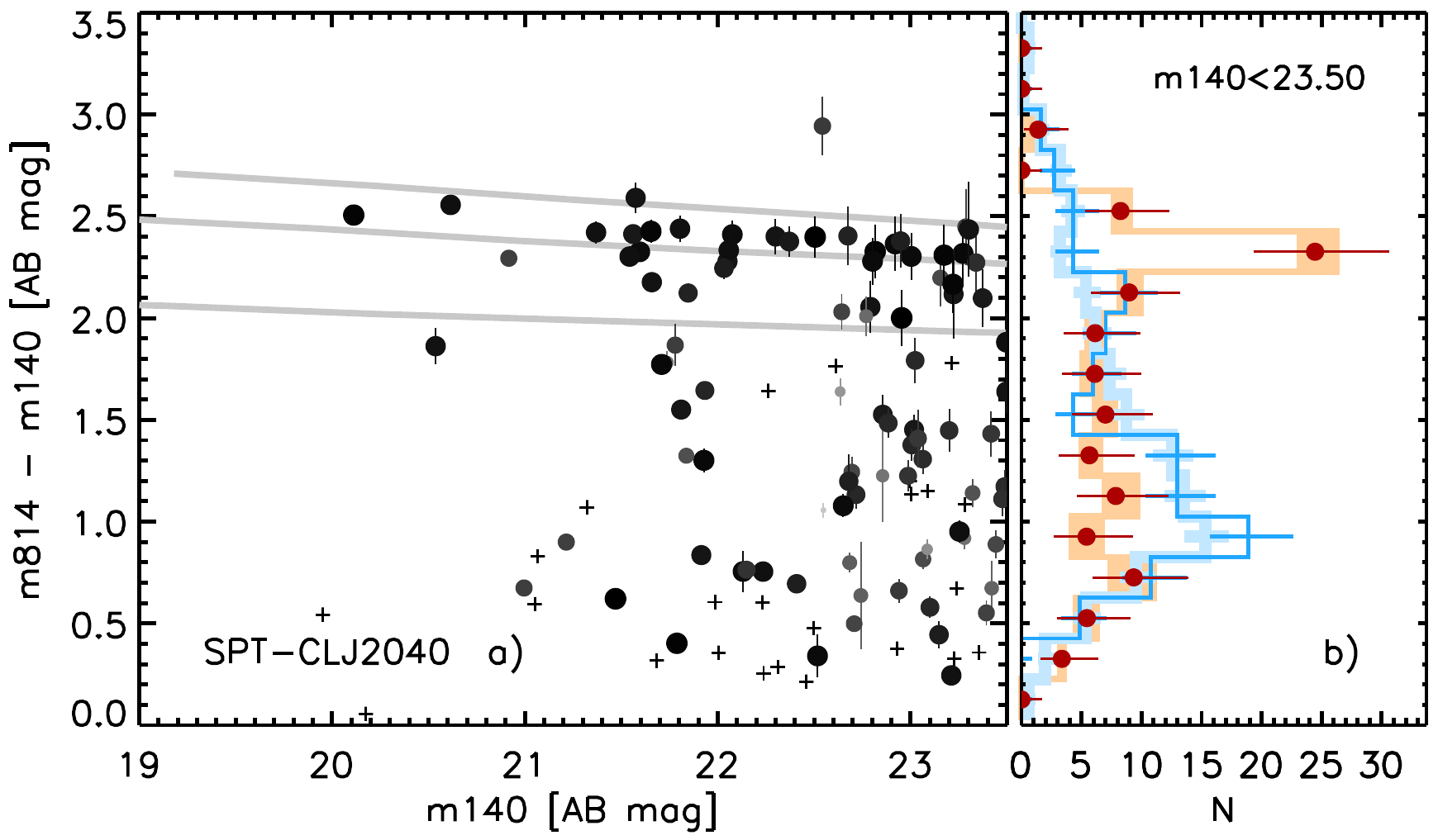}%
 \includegraphics[height=0.26\textwidth,viewport= 490 428 564 721, clip]{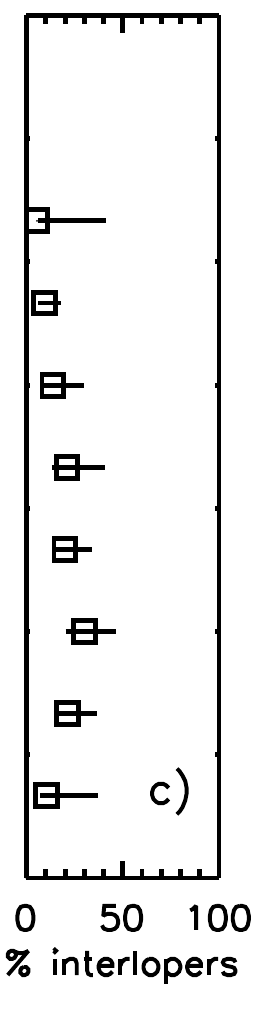}%
 \includegraphics[height=0.26\textwidth,viewport= 79 428 557 721, clip]{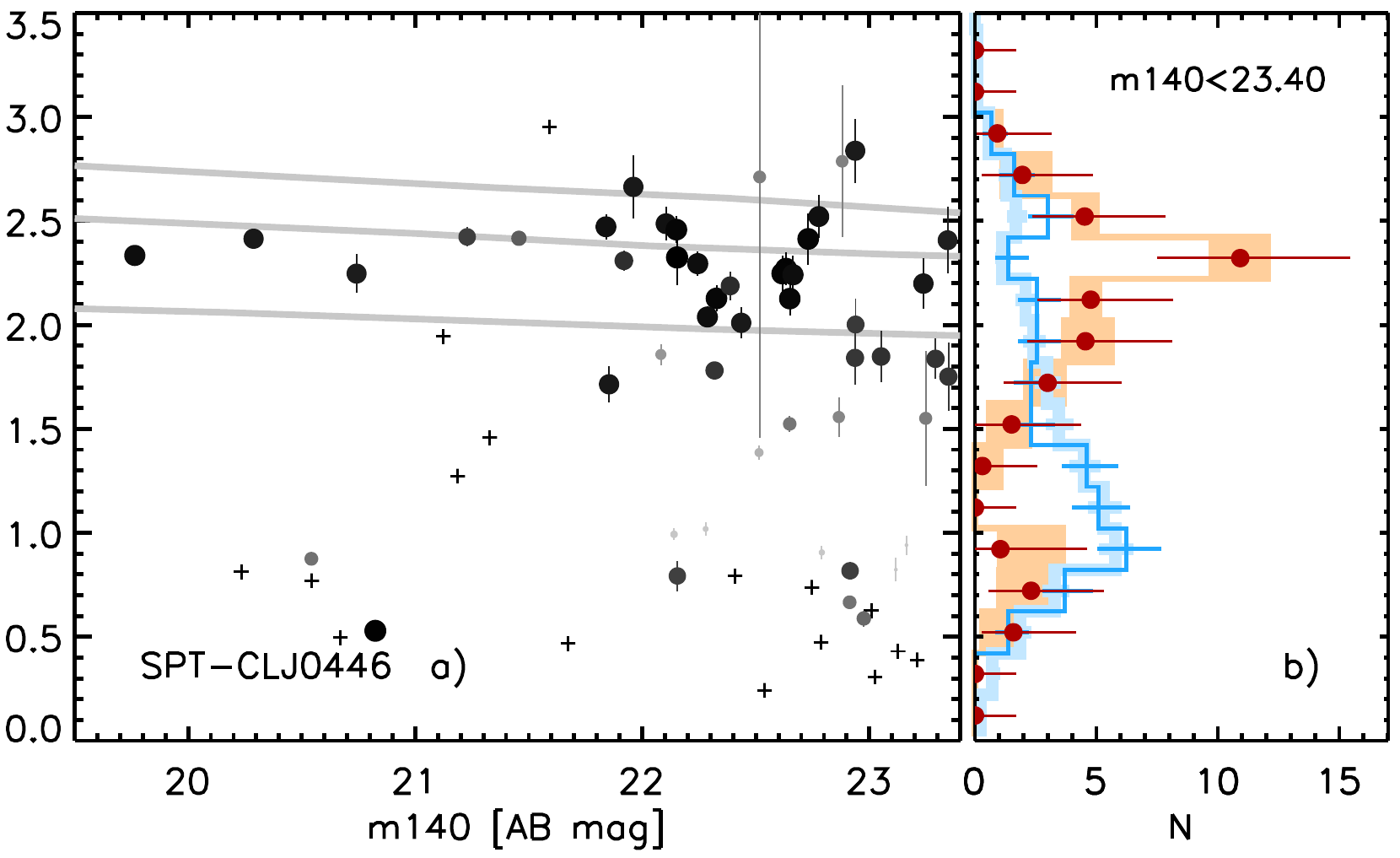}%
 \includegraphics[height=0.26\textwidth,viewport= 490 428 564 721, clip]{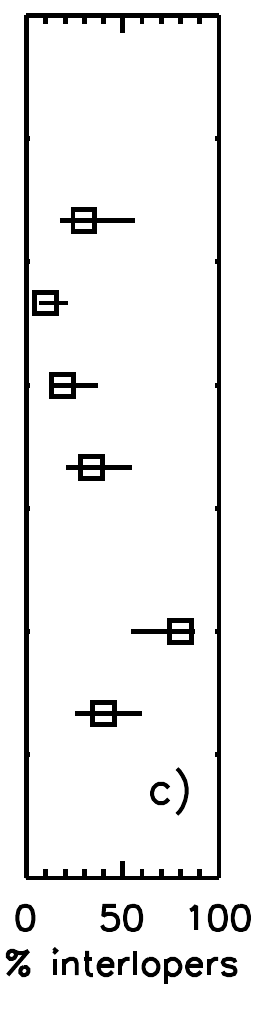}
 \includegraphics[height=0.26\textwidth,viewport= 56 428 557 721, clip]{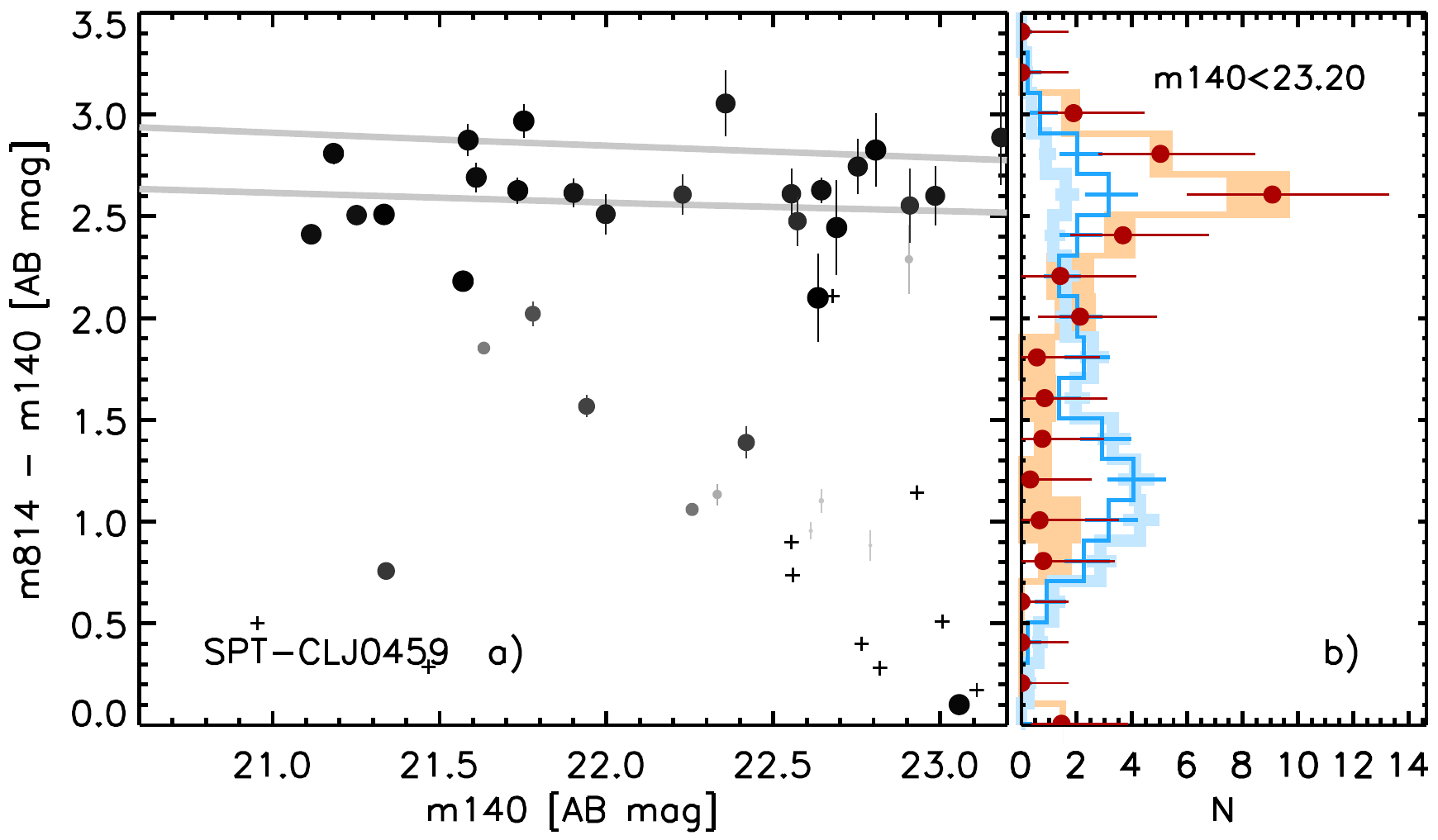}%
 \includegraphics[height=0.26\textwidth,viewport= 490 428 564 721, clip]{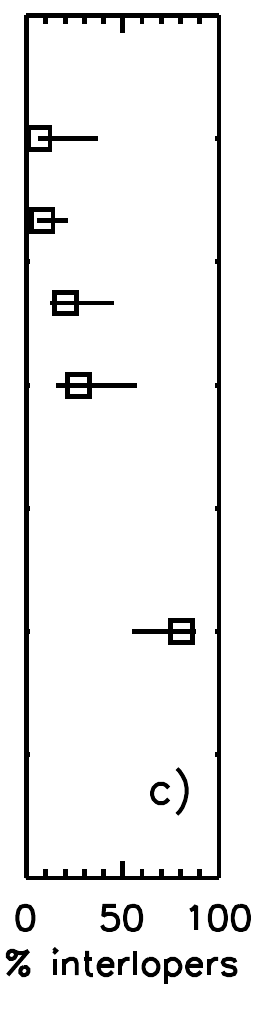}
 \includegraphics[height=0.26\textwidth,viewport= 100 505 557 715, clip]{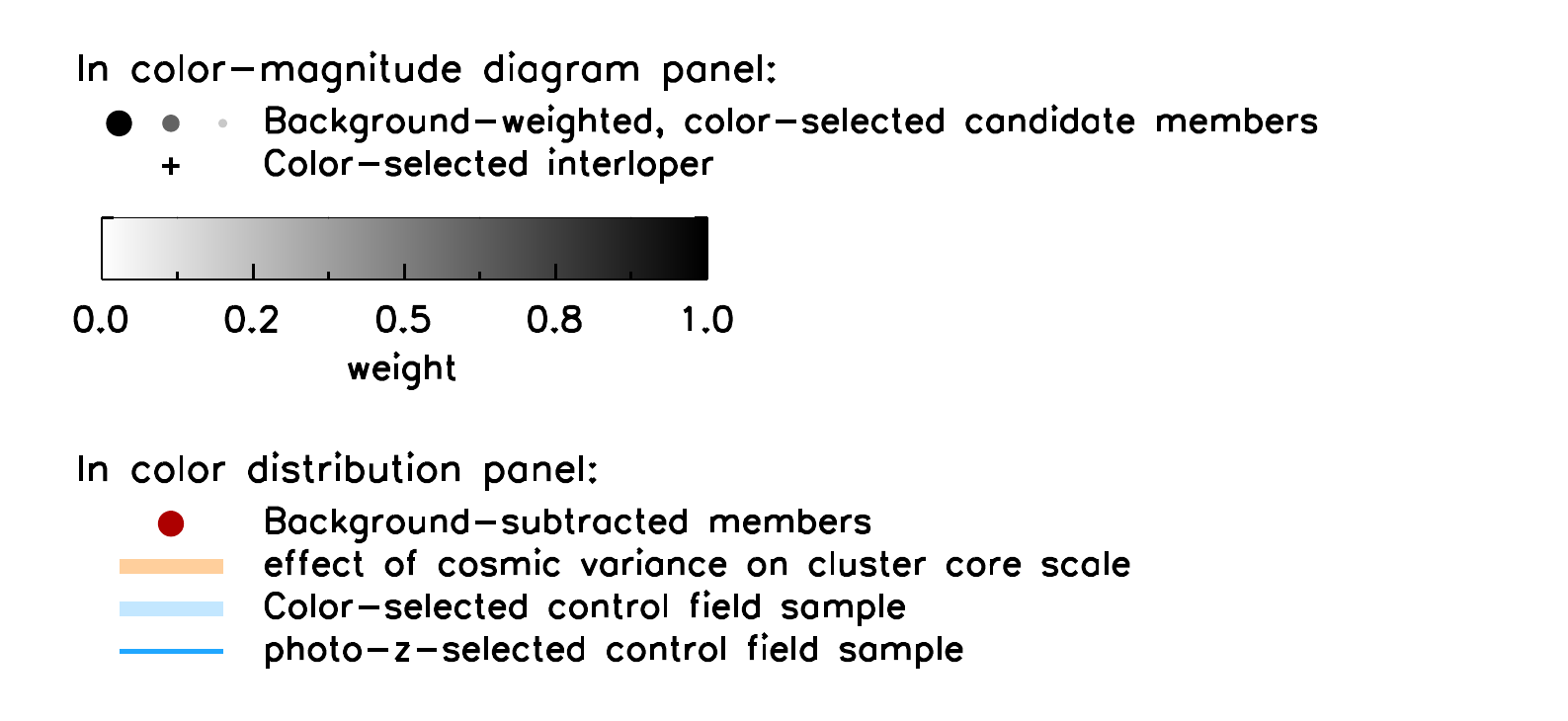}
 \caption{Color-magnitude diagrams and derived properties within the
   $r<0.7 r_{500}$ region of each cluster are shown in subfigures,
   each with three panels.  {\it Panel {\it a}) :} Visualization of
   the background-subtracted color-magnitude diagram, where size and
   color of each galaxy point scale according to its statistical
   background subtraction weight (see color bar in legend) determined
   in Sect.~\ref{sec:statsub}. All galaxies are shown, but color-rejected
   interlopers (see Sect.~\ref{sec:colselmembers}) are shown as
   crosses. Gray lines show \citet{kodamaearimoto} red-sequence (RS)
   models with formation redshifts $z_f =2, 3, 5$. The color range is
   the same for all clusters to facilitate direct comparison of RS
   colors.  {\it Panel {\it b}) : } Color distribution of
   background-subtracted and area-corrected cluster members (red
   points with error bars) down to the indicated m140 limit. The
   orange shaded area shows an estimate of the impact of cosmic
   variance on the scale of the cluster core field, as detailed in
   Sect.~\ref{sec:redpops}. The blue histograms show the color
   distribution (rescaled by total number of galaxies) in the
   control-field sample, using the same color selection as for cluster
   candidate members (light blue), or a photometric redshift selection
   within $\pm0.2$ of the cluster redshift (darker blue). All clusters
   show a clear excess of red galaxies with respect to the field
   distribution.  {\it Panel {\it c}) :} Estimated fraction of
   interlopers in the color-selected candidate member sample as a
   function of color (down to the indicated m140 limit), based on the
   weights in panel {\it a}. Contamination is low for RS galaxies but
   significant for blue galaxies.  Error bars show binomial confidence
   intervals (1$\sigma$) computed following \citet{cameron2011}.
  \label{fig:cmd}}
\end{figure*}

\section{Galaxies in the cluster fields}
\label{sec:samplesel}

\begin{figure}[]
\begin{center}
 \includegraphics[width=0.45\textwidth,viewport= 68 409 547 697,
   clip]{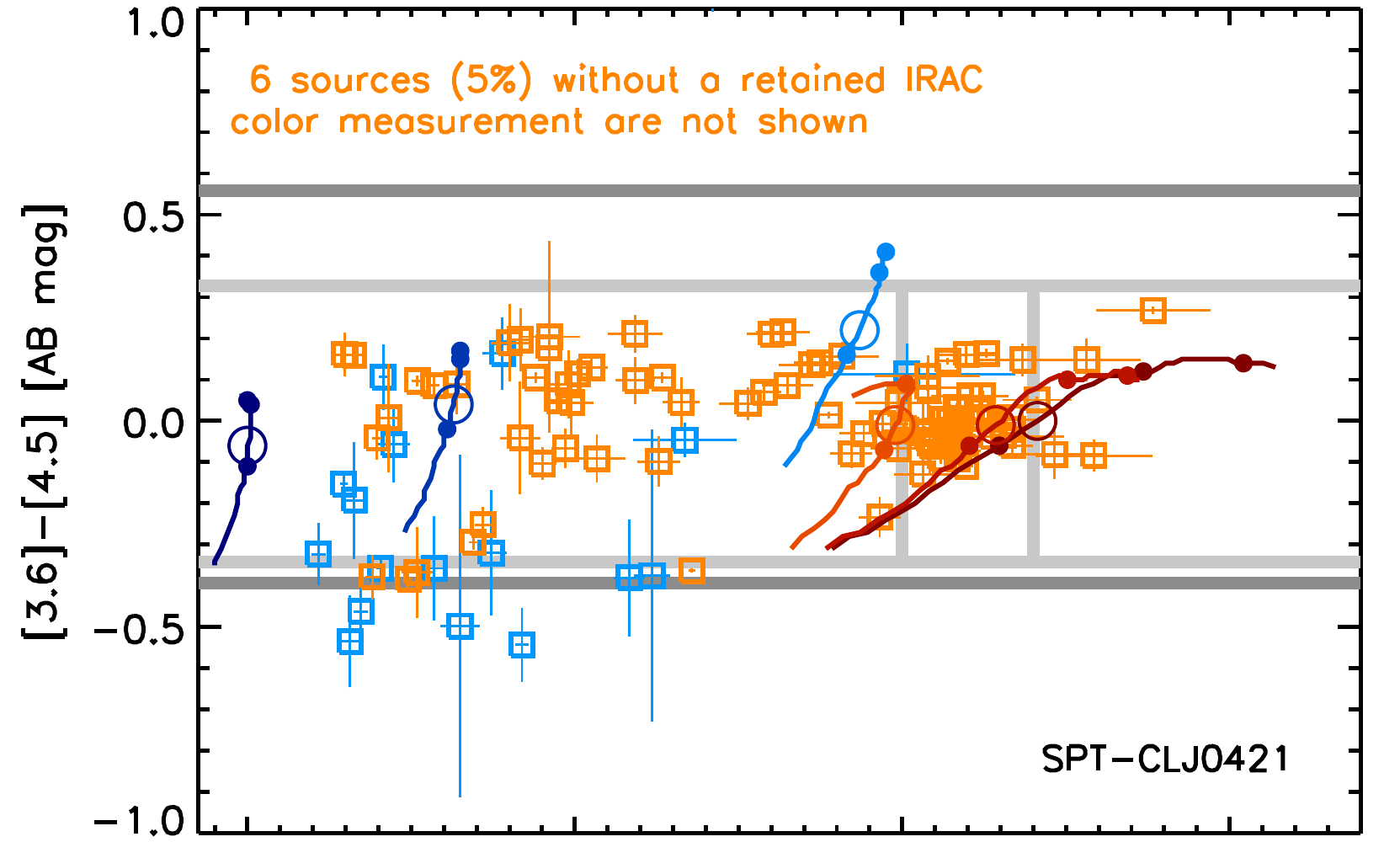}
 \includegraphics[width=0.45\textwidth,viewport= 68 368 547 697,
   clip]{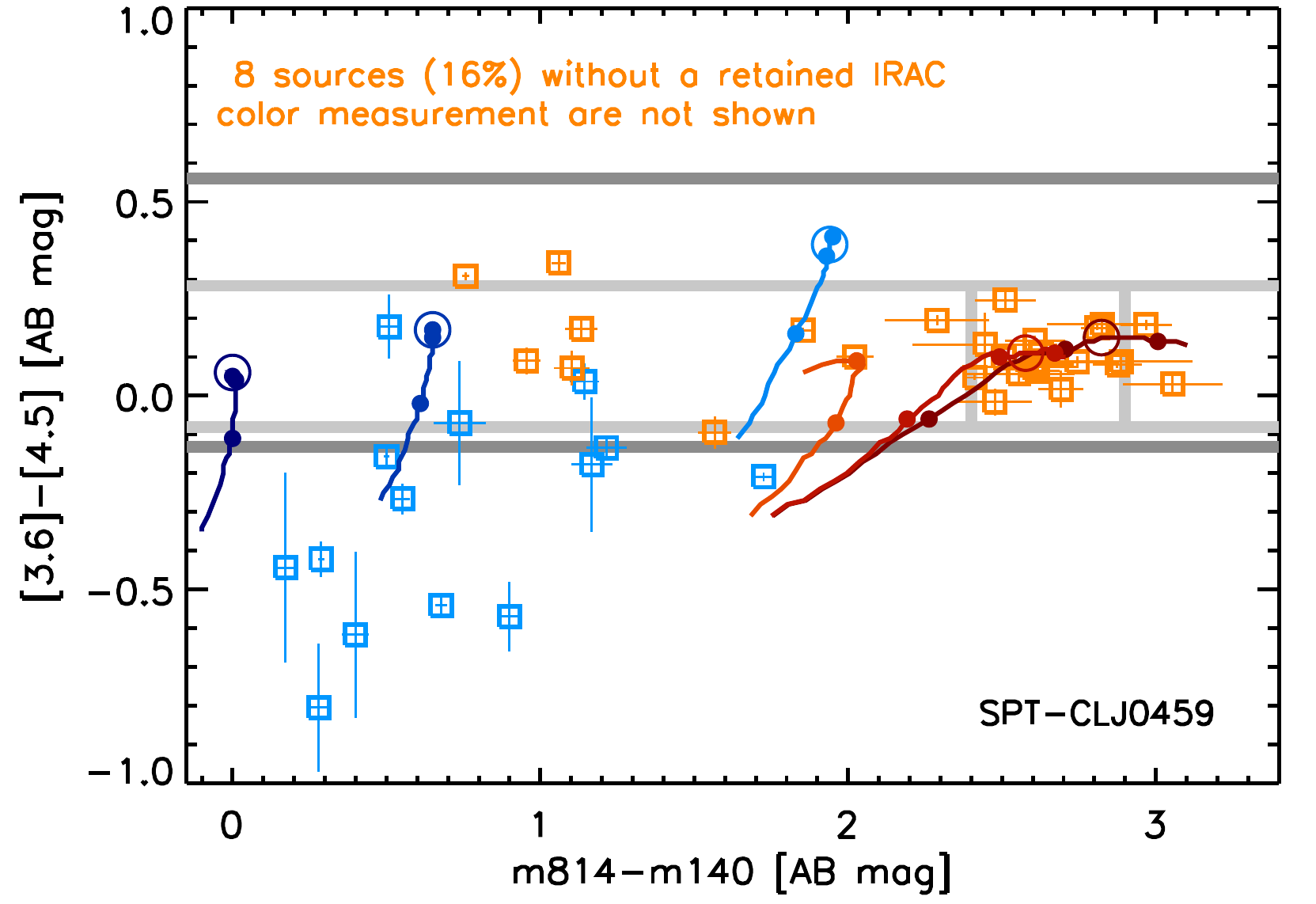}
\end{center}
\caption{Two examples of the selection of candidate cluster members
  (see Sect.~\ref{sec:colselmembers}).  Vertical gray lines bound the
  m814-m140 RS sample used for the initial definition of the IRAC
  color selection range, shown by the horizontal light-gray
  lines. Groups of red and blue tracks show, respectively, the colors
  as a function of redshift in the range $1<z<2$ of quiescent (SSPs
  with $z_{f}=2,3,5$) and actively star-forming (100~Myr old constant
  star formation, $A_{v}=0,1,3$) stellar populations, used to obtain
  the final adopted IRAC color selection range (dark-gray lines, see
  Sect.~\ref{sec:colselmembers} for full details). Small solid points
  along the tracks indicate z=1.3, 1.6, 1.9; the large circle
  indicates the cluster redshift.  Orange and blue squares mark the
  galaxies in the cluster field classified as candidate members or
  interlopers, respectively, with the full IRAC+{\it HST} color
  selection criteria described in Sect.~\ref{sec:colselmembers}.
\label{fig:samplesel} }
\end{figure}

  In this Section we describe the definition of the galaxy samples
  used in this work.  Figure~\ref{fig:cmd} (panels {\it a}, hereafter
  Fig.~\ref{fig:cmd}{\it a}) shows, for each cluster, the m140
  vs.\ m814-m140 (approximately V vs.\ U-V restframe) color-magnitude
  diagram of galaxies in the cluster central region. The details of
  our imaging data (in particular the $\sim 2 \times 2$~arcmin$^{2}$
  WFC3 field of view) limit our analysis to a homogeneously covered
  region reaching out to $r \lesssim r_{500}\sim1'\sim500$~kpc
  (proper, for this cluster sample). Given the cluster position in the
  HST images (as well as the occurrence of masked areas for some of
  the clusters), this $r\lesssim r_{500}$ region is not fully and
  homogeneously covered by our data (see
  Fig.~\ref{fig:sigmamaps}). Given the assumption of spherical
  symmetry of galaxy distribution and population properties, in the
  following we will correct for this by introducing a coverage weight
  factor for each galaxy, given by the fraction of covered area in a
  thin (5'') annulus at the galaxy's clustercentric distance.  For
  this reason, we limit our analysis to a clustercentric
  distance\footnote{See Sect.~\ref{sec:clustercenter} for the adopted
    definition of cluster center.} $r<0.7 r_{500}$, so that the
  maximum coverage weight factor that we need to apply is $\leq 2$ for
  all clusters, and the fraction of uncovered area is typically small
  ($\lesssim 10\%$ at the outskirts of the $r < 0.7 r_{500}$ area,
  with the exception of SPT-CLJ2040 for which it is $\sim$20\%).  For
  each cluster, we thus start by considering a sample of galaxies
  within a clustercentric distance $r <0.7 r_{500}$.

Furthermore, we focus in this work on the bright F140W-selected galaxy
sample down to m140$\sim$23.5 AB mag. More specifically, for each
cluster we limit ourselves to the m140 magnitude range where we can
measure the F814W-band aperture magnitudes for red sequence (RS)
galaxies (see Fig.~\ref{fig:cmd}{\it a}) with a
S/N$>$5. The  adopted m140 magnitude
  limits range from m140=23.2 to 23.5, as reported in
  Fig.~\ref{fig:cmd}{\it b}.

 Figure~\ref{fig:cmd}{\it a} thus shows the
  color-magnitude diagrams of galaxies in cluster fields within
  $r<0.7r_{500}$ of the cluster center, and down to m140$\sim$23.5 AB
  mag.  All galaxies in the probed field are shown, with different
  symbols referring to a visualization of the background-subtracted
  color-magnitude diagram that is discussed in Sect.~\ref{sec:statsub}.

We describe in the following the definition of the candidate cluster
member samples at the core of all the analyses reported in this work.
A description of Fig.~\ref{fig:cmd} in the context of cluster galaxy
population properties is given instead in Sect.~\ref{sec:galpops}. 
Given the very limited spectroscopic follow-up available for these
clusters, we cannot rely on spectroscopic redshifts for membership
determination of a representative sample of cluster
galaxies. Furthermore, this work is based on deep photometric coverage
in only a small number of bands, resulting in limited photometric
redshift (photo-z) accuracy. We thus adopt a color selection to
identify candidate cluster members (Sect.~\ref{sec:colselmembers}),
plus a subsequent statistical background subtraction
(Sect.~\ref{sec:statsub}) to account for the residual fore-/background
contamination.

\subsection{Color-selected candidate member samples}
\label{sec:colselmembers}

 For each cluster, we defined a color selection to identify a sample of
 candidate cluster members.  The m814-m140 vs.\ [3.6]$-$[4.5]
 color-color diagram in Figure \ref{fig:samplesel} shows an example of
 this selection, illustrating the main points of this approach for the
 two clusters at the lowest and highest redshift in our sample, as
 described in detail here below.

$\bullet$ Initial m140-selected sample -  Colored squares in
Fig.~\ref{fig:samplesel} show all galaxies in the inital m140-selected
sample with a measured [3.6]$-$[4.5] color (see
Sect.~\ref{sec:spitzerobs}). For the galaxy samples we are interested
in, that is at a clustercentric distance within $0.7 r_{500}$ (or $
0.45 r_{500}$ in part of the analysis) and down to m140=23.2 to 23.5
AB mag (depending on the cluster, see Sect.~\ref{sec:samplesel}), we
have at least one retained measured IRAC flux (3.6 or 4.5$\mu$m) for
on average 90\% (ranging from 84\% to 95\% for the different clusters)
of the initial sample, with both IRAC fluxes missing for typically
$\lesssim 5\%$ (or 8\% in the worse case) of the initial samples. To
favor completeness (and at the expense of purity) all galaxies for
which we do not have a [3.6]$-$[4.5] color measurement were initially
retained in the candidate member sample.

$\bullet$ Selection of a high-purity red-sequence candidate
  member sample -  As can be seen in Fig.~\ref{fig:cmd}{\it a}, all
clusters in this sample show a clear red (in the m814-m140 color,
$\sim$U-V restframe) galaxy population, with high contrast over the
background. We can thus easily identify a sample of red candidate
cluster members with low contamination from interlopers. For this
purpose, we consider here only RS galaxies within $<2 \sigma$ from the
red peak in the galaxy color distribution (see, e.g.,
Fig.~\ref{fig:cmd}{\it b}).  From our analysis of statistically
background-subtracted samples in Sect.~\ref{sec:statsub}, the overall
fore-/background contamination for RS galaxies in the probed region is
estimated to be typically at the 10\% level (Fig.~\ref{fig:cmd}{\it
  c}).  We note that this RS sample used here is not meant to be a
complete sample of RS galaxies, but rather to be a sample of cluster
members with the lowest possible contamination by interlopers. We thus
select galaxies in the m814-m140 color range where the contrast of the
cluster vs. background galaxy population is highest.  In both panels
of Fig.~\ref{fig:samplesel}, the concentration of sources at the RS
color (see for comparison Fig.~\ref{fig:cmd}{\it a}) is clearly
visible. The vertical light gray lines show the adopted selection of
the RS sample described above.

$\bullet$ First definition of IRAC color selection for candidate
    cluster members -  The [3.6]$-$[4.5] color (or also ``IRAC color''
  hereafter) at this redshift has a relatively weak dependence on
  stellar population properties (see Fig.~\ref{fig:samplesel}, and
  Sect.~\ref{sec:redshifts} below) as compared to optical colors
  spanning the 4000\AA~ break.  In fact, the similar IRAC colors of
  galaxies in a cluster result in a ``stellar bump sequence'' whose
  color mainly depends on the cluster redshift
  \citep[e.g.,][]{muzzin2013a}. We can thus tune, for each cluster,
  the IRAC color selection based on the high-purity RS sample, and
  then use it to identify a sample of candidate members including all
  galaxy types. In fact, to ensure the high completeness of such a
  sample, we will later adapt this first IRAC color selection (based
  on the RS sample alone) by considering a plausible IRAC color range
  for galaxies at the same redshift but with different stellar
  population properties, as detailed below.

For each cluster, we thus initially defined a first IRAC color
selection as the $\pm 3 \sigma$ range of [3.6]$-$[4.5] color of the RS
sample.  This [3.6]$-$[4.5] color range is much larger than the color
uncertainty of individual galaxies (see Fig.~\ref{fig:samplesel}),
which was thus not considered to define this color selection. As shown
in Fig.~\ref{fig:samplesel}, these RS galaxies have indeed very
similar IRAC colors - as expected, given the low background
contamination of the selected RS sample and the uniformity of IRAC
colors for galaxies at the same redshift, and in this case also of
very similar stellar population properties. We also note in
Fig.~\ref{fig:samplesel} the clear concentration of galaxies with
m814-m140 colors corresponding to the peak of the blue cloud (see
Fig.~\ref{fig:cmd}{\it a} and {\it b}) and with IRAC colors very
similar to those of the RS sample. The resulting first IRAC color
selection of candidate cluster members is shown by the horizontal
light gray lines in Fig.~\ref{fig:samplesel}.

$\bullet$ Refined definition of IRAC color selection for candidate
cluster members, accounting for galaxy populations bluer than the red
sequence - To extend this selection to the full sample of cluster
galaxies, we need to account for the possible color difference between
galaxies hosting different stellar populations. Using \citet{bc03}
models, we thus increased the IRAC color selection range by the
estimated difference between the color of passive galaxies (nominally
an SSP with a formation redshift $z_{f}=3$, but the difference between
plausible passive galaxy colors is very small, see
Fig.~\ref{fig:samplesel}) and the color of a stellar population at the
same redshift forming stars at a constant rate for 100~Myr (different
ages in a reasonable range do not increase further the color range of
our selection), with solar metallicity and a dust attenuation ranging
from $A_{v}=0$ to 3 mag (see Fig.~\ref{fig:samplesel}). This procedure
yields the final adopted color selection, shown by dark gray lines in
Fig.~\ref{fig:samplesel}.  The red and blue sets of tracks in
Fig.~\ref{fig:samplesel} show the color evolution of the adopted
passive (SSPs with $z_{f}=2,3,5$) and star-forming (with
$A_{v}=0,1,3$) models, respectively, in the redshift range $1<z<2$,
with small solid points indicating z=1.3, 1.6, 1.9 and the large
circle indicating the cluster redshift. In order to be representative
of a passive population, the $z_f = 2$ SSP track is only shown up to
z=1.7.

$\bullet$ Final definition of candidate member samples - All galaxies
(except those with unmeasured or unreliable IRAC fluxes, as discussed
above) outside of the adopted color selection were discarded as
interlopers. The rest of the galaxies were retained to form the
candidate cluster member sample. Finally, given the redshift
dependence of the [3.6]$-$[4.5] color, we expect that across the
redshift range of interest here ($z\sim1.4-1.8$), the IRAC
color-selected candidate member samples defined as above are affected
by some (variable level of) contamination from interlopers at
redshifts similar to the cluster, as well as contamination from
low-redshift sources \citep[e.g.,][]{muzzin2013a,papovich2008}. To
remove sources which are most likely low-redshift interlopers, we
further cleaned the candidate member sample by requiring that
m140-[3.6]$\geq$0, m814-[3.6]$\geq$0.4, m140-[4.5]$\geq$-0.5,
m814-[4.5]$\geq$0, and m814-m140$>$0, as calibrated on galaxies in the
GOODS-S field sample described in
Section~\ref{sec:massesandcontrolfield}. This is the candidate member
sample used in the following analysis.  In the example
Fig.~\ref{fig:samplesel}, this sample is shown with orange squares,
while rejected interlopers are shown with blue squares. We deal with
the residual contamination from interlopers at redshifts broadly
similar to the cluster in Sect.~\ref{sec:statsub}.

\subsection{Statistical subtraction of residual background contamination}
\label{sec:statsub}

Although the color selection applied to this point removes obvious
interlopers, the large IRAC color range used in the selection to
ensure a high completeness of the sample, and the intrinsically small
variation in IRAC color spanned by different populations in a
relatively broad redshift range, result in an expected significant
residual contamination from sources at redshifts broadly similar to
the cluster (see tracks in Fig.~\ref{fig:samplesel}, as well as
Fig.~\ref{fig:redshifts} below). This was accounted for by applying a
residual statistical background subtraction using a $\sim60$
arcmin$^2$ control field in the CANDELS GOODS-S field discussed in
Sect.~\ref{sec:massesandcontrolfield}, where we applied exactly the
same color selections as we do for our candidate cluster member
samples.

Fig.~\ref{fig:cmd}{\it a} shows for each cluster a visualization of
the resulting background-subtracted color-magnitude diagram, produced
with an approach similar to that of
\citet{vanderburg2016}. Specifically, we subtracted the residual
background contamination as follows: 1) We start from a candidate
cluster member sample obtained as discussed in
Sect.~\ref{sec:colselmembers}, and a control field sample from the
GOODS-S field (see Sect.~\ref{sec:massesandcontrolfield}) that is
selected in the same manner. 2) For each galaxy in the candidate
member sample, we calculate a ``weight'' that corresponds to the
statistical excess of the candidate member sample over the control
field density at the magnitude and colors of the given galaxy.
Weights are calculated as follows \citep[see][for a more detailed
  description]{vanderburg2016}: first, all candidate member weights
are initially set to 1. Then, for each galaxy in the control field
sample we subtract the corresponding ``background contamination'' from
the candidate member sample by appropriately reducing the weights of
all candidate members that lie within a distance in the color
(m814-m140) - color (m140 - [3.6]) - magnitude (m140) space given by
their photometric uncertainties (1$\sigma$), with a minimum distance
of 0.3~mag, effectively resulting in a smoothing of galaxy densities
in the color-color-magnitude space. If no galaxies are found within
this distance, we double the search distance, and then if necessary
increase the search distance to 1.3 times the distance to the closest
galaxy.  This criterion allows the full subtraction of the background
contamination estimated from all galaxies in the control field sample,
while reducing the weights of those candidate members that are more
similar to the galaxies in the control field. For each considered
control-field galaxy, the weights of all selected candidate members
identified with the above criterion are reduced so that the
contribution of the considered field galaxy (normalized by the areas
of the probed cluster region and of the control field) is removed. At
the end of this procedure, the contributions of all galaxies from the
field sample have been subtracted from the candidate member sample.

This approach effectively follows the procedure adopted in previous
analyses \citep[e.g.,][]{vanderburg2016,strazzullo2016}, except that
we apply the statistical background subtraction in
magnitude-color-color rather than in magnitude-color space, and that
we equally de-weight candidate members with similar
color-color-magnitude rather than de-weighting the single candidate
member closest in color-color-magnitude space to the given field
galaxy. The reason why we perform the
  statistical background subtraction in magnitude-color-color space,
is that by adopting colors close to the restframe U-V, V-J we expect
to account for differences in specific star formation activity and
dust attenuation (and thus also mass-to-light ratio) across galaxy
populations better than with a subtraction in color-magnitude space
(see e.g., Sect.~\ref{sec:uvj}).

The result of this procedure is that in regions of the
color-color-magnitude space that are more densely populated in the
control field sample, the weights of candidate cluster members are
more significantly reduced. At the end of the procedure, each galaxy
in the candidate member sample has an associated weight corresponding
to its statistical excess probability over the galaxy density in the
control field at that location in color-color-magnitude space.  We
stress that, as this excess probability results from a
\emph{statistical} background subtraction, it does not translate into
a membership probability on a single galaxy basis, and it is only
meant to describe the contribution of cluster galaxies as a population
across the color-color-magnitude space.  We also note that the
determination of the statistical background subtraction weight does
not include any dependence on clustercentric distance.  In
Fig.~\ref{fig:cmd}{\it a}, candidate cluster members are shown as
filled circles whose size scales with this weight.

\begin{figure*}
 \includegraphics[width=0.376\textwidth,viewport= 53 413 536 699, clip]{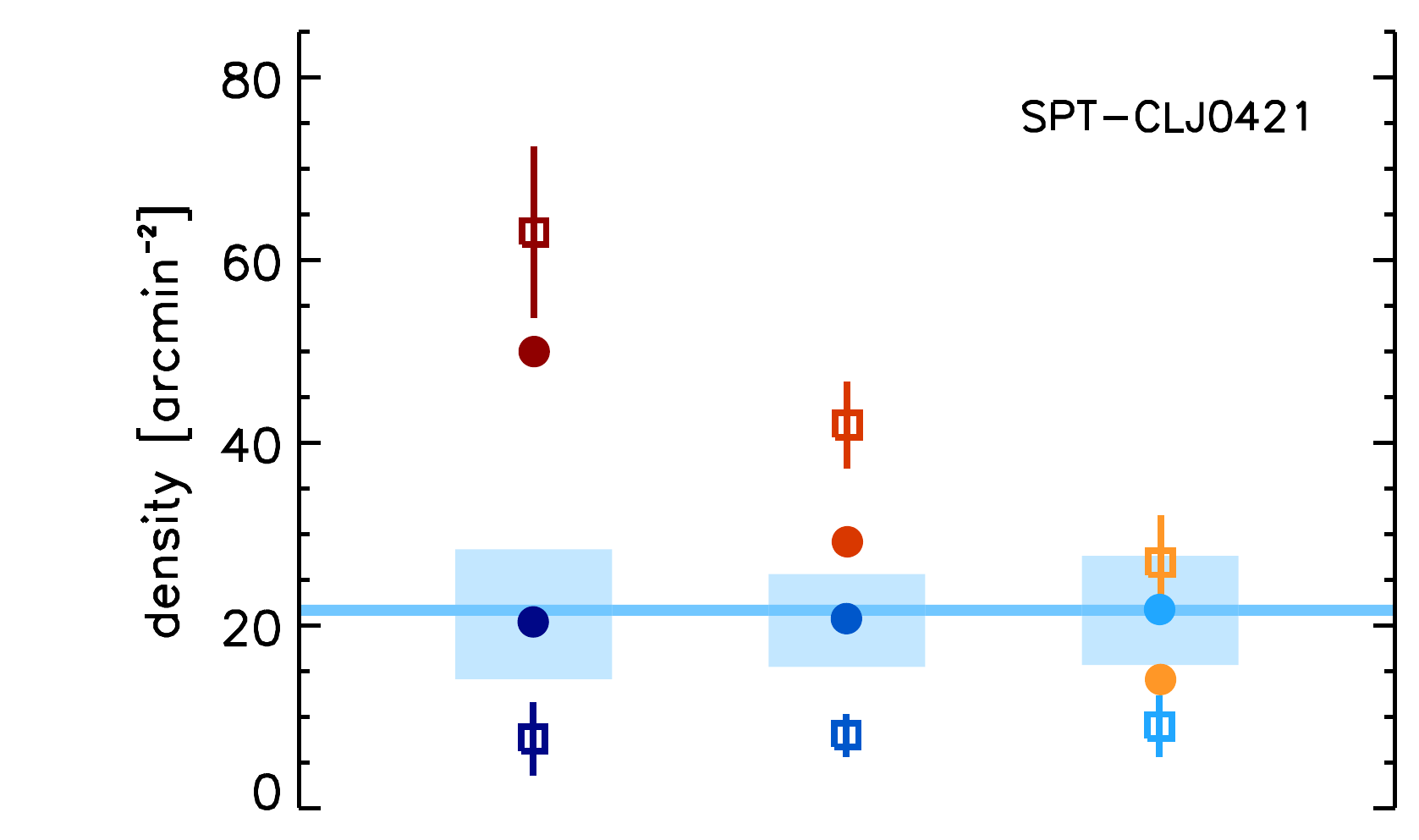}%
 \includegraphics[width=0.3\textwidth,viewport= 151 413 536 699, clip]{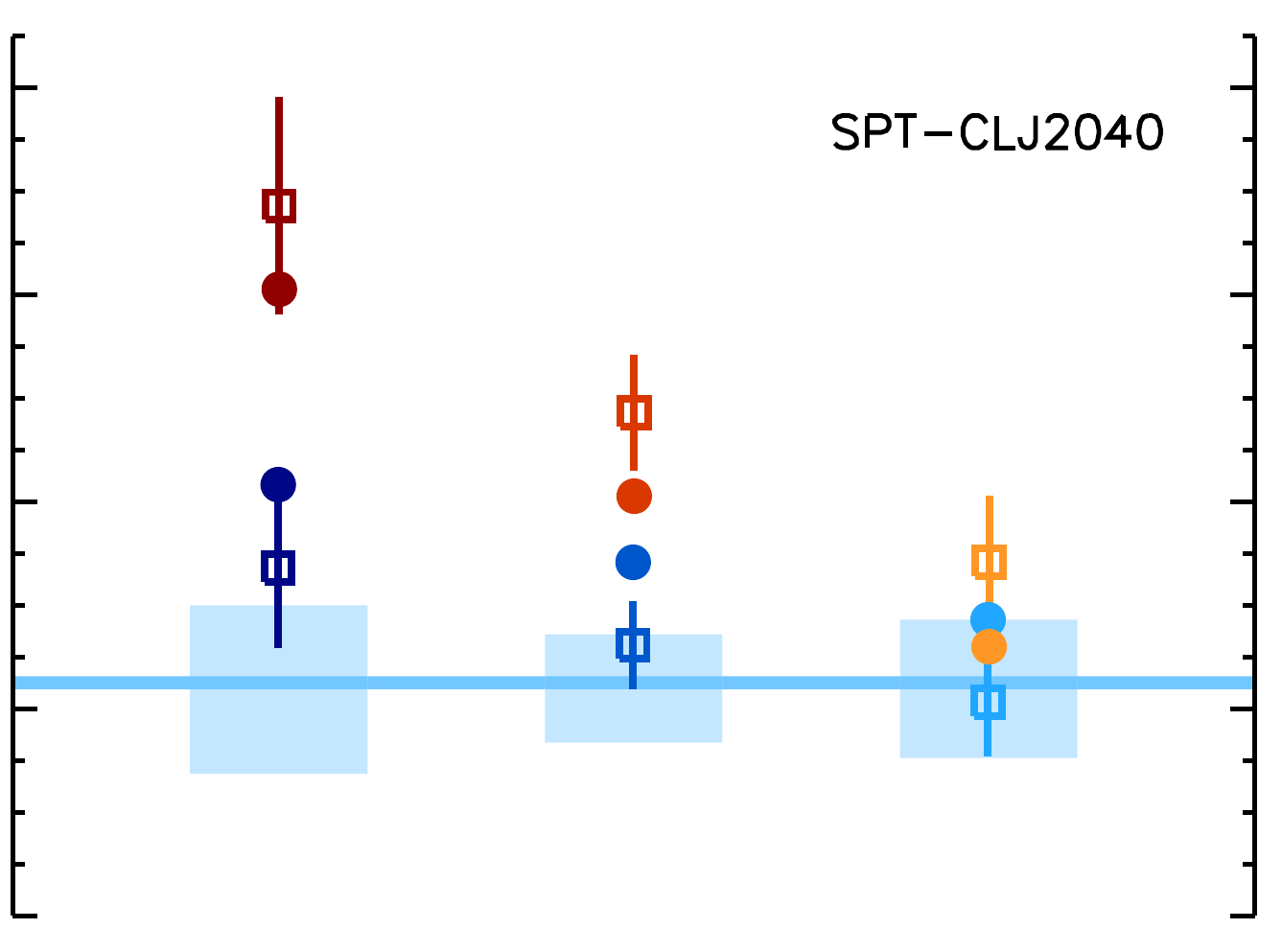}%
 \includegraphics[width=0.3\textwidth,viewport= 135 420 510 699, clip]{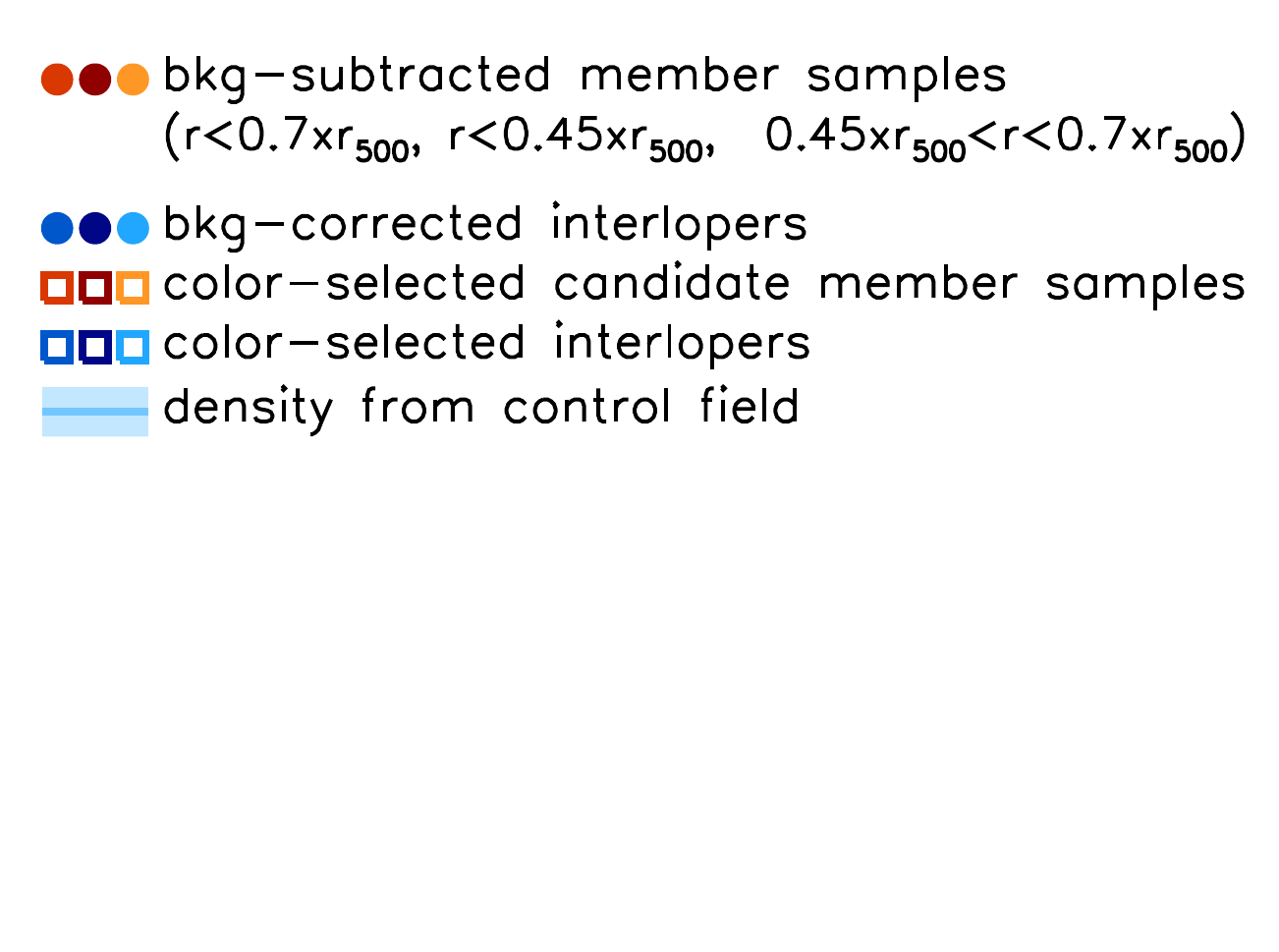}
 \includegraphics[width=0.376\textwidth,viewport= 53 368 536 699, clip]{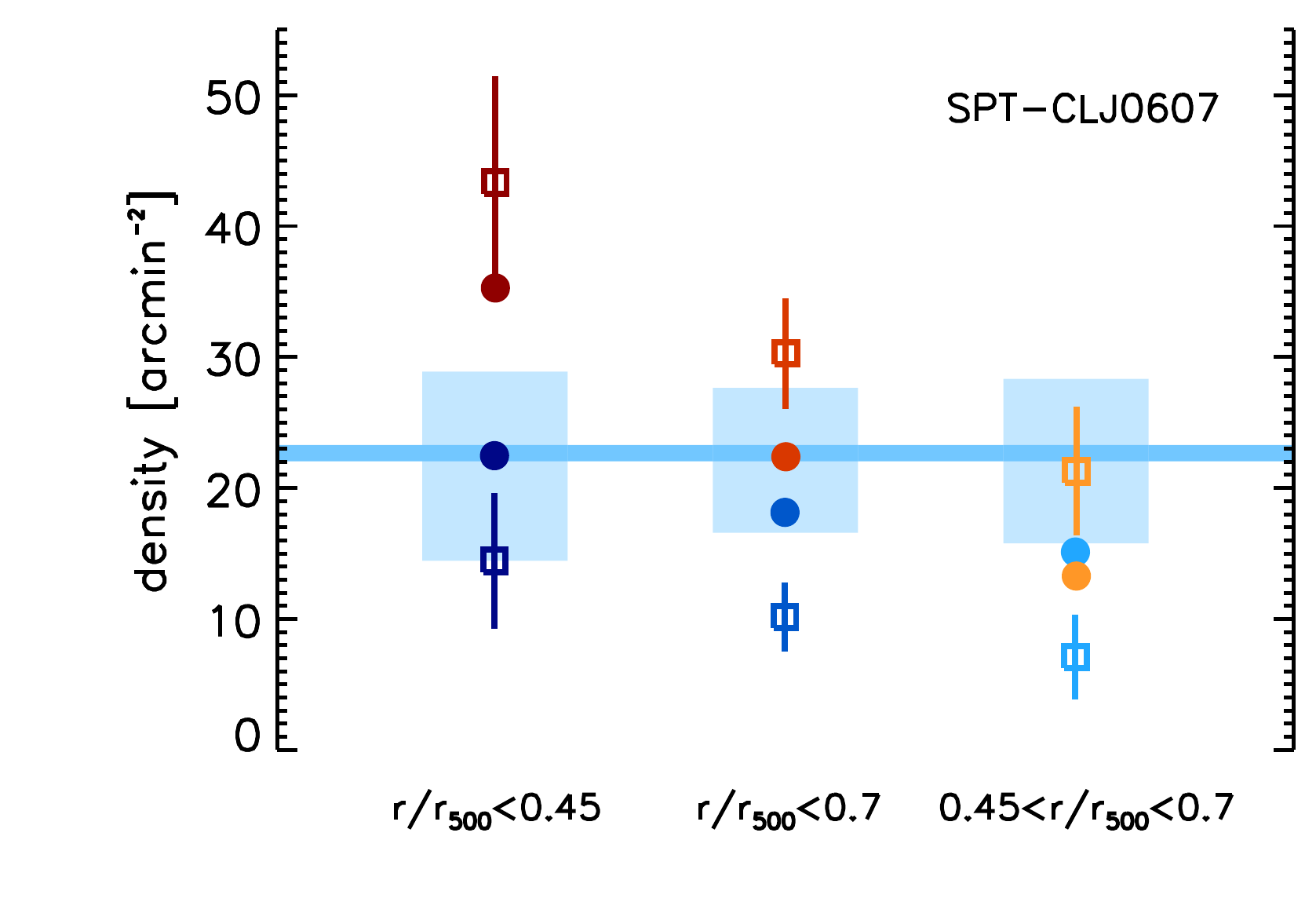}%
 \includegraphics[width=0.3\textwidth,viewport= 151 368 536 699, clip]{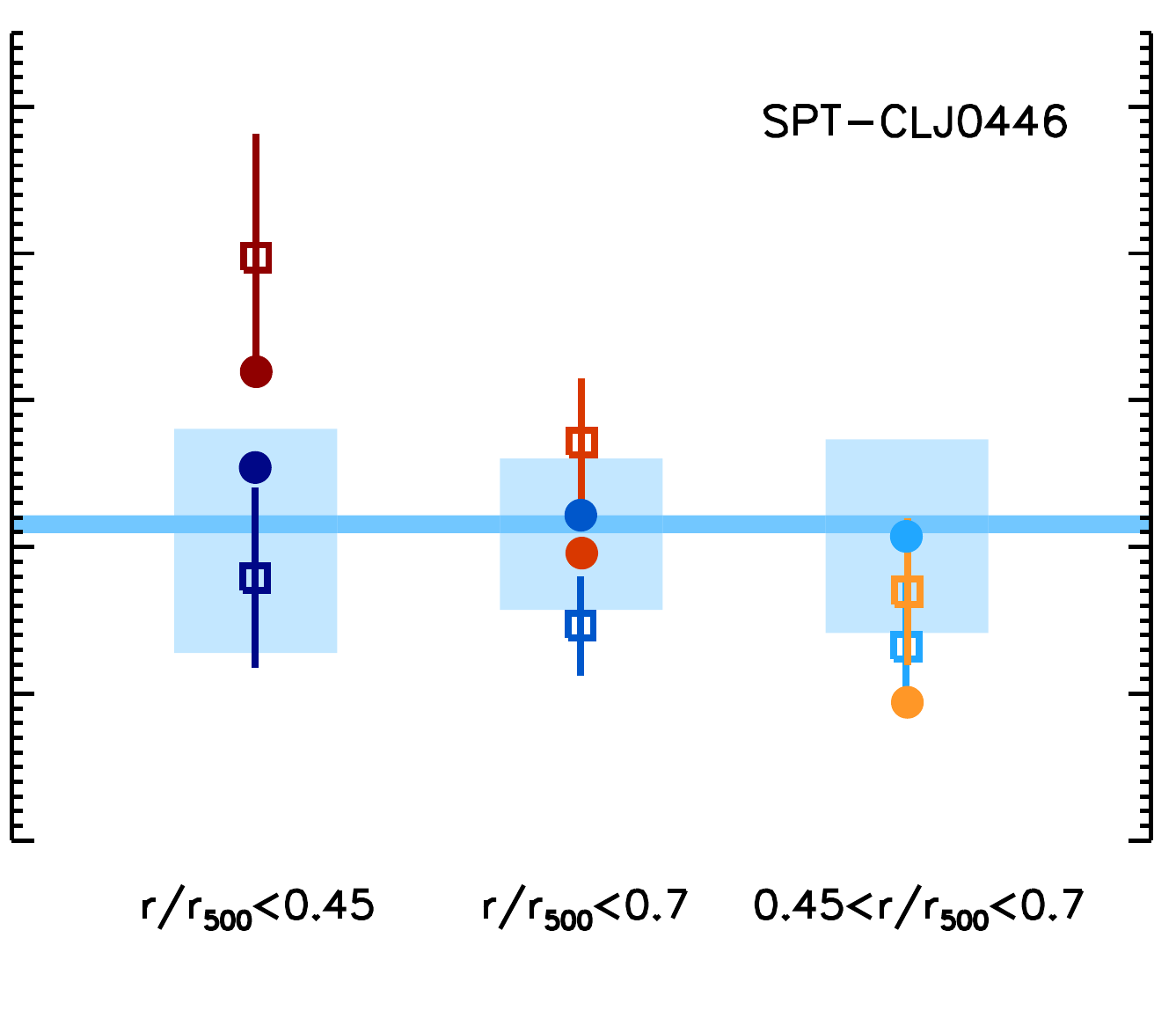}%
 \includegraphics[width=0.3\textwidth,viewport= 151 368 536 699, clip]{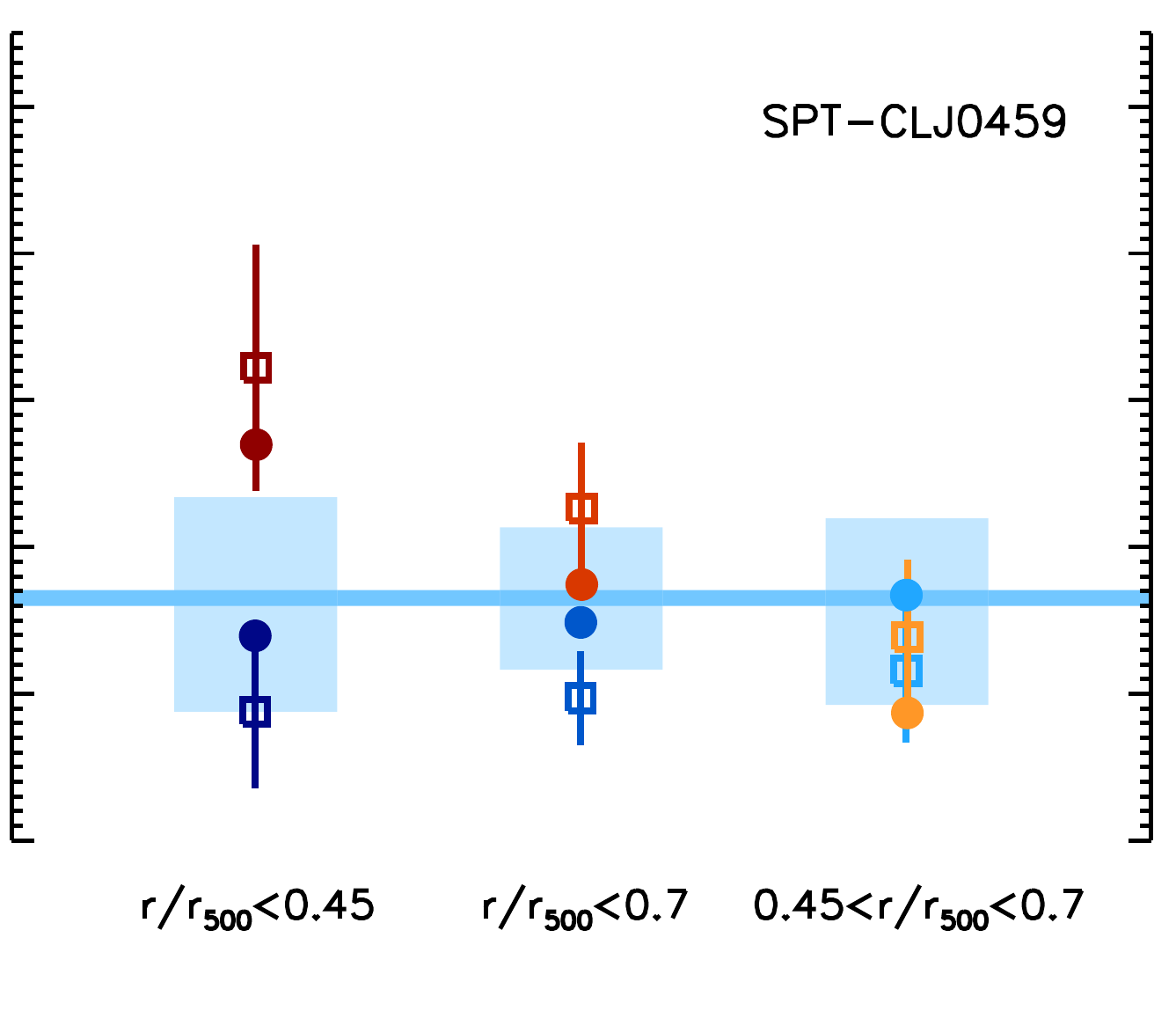}
\caption{Projected number density of candidate cluster members
  (shades of red) and interlopers (blue) in the five clusters as
  measured in three cluster regions, as indicated. Empty squares show
  densities for color-selected samples (see
  Sect.~\ref{sec:colselmembers}), and filled circles show how member
  number density decreases -- and interloper density increases --
  after applying the statistical subtraction of residual background
  contamination described in Sect.~\ref{sec:statsub}.  The blue line
  shows the average density of control field galaxies in the same
  magnitude range adopted for the cluster, with the light-blue shaded
  areas showing the density variation ($16^{th}-84^{th}$ percentiles)
  in apertures corresponding to the three probed regions.  The
  relative flatness of the interloper density profiles constrains the
  cluster member sample incompleteness due to member misclassification
  (see Sect.~\ref{sec:complcontsamples}).
  \label{fig:densities}  }
\end{figure*}

\subsection{Completeness and contamination of candidate member samples}
\label{sec:complcontsamples}

 Given the statistical background subtraction weights calculated in
 Sect.~\ref{sec:statsub}, we can further examine the color-selected
 candidate member samples defined in Sect.~\ref{sec:colselmembers},
 that are the basis of all the following analyses.

Based on these weights, we show in Fig.~\ref{fig:cmd}{\it c} the
expected fraction of interlopers in the candidate member sample as a
function of m814-m140 color. We estimate a very low contamination (as
 expected) from fore-/background galaxies in the red candidate member
 sample, but a non-negligible contamination for blue galaxies. The
 background subtraction weights allow us to account for this
 contamination in the analyses described below.

In Fig.~\ref{fig:densities} we show the projected densities of cluster
candidate members and interlopers resulting from the procedure in
Sect.~\ref{sec:colselmembers}. The figure shows projected densities
for the two regions in each cluster used in the following analyses,
the inner core region $r<0.45 r_{500}$, and a larger region out to
$r<0.7 r_{500}$. (Because the $r<0.45 r_{500}$ region is included in
the $r<0.7 r_{500}$ one, the respective measured densities shown in
Fig.~\ref{fig:densities} are not independent.)  For comparison, we
also show the region corresponding to the annulus
$0.45<r/r_{500}<0.7$, although given the expected more significant
background contamination this annulus is not investigated in detail in
the following. We note that both the $r<0.7 r_{500}$ and the outer
annulus regions are not completely covered by our homogeneous
catalogs, and that the inner $r<0.45 r_{500}$ region is affected by
masked areas for SPT-CLJ2040 and SPT-CLJ0607
(Fig.~\ref{fig:sigmamaps}). In the analysis below we will correct for
the uncovered portions of the probed regions assuming spherical
symmetry (see Sect.~\ref{sec:samplesel}), but for the purpose of this
figure we use densities computed in the actual covered area. The red-
and blue-shade squares show, respectively, the galaxy projected number
density for the samples of candidate members and interlopers as
determined with the color selection described in
Sect~\ref{sec:colselmembers}.

Because, as discussed in Sects.~\ref{sec:colselmembers} and
\ref{sec:statsub}, the candidate member samples are affected by
residual background contamination, the projected number densities of
both candidate members and interlopers will also be affected. In
particular, the number density of candidate cluster members will
be an overestimate of the actual density of cluster members, while the
density of color-selected interlopers will be an underestimate of the
actual interloper density. By applying the statistical residual
background subtraction weights from Sect.~\ref{sec:statsub}, we can
estimate the actual (background corrected) projected densities of
cluster members and interlopers. These are shown as filled circles in
Fig.~\ref{fig:densities}.

The blue line in Fig.~\ref{fig:densities} shows the average density of
galaxies in the selected m140 range in the GOODS-S control field,
where the thickness of the line shows the uncertainty on this average
value. The light-blue shaded areas show, as an estimate of the
variance on the scales of the considered cluster regions, the
$16^{th}-84^{th}$ percentile range of projected densities obtained at
100 random positions in the control field, within apertures of the
same angular size as the adopted cluster regions.

The estimated density of interlopers for each cluster is consistent
with being flat across the three (partially overlapping) samples at
different clustercentric distances, with a marginal increase of the
estimated interloper density towards the cluster center for
SPT-CLJ0607 and SPT-CLJ2040 that is nevertheless still consistent with
the outer value. In particular for SPT-CLJ2040, the interloper density
tends to also be marginally higher than the typical expectations from
the control field (Fig.~\ref{fig:densities}). An interloper density
higher than expected and, in particular, rising towards the cluster
center, might indicate a misclassification of a fraction of candidate
members as interlopers, possibly due to inaccurate IRAC fluxes because
of contamination from bright neighbors, or biased color selection.
Other explanations such as weak gravitational lensing magnification
that is stronger in this cluster or a truly higher density of
fore-background sources are also possible. In this respect, we note
that a very bright and extended foreground source lies along the line
of sight very close to the center of SPT-CLJ2040.

If misclassification of candidate members as interlopers is indeed
occurring, Fig.~\ref{fig:densities} provides an estimate of the
incompleteness of the member sample. Assuming that the interloper
density in the outer annulus is actually correct\footnote{We recall
  that for all clusters the interloper density in the outer annulus is
  consistent with expectations from the average density in the control
  field. For SPT-CLJ2040, which has the highest interloper density in
  the outer annulus and also the highest difference with respect to
  the control field, the completeness levels of the $r<0.7 r_{500}$
  and $r<0.45 r_{500}$ member samples would be 78\% and 76\%,
  respectively, if assuming the control field average value for the
  reference interloper density rather than the measured value in the
  outer annulus. }, and that the interloper density is flat across the
probed cluster field, we can estimate the expected number of
interlopers in the cluster central fields ($r/r_{500}<0.45, 0.7$) from
the interloper density in the outer annulus and the projected area of
the central fields. By comparing this number to the number of actually
identified interlopers in the central fields we can estimate how many
cluster members were possibly misclassified as interlopers. This
calculation yields an estimate of the completeness for the member
sample in the $r<0.7 r_{500}$ region of 88\% for SPT-CLJ0607, 93\% for
SPT-CLJ0446, 88\% for SPT-CLJ2040, and 100\% for SPT-CLJ0421 and
SPT-CLJ0459. Because of the low number statistics affecting these
completeness estimates, these percentages are affected by significant
uncertanties, with errors of $\pm$40\%. In the central $r<0.45
r_{500}$ region, the estimated completeness would be 83\% for
SPT-CLJ0607, 87\% for SPT-CLJ0446, 82\% for SPT-CLJ2040, and 100\% for
SPT-CLJ0421 and SPT-CLJ0459 (with errors of $\pm$30\%).  Given the
significant uncertainties (Fig.~\ref{fig:densities}), all samples are
consistent with being formally 100\% complete (meaning that the number
of cluster members that we estimate might be misclassified as
interlopers is always consistent with zero).

\subsection{Projected galaxy density maps and cluster center definition}
\label{sec:clustercenter}

\begin{figure*}
 \includegraphics[width=0.33\textwidth,viewport= 56 365 458 718, clip]{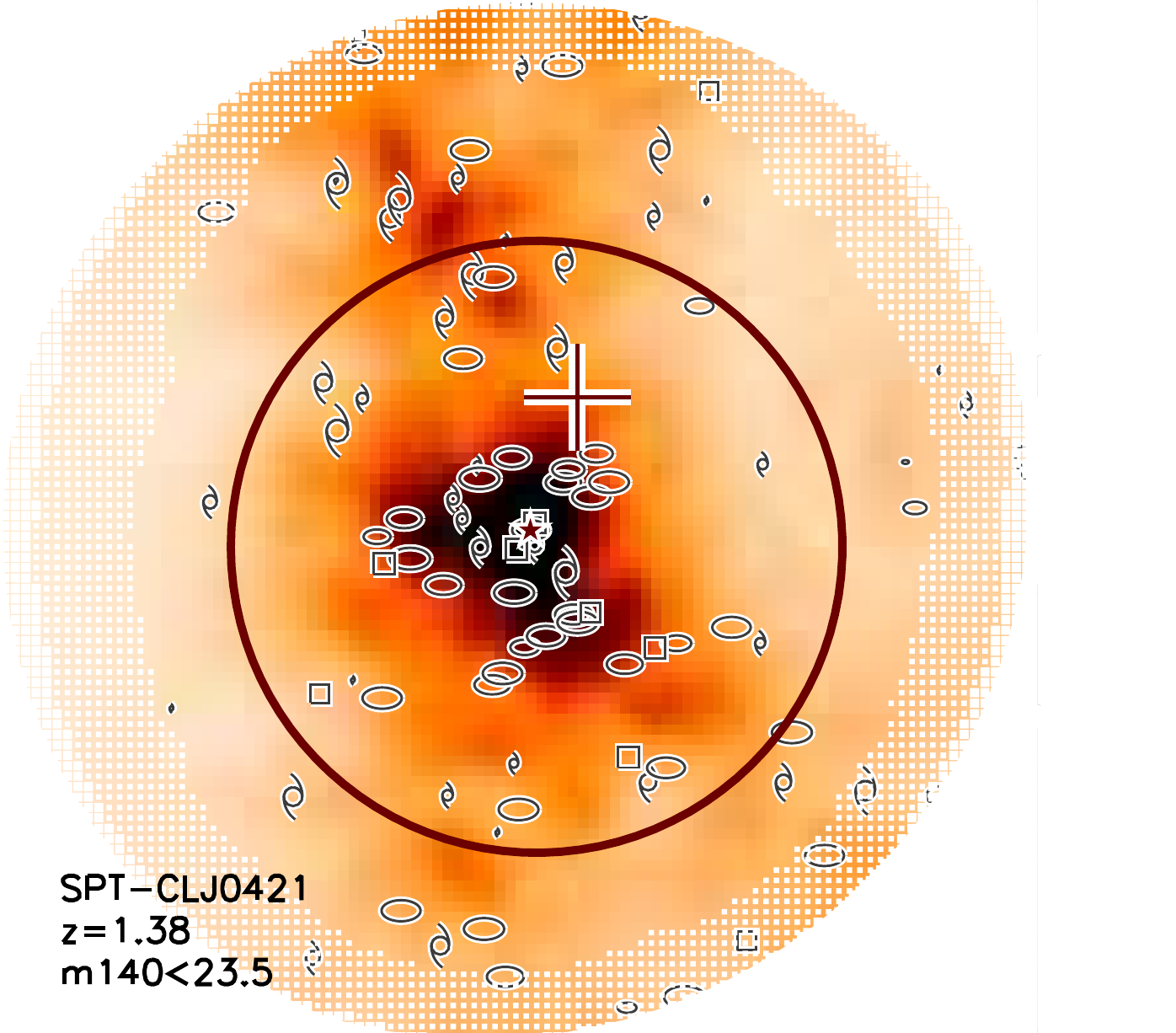}%
 \includegraphics[width=0.33\textwidth,viewport= 56 365 458 718, clip]{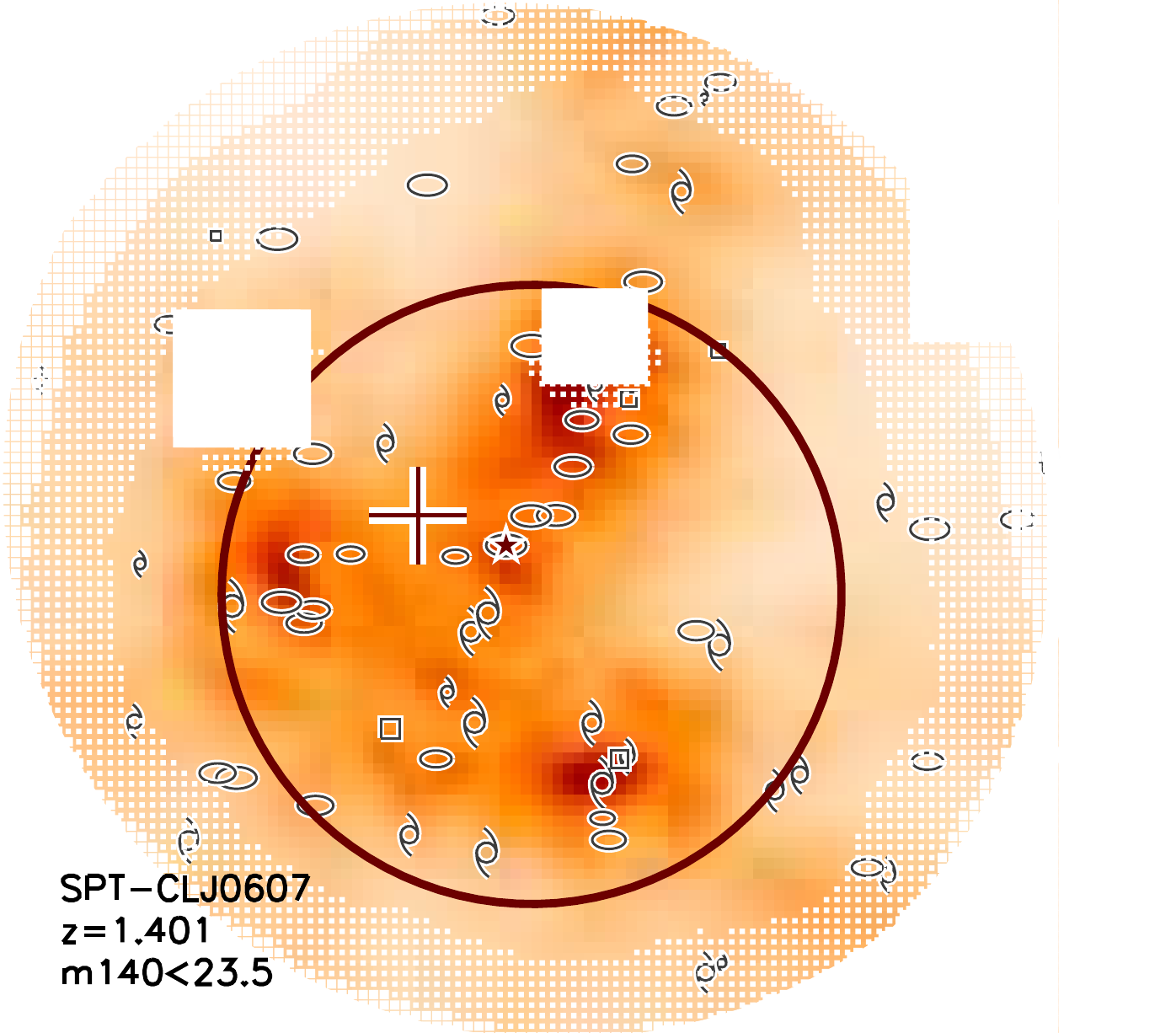}%
 \includegraphics[width=0.35\textwidth,viewport= 145 500 479 720, clip]{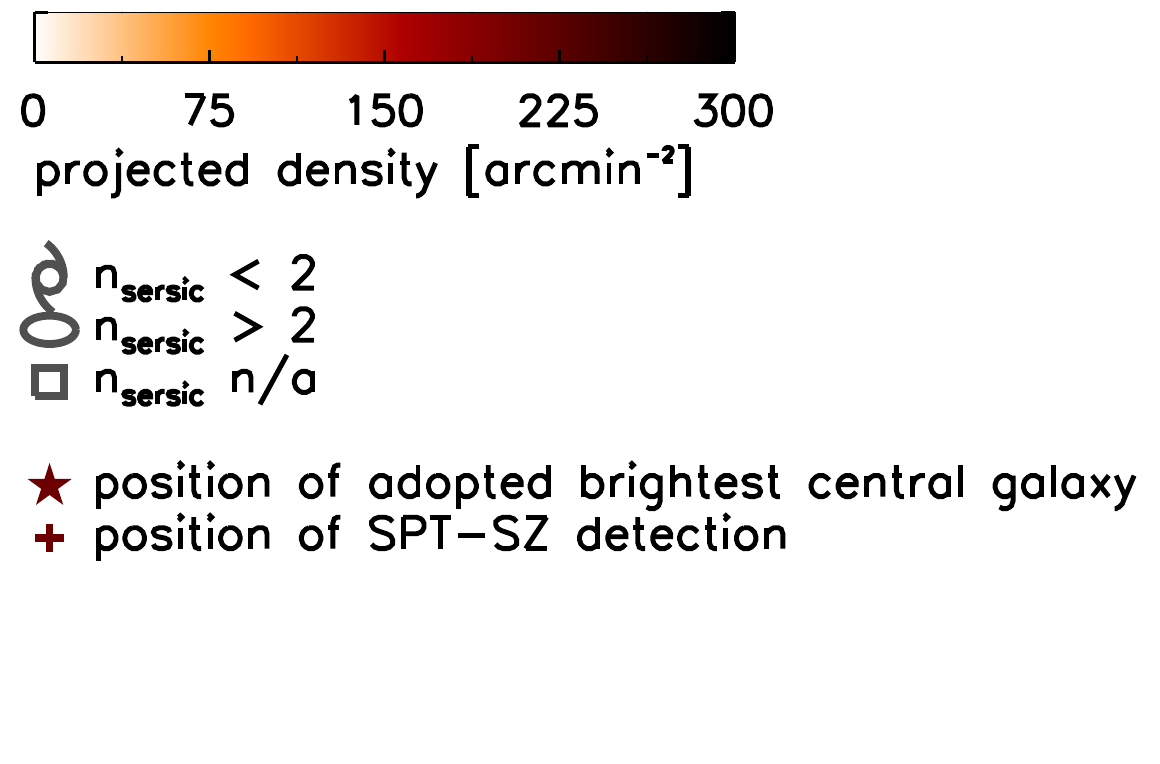}
 \includegraphics[width=0.33\textwidth,viewport= 56 365 458 718, clip]{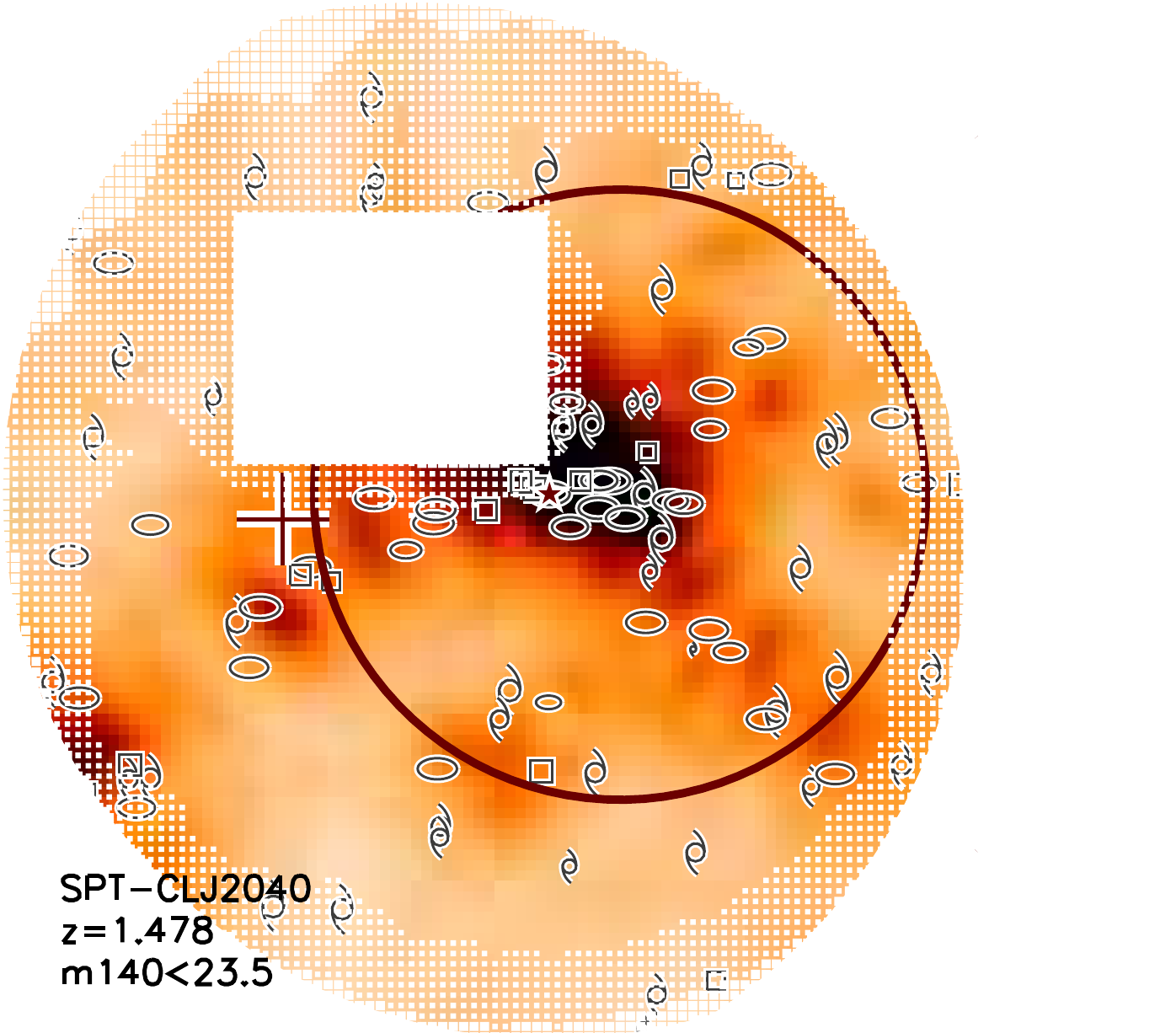}%
 \includegraphics[width=0.33\textwidth,viewport= 56 365 458 718, clip]{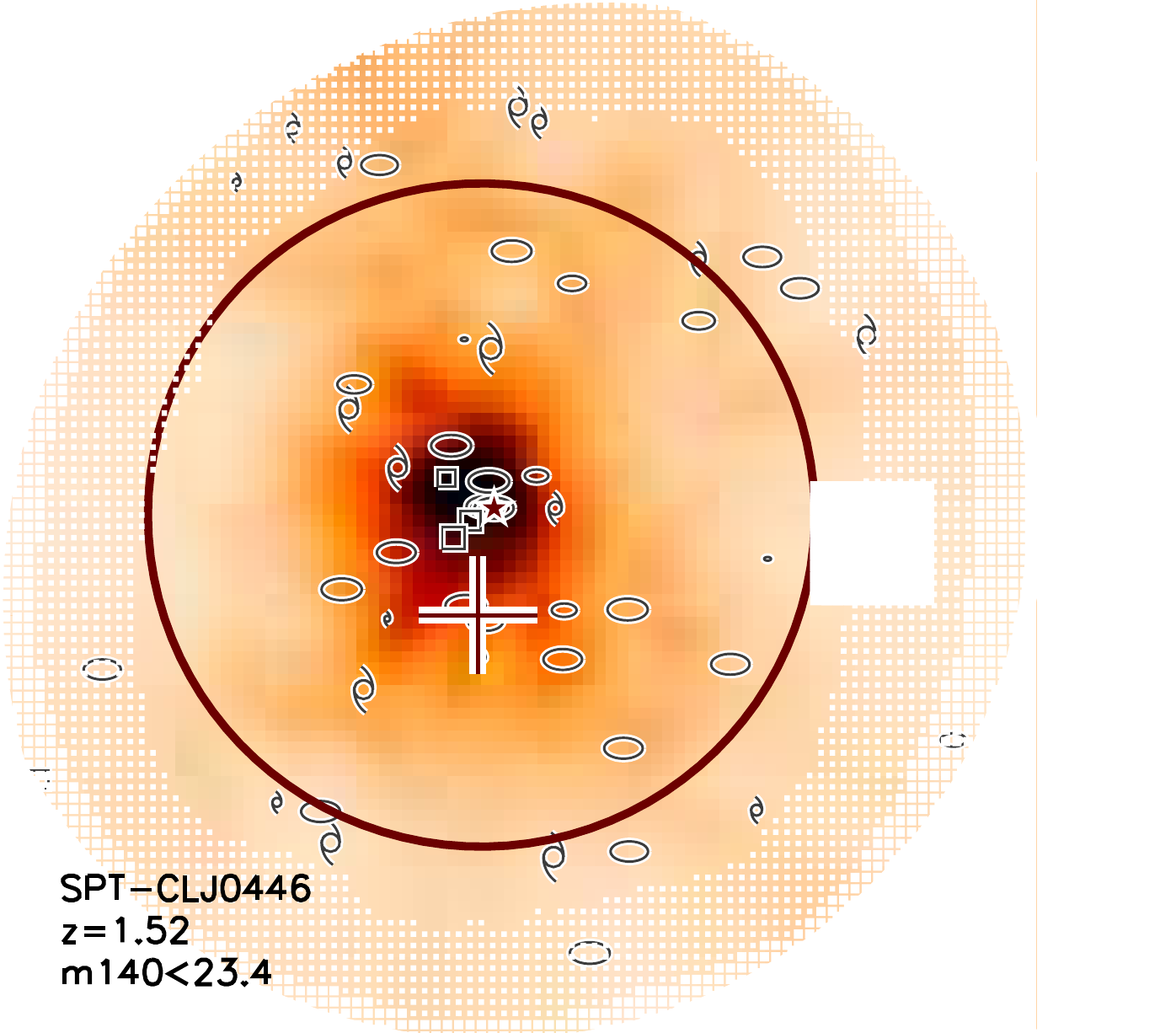}%
 \includegraphics[width=0.33\textwidth,viewport= 56 365 458 718, clip]{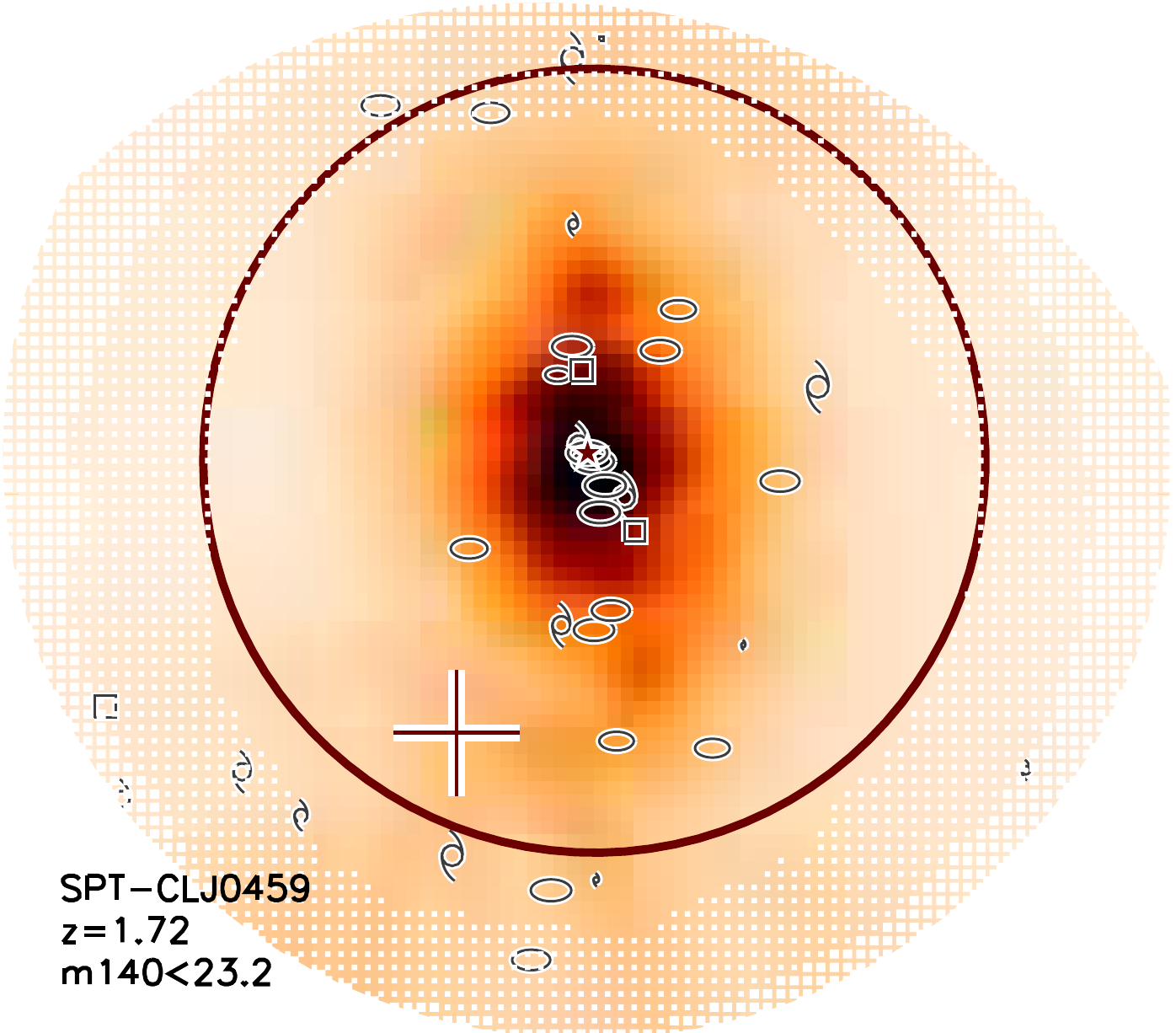}
\caption{Projected density maps (see color bar) of candidate cluster
  members for the five clusters. Gray empty symbols show the position
  of individual candidate members, where symbol size scales with the
  galaxy statistical background subtraction weight
  (Sect.~\ref{sec:statsub}), and symbol shape reflects galaxy
  morphology. The cross marks the coordinates of the SPT-SZ detection,
  with the cross size showing the estimated positional uncertainty
  (1$\sigma$, see \ref{sec:clustercenter}).  The position of the
  adopted brightest central galaxy, defining the cluster center in the
  subsequent analysis, is indicated by a star. The circle is centered
  at the median coordinates of the 5\% highest-density points in the
  map, and has a radius of 250~kpc at the cluster redshift (for
  comparison $r_{500}\sim 70'' \sim 600$~kpc for these clusters).
  Masked areas are blanked out and white gridding marks areas affected
  by edge effects.
  \label{fig:sigmamaps}  }
\end{figure*}

Figure~\ref{fig:sigmamaps} shows for all clusters the 5$^{th}$ nearest
neighbour density map of candidate cluster members, down to the m140
limit indicated in each panel.  Individual candidate members are also
shown with gray symbols, with symbol shape reflecting galaxy
morphology (Strazzullo et al., in prep.) as indicated, and symbol size
scaling with the galaxy statistical background subtraction
weight. Because clustercentric distance is not considered in the
weight determination (Sect.~\ref{sec:statsub}), these density maps are
based on the full color-selected candidate member sample, and thus
include residual background contamination.

Figure~\ref{fig:sigmamaps} shows the very prominent concentrations of
massive galaxies associated with the clusters, with the exception of
SPT-CLJ0607 that seems to exhibit a milder or less concentrated galaxy
overdensity at least in the magnitude range probed here (see also
Fig.~\ref{fig:cmd}).  We note that \citet{khullar2018} suggest the
possible presence of a background structure close to the line of sight
of SPT-CLJ0607 and at slightly higher (spectroscopic) redshift
$z=1.48$. Given the redshift difference, it is possible that this
background structure contaminates our candidate member sample, and
thus also the projected density distribution in
Fig.~\ref{fig:sigmamaps}.  On the other hand, the X-ray based (Fe-K
emission line) redshift estimate for SPT-CLJ0607, $z\sim 1.37\pm0.02$
(Bulbul et al., in prep.), is consistent with the $z=1.40$ redshift
from \citet{khullar2018}, suggesting that the $z\sim1.48$ background
structure is likely sub-dominant with respect to the more massive
$z=1.40$ cluster.

In this work, we adopt as the cluster center the position of the
brightest red galaxy (based on the observed color-magnitude diagram in
Fig.~\ref{fig:cmd}{\it a}; hereafter referred to as the brightest
central galaxy, BCG) within 100~kpc (proper) of the projected number
density peak of candidate members (Fig.~\ref{fig:sigmamaps}). The
adopted BCG is the brightest red (and most massive) galaxy lying at
the center of the red galaxy concentration associated with the
cluster. It is generally also the brightest red galaxy in the cluster
core (at least out to $r\sim250$~kpc), and the most massive galaxy of
the whole candidate member sample (Sect.~\ref{sec:colselmembers}),
with the exception of SPT-CLJ0459 for which the brightest red galaxy
is $\sim 120$~kpc away from the galaxy overdensity peak, and six
massive sources formally more massive (by a factor ranging from 5\% to
70\%) than the adopted BCG are spread along the overdensity described
by the full candidate member population.  Interestingly, despite the
less prominent galaxy overdensity of SPT-CLJ0607, its adopted BCG is
instead the largest, most massive central galaxy across the cluster
sample (see e.g., Figs.~\ref{fig:cmd}{\it a}, \ref{fig:colormass}).

In Fig.~\ref{fig:sigmamaps} the position of the adopted cluster center
(BCG; marked by a star) is compared to the candidate member
overdensity, and to the original cluster coordinates from the position
of the SZE detection (marked by a cross; the cross size shows the SPT
beam FWHM divided by the S/N of the cluster SZE detection, as an
indication of the positional uncertainty).

The distance between the adopted BCG and the center of the candidate
member overdensity ranges from $\sim$10~kpc ($\sim 1\%$ of the virial
radius $r_{200}$; SPT-CLJ0421, SPT-CLJ0459, SPT-CLJ0446) to
$\sim$60~kpc ($\sim 6\%$ of the virial radius; SPT-CLJ0607,
SPT-CLJ2040). In this respect, we note that for SPT-CLJ2040 the center
of the candidate member overdensity may be biased by the masked area
close to the cluster center (see Fig.~\ref{fig:sigmamaps}), while for
SPT-CLJ0607 it could be affected by other structure close to the line
of sight as mentioned above.

The distance between the adopted BCG and the centroid of the SZE
detection ranges instead from $\sim$80-110~kpc ($\sim10\%$ of
$r_{200}$; SPT-CLJ0421, SPT-CLJ0607, SPT-CLJ0446) to $\sim$200~kpc
($\sim20\%$ of $r_{200}$; SPT-CLJ0459, SPT-CLJ2040). Given the S/N of
the SZE detection for these clusters (Table~\ref{tab:sample}), and the
SPT beam FWHM ($\sim1.1'$ at 150~GHz), the formal positional
uncertainty ($1\sigma$) of the SZE centroid would correspond to
$\sim35-40$~kpc. This suggests that the distribution of positional
offsets between the adopted BCG and the SZE center cannot be
attributed to positional uncertainties alone. With a larger sample of
SPT-SZ clusters spanning the range $0.1 \lesssim z \lesssim 1.3$
(median redshift $\sim0.6$), \citet{song2012b} indeed concluded that
68\% (95\%) of these clusters show a BCG vs.\ SZE centroid offset of
$<0.17 r_{200}$ ($<0.43 r_{200}$, respectively), suggesting an
intrinsic positional offset of BCGs from the ICM based cluster
centroid similar to what has been estimated using X-ray selected
samples of low-redshift clusters \citep[e.g.][]{lin2004}. For higher
redshift clusters, significantly larger BCG offsets from the X-ray
centroid (up to $\sim300$~kpc, with a median offset of $\sim50$~kpc)
have been suggested \citep{fassbender2011}.

Adopting the BCG position defined as described above as the cluster
center, implicitly focuses the analysis presented in this work on the
cluster region with the highest galaxy density. Nonetheless, this
choice of the cluster center has a limited relevance with respect to
our results, given the density maps shown in Fig.~\ref{fig:sigmamaps}
and the relatively small BCG vs.\ SZE centroid distances of
$\lesssim$100~kpc for SPT-CLJ0421, SPT-CLJ0607,
SPT-CLJ0446. Concerning the two clusters with more significant offsets
of $\sim200$~kpc (SPT-CLJ0459, SPT-CLJ2040), in both cases X-ray
imaging is available and the centroid of the X-ray emission (Bulbul et
al. in prep., Mantz et al. in prep.) is closer to the adopted BCG than
to the SZE centroid, with BCG vs.\ X-ray centroid offsets of $\sim$30
and 90~kpc for the two clusters, respectively. A more detailed
investigation of the X-ray vs.\ SZE vs.\ galaxy overdensity center
offsets is beyond the scope of this work. As we only have X-ray data
on three clusters (SPT-CLJ0607, SPT-CLJ0459, SPT-CLJ2040) we conclude
in the interests of homogeneity that adopting the BCG position as the
cluster center is the most reasonable choice for this work.

\section{Cluster redshift constraints}
\label{sec:redshifts}

Out of five clusters in our sample, spectroscopic redshifts have been
obtained for SPT-CLJ2040 and more recently for SPT-CLJ0607
\citep{bayliss2014, khullar2018}. Future spectroscopic campaigns are
planned for the remaining three clusters. The new observations
presented in this work allow us to improve upon the originally
published \citep{bleem2015} photometric redshifts of these systems,
which were used for our sample definition. We employ both the RS color
m814-m140, and the IRAC color [3.6]$-$[4.5], each offering a handle on
the cluster redshift with different limitations.

\begin{figure}[]
\begin{center}
 \includegraphics[width=0.5\textwidth,viewport= 68 409 547 697, clip]{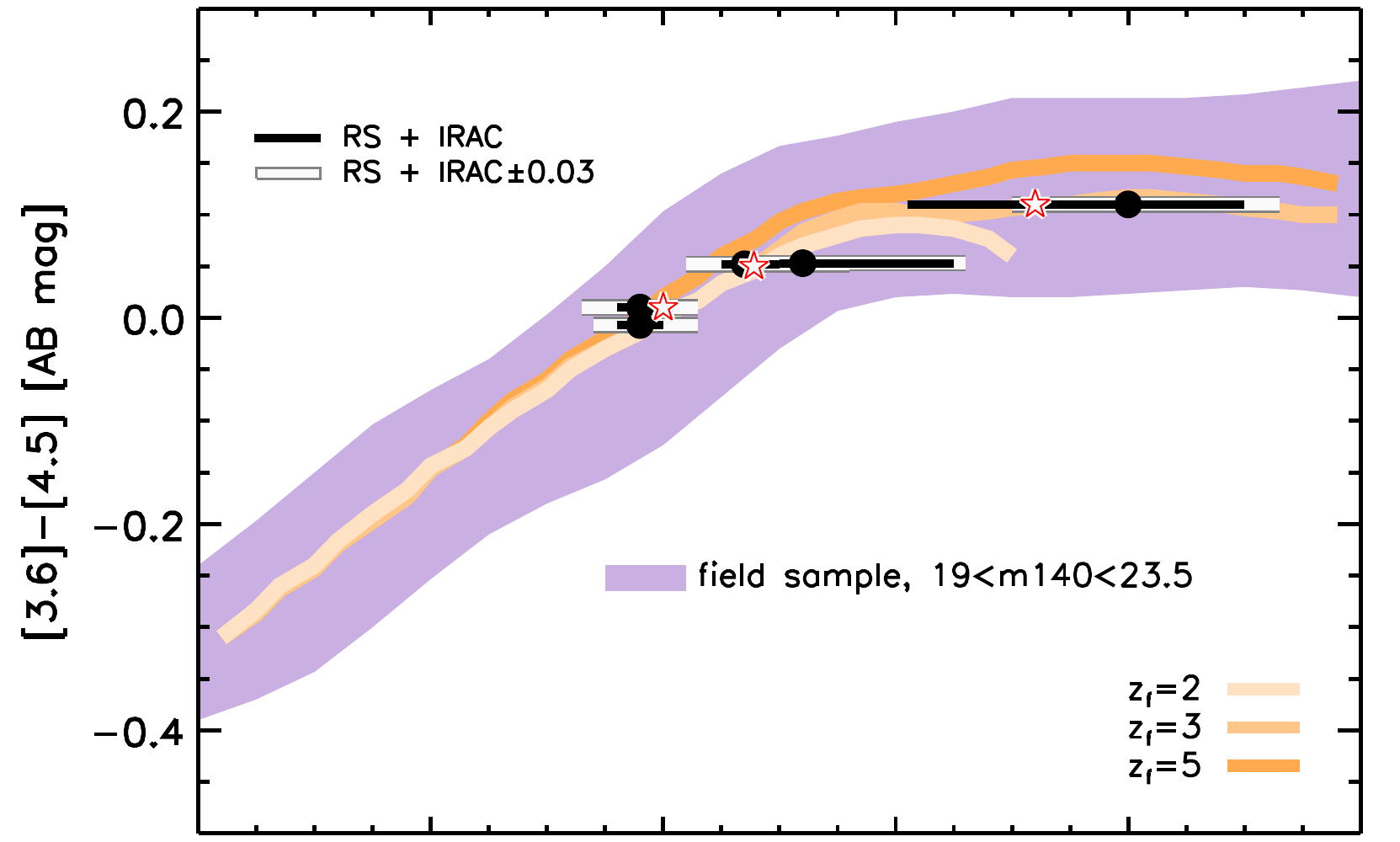}
 \includegraphics[width=0.5\textwidth,viewport= 68 368 547 697, clip]{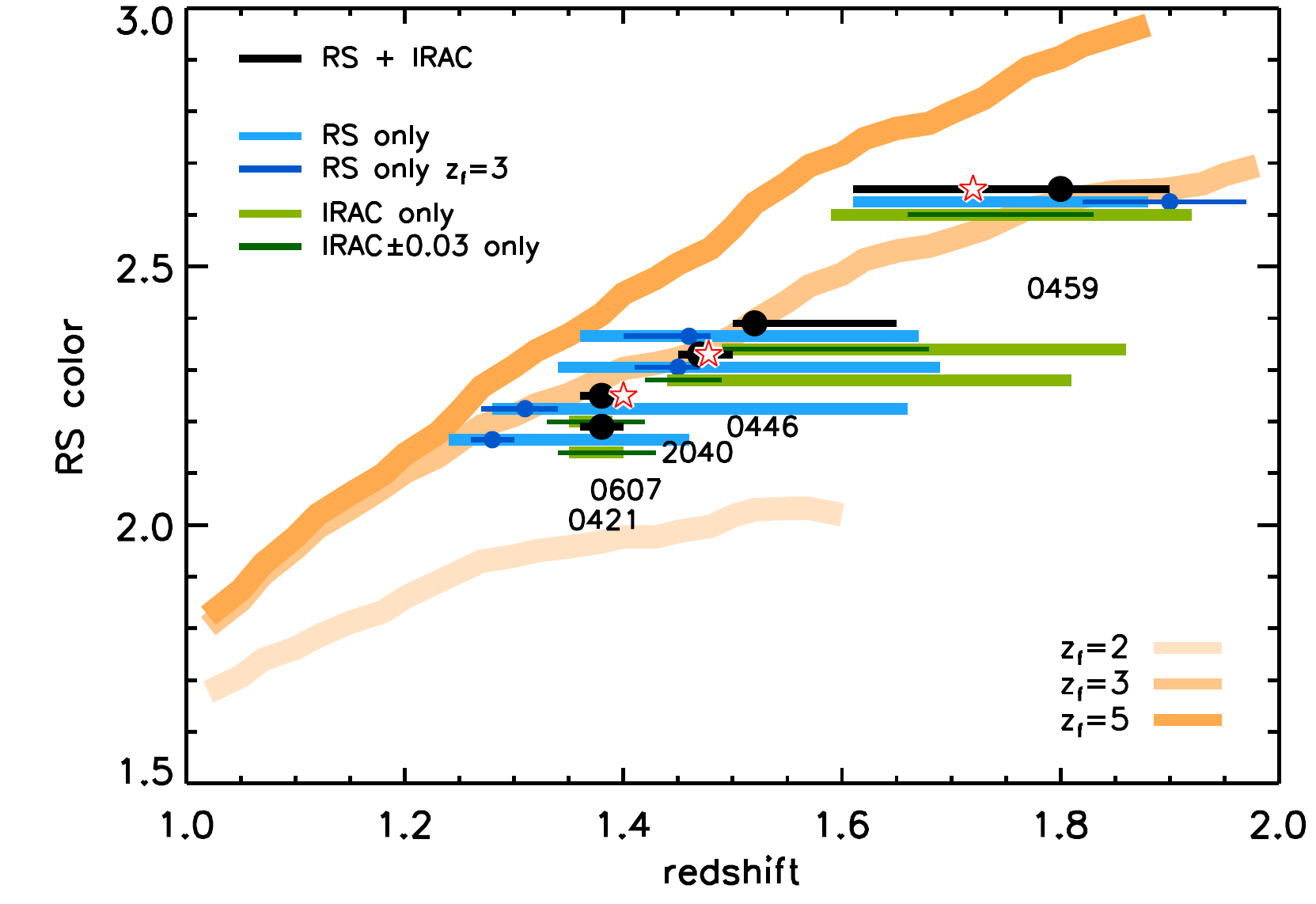} 
\end{center}
\caption{Photometric redshift constraints for all clusters (black
  points with 1$\sigma$ errorbars reported in both panels) as
  determined with the simultaneous modeling of the red sequence (RS)
  and [3.6]$-$[4.5] color in Sect.~\ref{sec:redshifts}. Red stars show
  the spectroscopic redshifts for SPT-CLJ2040 and SPT-CLJ0607, as well
  as the best current constraint for SPT-CLJ0459 (see
  Sect.~\ref{sec:redshifts}).  {\it Bottom panel -- } Adopted model RS
  m814-m140 color at M$^*$ vs.\ redshift, for formation redshifts
  $2<z_{f}<5$ (orange lines). Light blue bars show the $1\sigma$
  constraints from modeling only the RS color (dark blue assuming
  $z_f=3$). Light green bars show constraints from the [3.6]$-$[4.5]
  color alone.  {\it Top -- } Adopted model [3.6]$-$[4.5] color
  vs.\ redshift, for $2<z_{f}<5$ as in the bottom panel.  The shaded
  area shows for reference the $16^{th}-84^{th}$ percentile range of
  [3.6]$-$[4.5] color vs. redshift for field galaxies in the magnitude
  range of our samples (m140$\lesssim$23.5).  Dark green (bottom
  panel) and white (top panel) bars show the impact on the best-fit
  redshift (from IRAC color alone and RS + IRAC color, respectively)
  of a $\pm$0.03~AB~mag systematic offset on the [3.6]$-$[4.5] color.
\label{fig:redshifts} }
\end{figure}

The zeropoint and slope of the red sequence can be used to constrain
the cluster redshift with fairly good accuracy up to $z\sim1$, and
indeed this is the primary approach used for SPT clusters
\citep{song2012b,bleem2015}.  However, given the high redshifts of the
clusters studied here, we need to consider the effect of different
formation redshifts or formation histories, as shown for instance in
Fig.~\ref{fig:redshifts}.

The [3.6]$-$[4.5] color used as a probe of the 1.6$\mu$m ``stellar
bump'' \citep[e.g.,][]{sawicki2002} is much less sensitive to star
formation history, which is an advantage for use as a photometric
redshift indicator \citep[e.g.,][]{sawicki2002,papovich2008}. In spite
of the small color range spanned by galaxy populations in
[3.6]$-$[4.5] color with respect to colors probing the 4000\AA~break
(see also Fig.~\ref{fig:redshifts}), the ``stellar bump sequence'' due
to the very similar [3.6]$-$[4.5] colors of galaxies in a cluster has
indeed been shown to be a potentially effective redshift indicator for
distant clusters where the observed [3.6]$-$[4.5] color uniquely maps
onto redshift \citep[$0.7<z<1.7$ e.g.,][]{muzzin2013a}. On the other
hand, the small [3.6]$-$[4.5] color range makes this approach more
sensitive to even small systematics in the color measurement (see
Fig.~\ref{fig:redshifts}), and the flattening of the [3.6]$-$[4.5]
color vs.\ redshift at $z\sim1.6$ reduces its usefulness at the
high-redshift end of our sample.

To estimate cluster photometric redshifts, we identify a red galaxy
sample by modeling the red peak in the color distribution with a
Gaussian and selecting all galaxies within 3$\sigma$ of the peak. As
mentioned in Sect.~\ref{sec:samplesel}, here we only consider a F140W
magnitude range where we can still measure the 1'' aperture magnitude
in the F814W band with a S/N$\gtrsim$5. This criterion, together with
the high contrast of the red sequence over the field, makes our
selection of the RS sample largely insensitive to the details in the
adopted criteria, as well as to background contamination.

Although not necessarily required for a redshift estimate based on the
[3.6]$-$[4.5] color, adopting the RS sample allows us to use a
cleaner, low-contamination sample of cluster galaxies, and to
simultaneously model both the RS (m814-m140 vs. m140) and
[3.6]$-$[4.5] colors as a function of redshift in a consistent way
(that is, the RS and stellar bump sequence colors correspond to the
same galaxy population, and are thus modeled for the same formation
redshift).  For the RS color, we use RS models by
\citet{kodamaearimoto}. IRAC colors are derived using \citet{bc03}
solar-metallicity SSP models. In our modeling, we account for
photometric uncertainties and allow for a variable intrinsic scatter
in both the red sequence and stellar bump sequence.  Color
uncertainties of individual galaxies and the intrinsic scatter -
modeled as a free parameter - are summed in quadrature to give a total
observed scatter both for the red sequence and stellar bump
sequence. We account for uncertainties associated with the formation
redshift $z_{f}$ of stellar populations in the red sample by adopting
a flat prior on the formation time $t_{f}$ (the age of the Universe at
redshift $z_{f}$) corresponding to the range $2<z_{f}<5$. The
parameter space is explored with an MCMC approach.  The free
parameters are the cluster redshift, the formation time $t_{f}$, and
the intrinsic scatters in the red sequence and in the stellar bump
sequence. For each cluster, uncertainties on the estimated photometric
redshift (1$\sigma$) are quoted after marginalization over all the
other parameters.

Figure~\ref{fig:redshifts} shows the derived redshift constraints from
the full modeling of the RS and IRAC color simultaneously (black
points and error bars reported in both panels) together with the
models used for the redshift estimation.  The bottom panel of
Fig.~\ref{fig:redshifts} shows the expected \citep{kodamaearimoto} RS
color at the characteristic magnitude M$^{*}$ for different formation
redshifts, while the top panel shows the adopted model [3.6]$-$[4.5]
color vs.\ redshift for the same formation redshifts. As discussed
above and as shown in the top panel of Fig.~\ref{fig:redshifts}, the
expected [3.6]$-$[4.5] color range of the red sample for the different
assumed formation redshifts is very small as compared to the
corresponding expected range of RS colors. Considering the expected
size of possible systematics on the IRAC zero-point, as well as the
color difference between plausible models for RS galaxies with
different $z_{f}$, and the fact that stellar population models are
expected to have larger uncertainties in the restframe NIR, we also
estimate the impact of a $\pm$0.03~AB~mag systematic bias on the
expected (model) IRAC color. The resulting effect on the redshift
estimates is also shown in Fig.~\ref{fig:redshifts} (white error bars
in the top panel). For reference, the top panel also shows (shaded
area) the $16^{th}-84^{th}$ percentile range of [3.6]$-$[4.5] color vs.
redshift for galaxies in the GOODS-S control field in the same m140
magnitude range adopted for the cluster samples (m140$\lesssim 23.5$).

The bottom panel of Fig.~\ref{fig:redshifts} also shows for reference
the constraints derived using only the RS or IRAC colors, respectively
(blue and green symbols, as indicated). In particular, following
results from previous investigations of the formation epoch of bright
RS galaxies in massive clusters in this redshift range
\citep[e.g.,][]{strazzullo2010b,andreon2014,cooke2016,beifiori2017,prichard2017},
the RS-only redshift estimate obtained assuming a formation redshift
$z_{f}=3$ as typically suggested by such studies is explicitly marked
in the figure. Finally, in both panels we also show (red stars) the
available spectroscopic redshifts for the clusters SPT-CLJ2040
\citep[$z=1.478$;][]{bayliss2014} and SPT-CLJ0607
\citep[$z=1.401$;][]{khullar2018}, and the current best redshift
constraint for SPT-CLJ0459 (see discussion below).

Overall, Fig.~\ref{fig:redshifts} clearly shows the discussed
limitations of the RS and IRAC colors as cluster redshift indicators
in this redshift range, and in particular the redshift - formation
epoch degeneracy for the RS color, and the lack of redshift
sensitivity at $z\gtrsim1.6$ for the IRAC color. At the same time,
Fig.~\ref{fig:redshifts} shows to what extent we may improve cluster
photometric redshift constraints by combining both indicators in our
simultaneous modeling of the RS and stellar bump sequence colors. Up
to $z\sim1.5$, the combined modeling effectively reduces the impact of
the degeneracy between cluster redshift and formation redshift that
otherwise significantly affects redshift estimates from the RS color
only. However, at $z>1.5$ the flattening of the IRAC color
vs. redshift relation significantly limits such improvement.

All clusters' estimated redshifts are consistent with our initial
$z>1.4$ selection \citep{bleem2015}. As expected given the redshift
distribution of massive clusters, most clusters lie at the low end of
the probed $z\gtrsim1.4$ range. The highest-redshift cluster is
SPT-CLJ0459, with a photometric redshift $z=1.8^{+0.10}_{-0.19}$ from
the combined RS+IRAC color modeling.  Modeling the RS and stellar bump
colors independently yields two completely consistent redshift
estimates placing SPT-CLJ0459 at $1.6 \lesssim z \lesssim 1.9$
(Fig.~\ref{fig:redshifts}, $z_{RS}=1.76^{+0.10}_{-0.17}$,
$z_{IRAC}=1.81^{+0.11}_{-0.20}$ ). Given the mentioned limitations of
the redshift estimates from RS and stellar bump colors in this
redshift range (which indeed reflect in the quoted redshift
uncertainties), we also note that modeling the RS color limiting the
formation redshift range by plausibility arguments\footnote{Given its
  RS and IRAC colors, SPT-CLJ0459 may be reasonably expected to be at
  a redshift at least higher - and likely significantly so -
  than SPT-CLJ2040 which is confirmed at $z=1.48$. A $z_{f} = 2$,
  though still potentially appropriate for galaxies more recently
  accreted onto the red sequence, is thus very likely too low for the
  bulk of the established RS population in this cluster. Furthermore,
  as already discussed earlier in this section, most studies of the RS
  population in massive clusters at $1.4\lesssim z \lesssim1.8$
  generally favor formation redshifts around $z_{f} \sim 3$. Even in
  the most distant such massive cluster studied thus far
  \citep{andreon2014}, the formation redshift of RS galaxies is rarely
  exceeding $z_{f} = 3.5$. Therefore, a range $2.5 \leq z_{f} \leq 4$
  is plausibly appropriate for a cluster like SPT-CLJ0459. On the
  other hand, the bulk of our analysis in this section purposedly
  avoids assumptions on formation redshifts, to minimize any bias on
  our conclusions and to account for the largely unexplored potential
  diversity of cluster galaxy populations at this redshift \citep[see
    e.g. the higher $z_{f}$ estimated by][]{zeimann2012}.} to $2.5
\leq z_{f} \leq 4$ yields a redshift of 1.76$^{+0.14}_{-0.08}$. An
independent and fully consistent redshift estimate for SPT-CLJ0459 was
also derived from the modeling of the 6.7 keV Fe-K emission line
complex from {\it Chandra} observations \citep[$z=1.84\pm0.12$,
  A. Mantz priv. comm., see also][]{mcdonald2017}, and has very
recently been refined with new {\it XMM} observations to
$z=1.72\pm0.02$ (preliminary, Mantz et al. in prep.). The agreement
between these independent estimates strongly suggests that SPT-CLJ0459
is not only the most distant cluster in the SPT-SZ survey, but also
likely the most distant massive (M$_{500}>10^{14}$M$_{\odot}$)
ICM-selected cluster discovered thus far, and one of only three known
such massive systems at $z>1.7$, irrespective of selection
\citep{andreon2009,andreon2014,stanford2012,newman2014}. Given that
the much tighter constraints derived from the {\it XMM} observations
are fully consistent with results from our modeling, in the following
we adopt for SPT-CLJ0459 a redshift of $z=1.72$.

As shown in Fig.~\ref{fig:redshifts} and Table~\ref{tab:sample}, for
the two (or three, including SPT-CLJ0459) clusters for which a
spectroscopic redshift is known, the photometric redshifts derived
above are fully consistent ($\leq1\sigma$ with uncertainties from the
default combined modeling) with the spectroscopic determination,
suggesting a likely minor impact of the possible systematics on the
redshifts of the two remaining clusters as well.

\section{Galaxy populations in the central cluster region}
\label{sec:galpops}

Studies of (massive) galaxy populations in the field observe a
significant drop in the fraction of quiescent galaxies beyond $z\sim1.5$
\citep[e.g., ][]{muzzin2013c, ilbert2013}. As discussed in
Sect.~\ref{sec:intro}, the evolution of quiescent and star-forming galaxy
populations in dense environments and in particular massive clusters
is still debated. In this section we focus on the prevalence of
massive quiescent galaxies in the central cluster regions. We first
investigate the predominance of (optically) red galaxies in the
cluster vs.\ control fields (Sect.~\ref{sec:redpops}).  Because optical
colors of massive star-forming galaxies are also significantly
dust-reddened, it is important to investigate the nature of the
stellar populations in massive red galaxies (Sect.~\ref{sec:uvj}), to
properly estimate the relevance of the quiescent (vs.\ star-forming)
population and to quantify the role of the cluster environment in
suppressing star formation
(Sects.~\ref{sec:passfrac},~\ref{sec:conclusions}).

\subsection{The red population}
\label{sec:redpops}

Fig.~\ref{fig:cmd} (panels {\it a}) shows the candidate member sample
at $r<0.7 r_{500}$ in the color-magnitude diagram for each cluster (as
described in Sect.~\ref{sec:samplesel}). As a reference, the expected
location of the red sequence \citep[according to][]{kodamaearimoto} at
the cluster redshifts derived in Section~\ref{sec:redshifts} is also
shown, for different formation redshifts in the range $2 \leq z_{f}
\leq 5$ (the $z_f = 2$ model is not shown for SPT-CLJ0459 as it would
correspond to a stellar population younger than $\sim$0.5~Gyr at the
cluster redshift). The colors and slopes of the observed red sequences
are in line with the expectations for a formation redshift around
$z\sim3$ as often found in previous studies of similar systems (see
Sect.~\ref{sec:redshifts}). When comparing the color-magnitude
diagrams in Fig.~\ref{fig:cmd}{\it a}, we stress that because of our
choice of the m140 limit described in Sect.~\ref{sec:samplesel},
different clusters are probed in this Figure down to different depths
ranging from $\sim$M$^{*}$+2 at $z\sim1.4$ to $\sim$M$^{*}$+1.3 at
$z\sim1.7$. In terms of galaxy stellar mass, the adopted m140 limits
translate into stellar mass completeness limits for the red population
ranging from log(M/M$_{\odot}$)$\sim$10.55 at $z\sim1.4$ to
log(M/M$_{\odot}$)$\sim$10.85 at $z\sim1.72$. For this reason, the
appearance of the color-magnitude diagrams in Fig.~\ref{fig:cmd}{\it
  a} is bound to be different because at lower redshifts we reach
fainter, lower mass galaxies.  Nonetheless, even the color-magnitude
diagrams of clusters at very similar redshifts show some noticeable
differences. For instance, SPT-CLJ0421 and SPT-CLJ0607 at $z\sim1.4$
show an overall different galaxy overdensity in their color-magnitude
diagram (see related discussion in Sect.~\ref{sec:clustercenter}), and
in particular a different population of the red sequence and a
different color distribution (see also Fig.~\ref{fig:cmd}{\it
  b}). Similarly, there are noticeable differences between the
color-magnitude diagrams of SPT-CLJ2040 and SPT-CLJ0446, that are also
estimated to be at similar redshift $z\sim1.5$.  SPT-CLJ2040 shows a
stronger overall galaxy overdensity, and in particular a well
populated red sequence, but at the same time also exhibits a
significant overdensity of blue galaxies. These differences are
discussed further below.

\begin{figure*}
\begin{center}  
\includegraphics[width = 0.95\textwidth,viewport= 17 251 1013 766, clip]{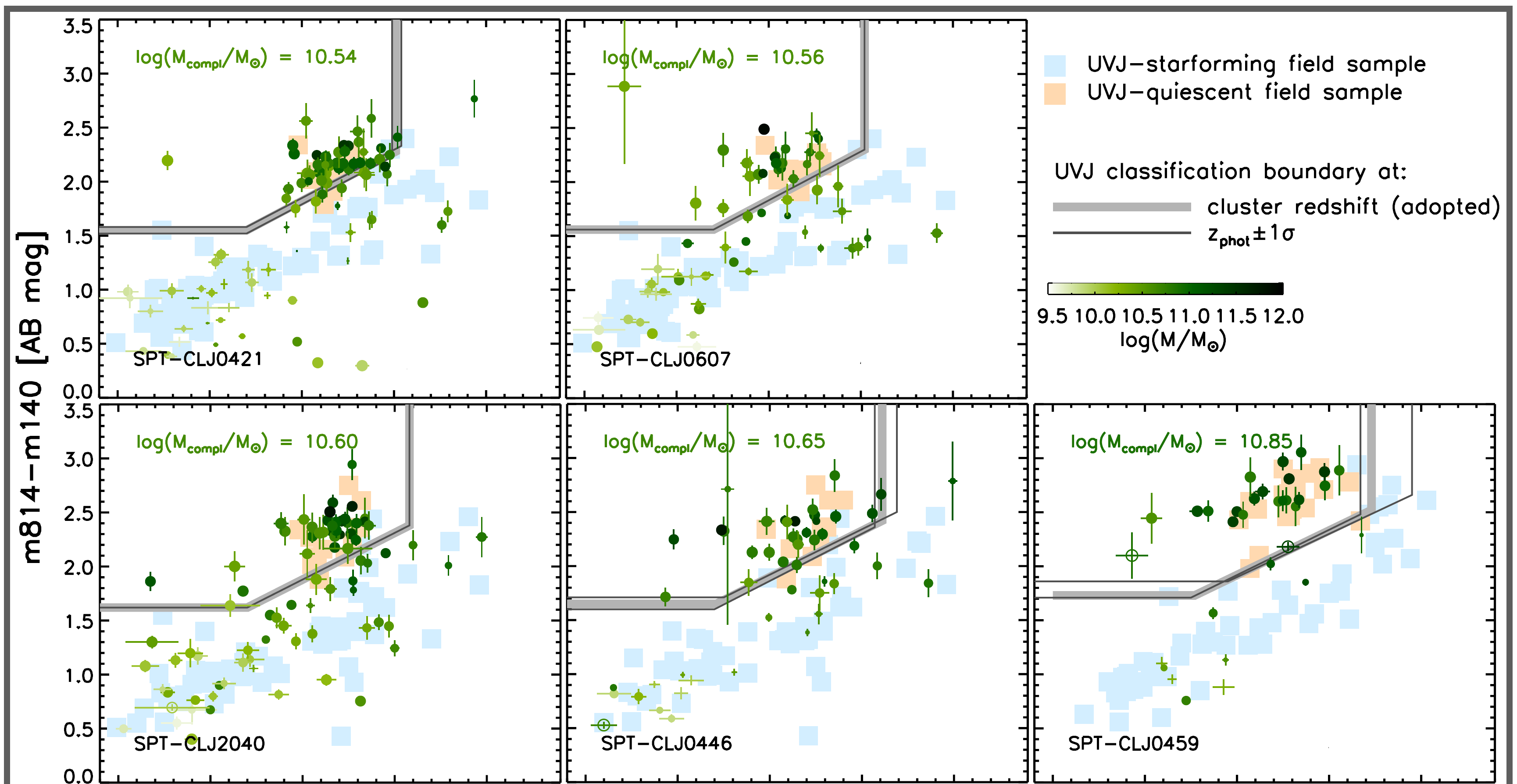}
\includegraphics[width=0.95\textwidth,viewport= 17 251 1013 762, clip]{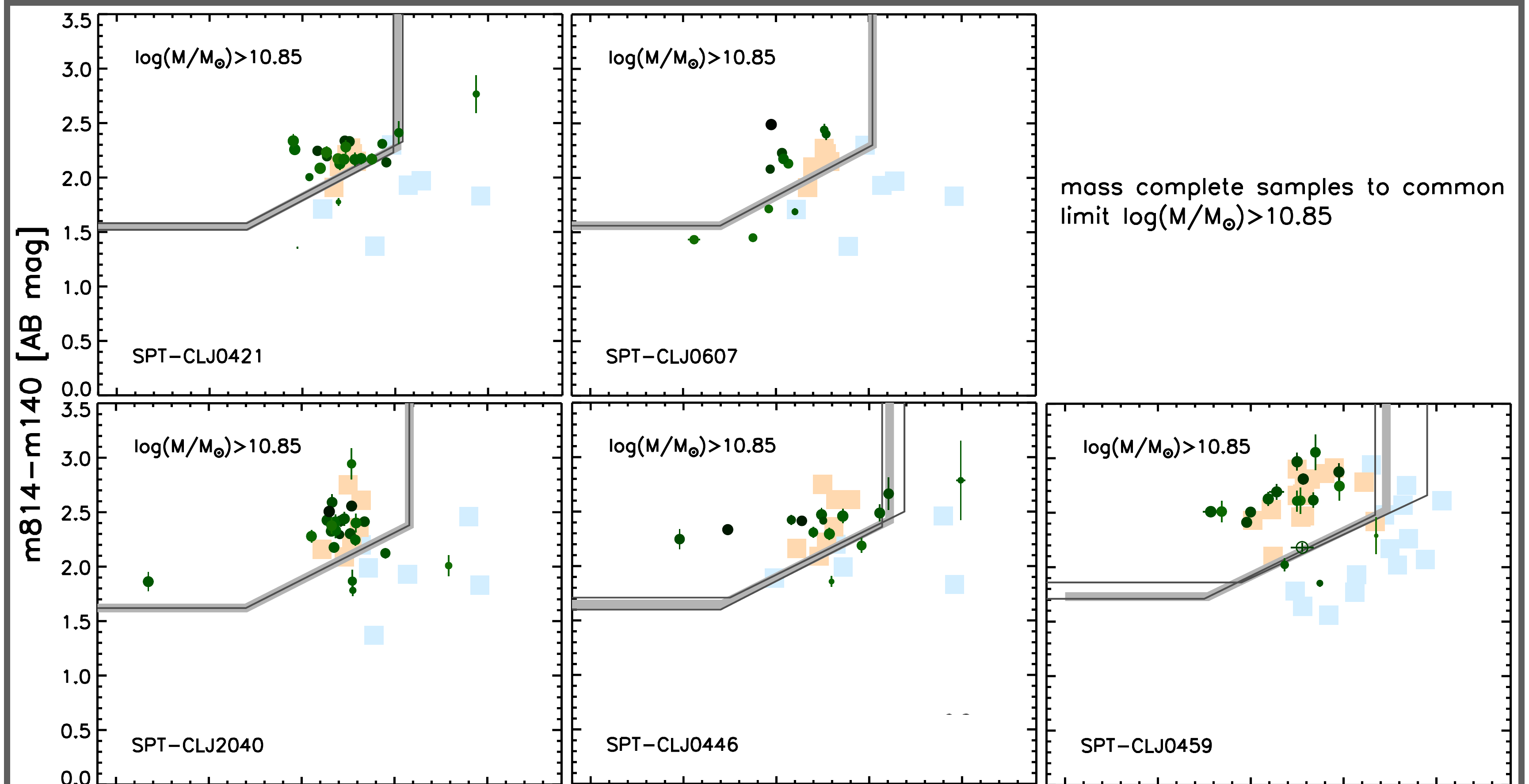}
\includegraphics[width = 0.95\textwidth,viewport= 17 190 1013 251, clip]{fig_uvj_a.pdf}
\end{center}
\caption{Adopted quiescent vs.\ star-forming galaxy
  classification. The observed m814-m140 vs.\ m140-[3.6] color-color
  diagram of candidate members (green points) is shown for all
  clusters in the $r/r_{500}<0.7$ region (the effective area covered
  for each cluster is indicated). The upper set of panels show
  m140-selected (not mass complete) candidate member samples (the mass
  completeness limit for each cluster is indicated). The lower set of
  panels show galaxies down to the common stellar mass completeness
  limit for all clusters, log(M/M$_{\odot}$)=10.85. Symbol size of
  green points scales with the statistical background-subtraction
  weights as in Fig.~\ref{fig:cmd}. Symbol color scales with stellar
  mass as shown by the color bar. Empty symbols show points for which
  a m140-[3.6] color was inferred using the 4.5$\mu$m flux (see
  Sect.~\ref{sec:uvj}).  Light red and blue squares show,
  respectively, UVJ-quiescent and UVJ-starforming galaxies from the
  control field sample with the same magnitude (top panels) or mass
  (bottom panels) threshold as candidate cluster members, and with a
  photometric redshift within $\pm0.1$ from the cluster redshift. The
  thick light-gray line shows the adopted quiescent vs.\ star-forming
  separation in the observed m814-m140 vs. m140-[3.6] color plane. For
  clusters without a final spectroscopic redshift confirmation (see
  Sect.~\ref{sec:redshifts}), the thin dark-gray lines show the
  separation that would be adopted if assuming a redshift at the edges
  of the black+white error bars in Fig.~\ref{fig:redshifts} (top).
\label{fig:uvj}}
\end{figure*}

For each cluster, Fig.~\ref{fig:cmd}{\it b} also shows the color
distribution of cluster galaxies (red symbols), where candidate
members are weighted according to the residual statistical background
subtraction discussed in Sect.~\ref{sec:statsub}, as well as by the area
coverage weight factor discussed in Sect.~\ref{sec:samplesel}.
To the extent that the control field is representative of the local
cluster background and that there are no strong asymmetries around the
cluster in galaxy population properties, the background and area
coverage corrected color distributions in Fig.~\ref{fig:cmd} will be
representative of the actual cluster galaxy population within $r<0.7
r_{500}$. As an estimate of the impact of cosmic variance on the scale
of the probed cluster core field, the orange shaded area shows the
$16^{th}-84^{th}$ percentile range of color distributions obtained by
performing the residual statistical background subtraction
(Sect.~\ref{sec:statsub}) considering, rather than the full control field,
100 fields of size $r=0.7 r_{500}$ located at random positions in the
GOODS-S field (see Sect.~\ref{sec:complcontsamples}).

For comparison, the figure also shows the color distribution of
galaxies in the control field at similar redshift (blue histograms,
scaled to the total number of galaxies in the background and area
coverage corrected cluster sample). The darker blue histogram refers
to a field sample selected by photometric redshifts within a range
$\pm 0.2$ from the cluster redshift, and in the same m140 magnitude
range as the candidate member sample. The lighter blue histogram
refers instead to a field sample obtained with the same magnitude and
color criteria as the candidate member sample.

For each cluster in our sample, a marked excess of red galaxies,
typically in a tight color range (the intrinsic scatter in the
observed m814-m140 color, as derived from the MCMC modeling in
Sect.~\ref{sec:redshifts}, is in the range $0.1 \pm 0.02$ to $0.15 \pm
0.04$), makes the color distribution of cluster galaxies in the probed
magnitude range clearly different from that of the field analogs.

\begin{figure*}
 \includegraphics[width=0.345\textwidth,viewport= 53 350 411 720, clip]{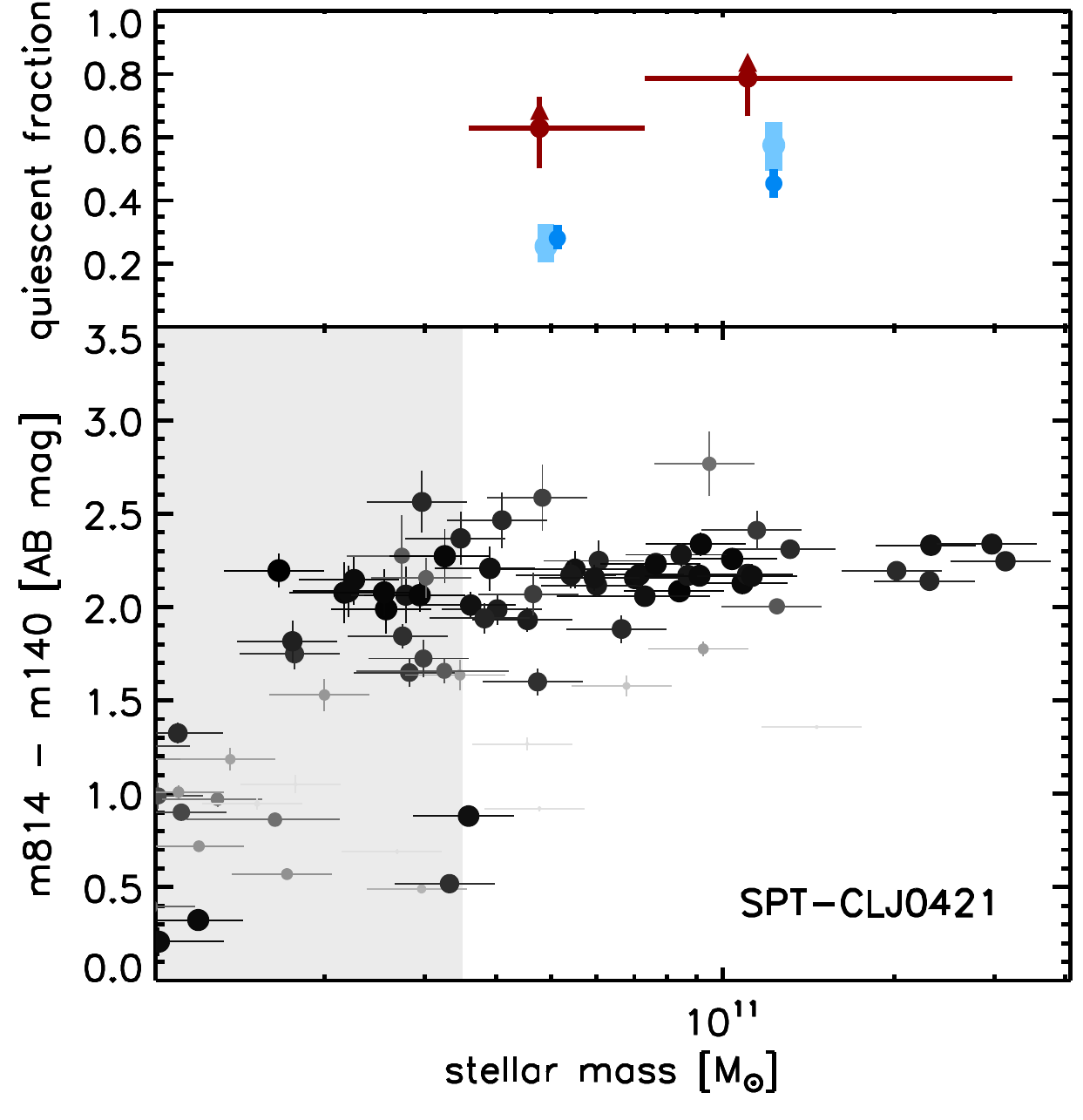}%
 \includegraphics[width=0.3\textwidth,viewport= 100 350 411 720 , clip]{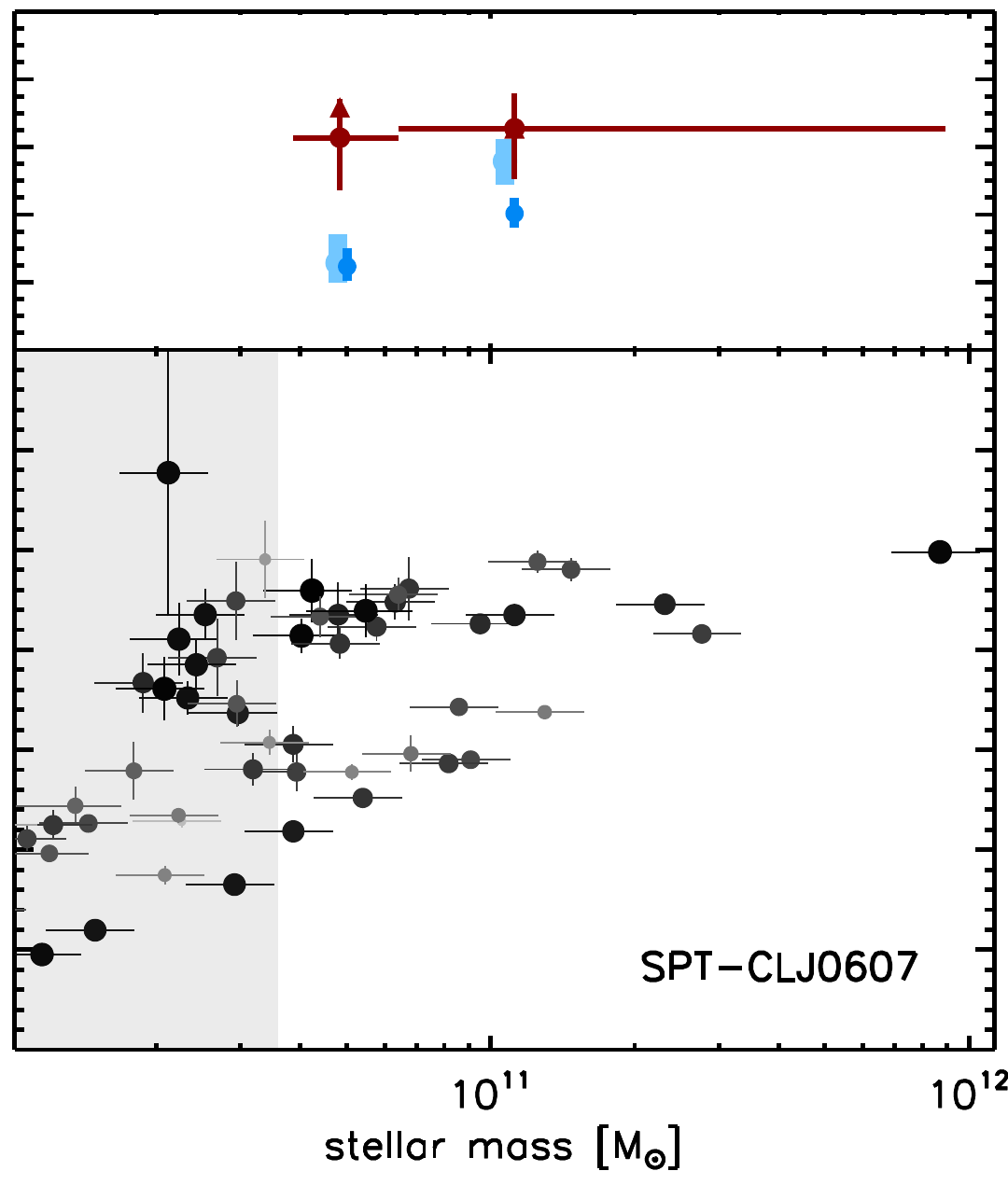}%
 \includegraphics[width=0.35\textwidth,viewport= 100 425 395 720, clip]{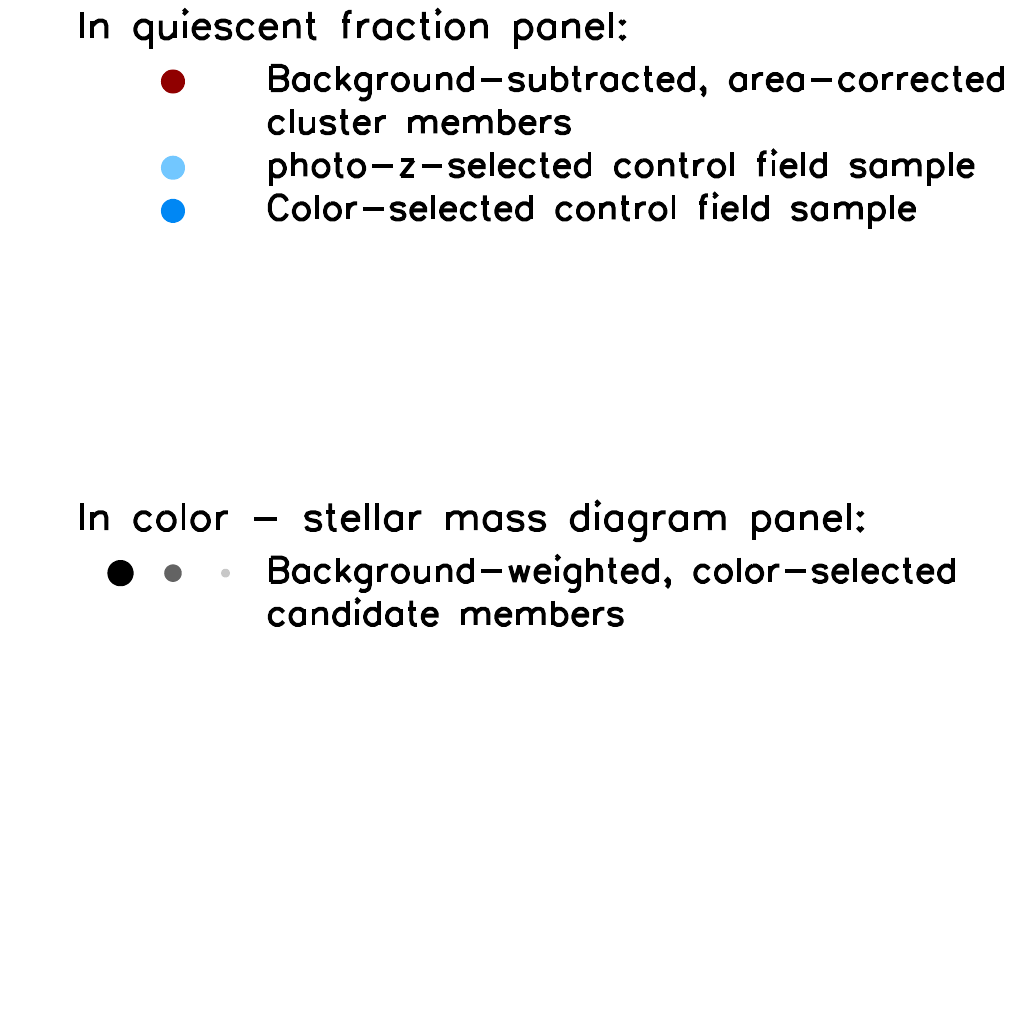}
 \includegraphics[width=0.345\textwidth,viewport= 53 350 411 720, clip]{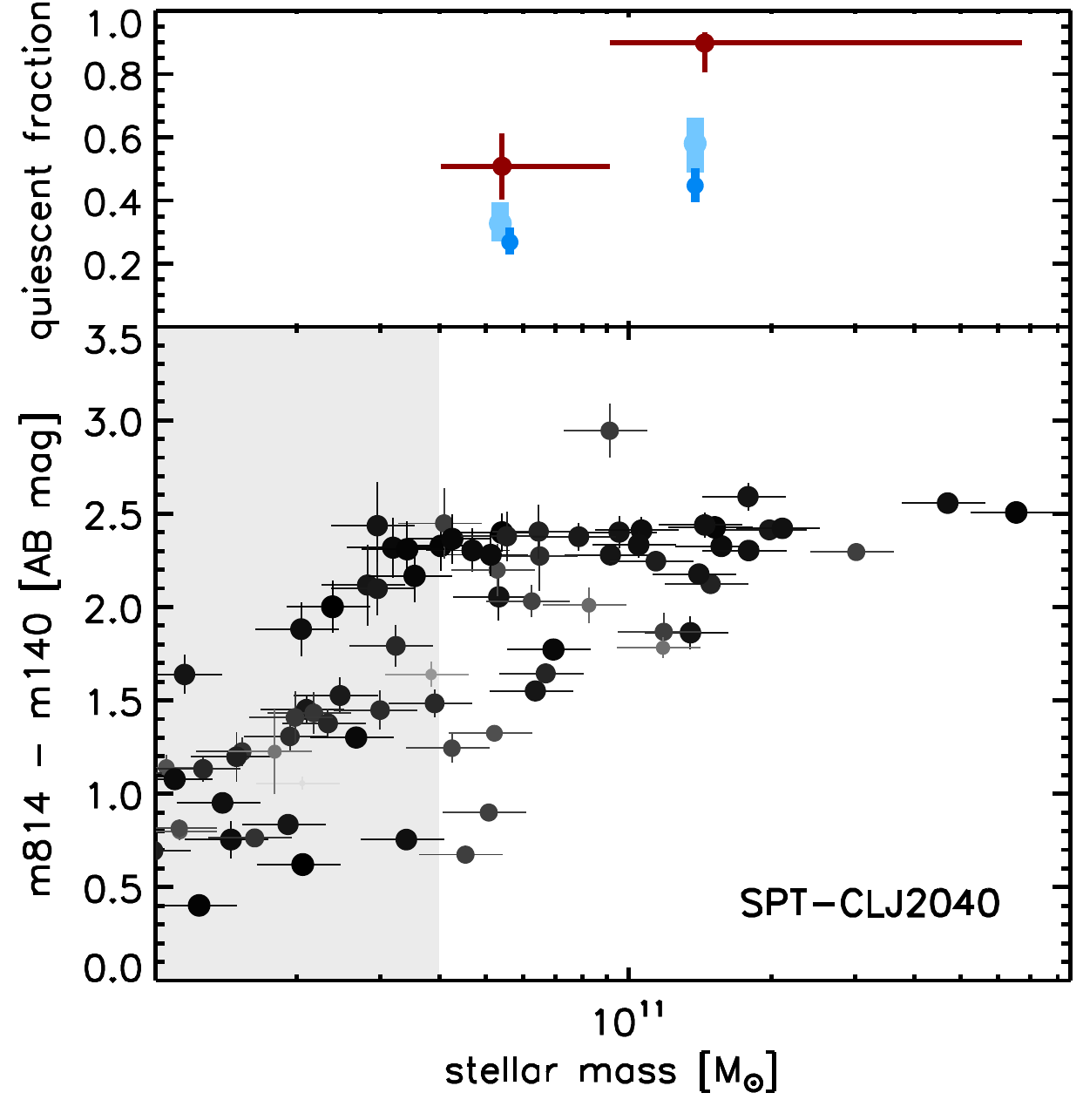}%
 \includegraphics[width=0.3\textwidth,viewport= 100 350 411 720, clip]{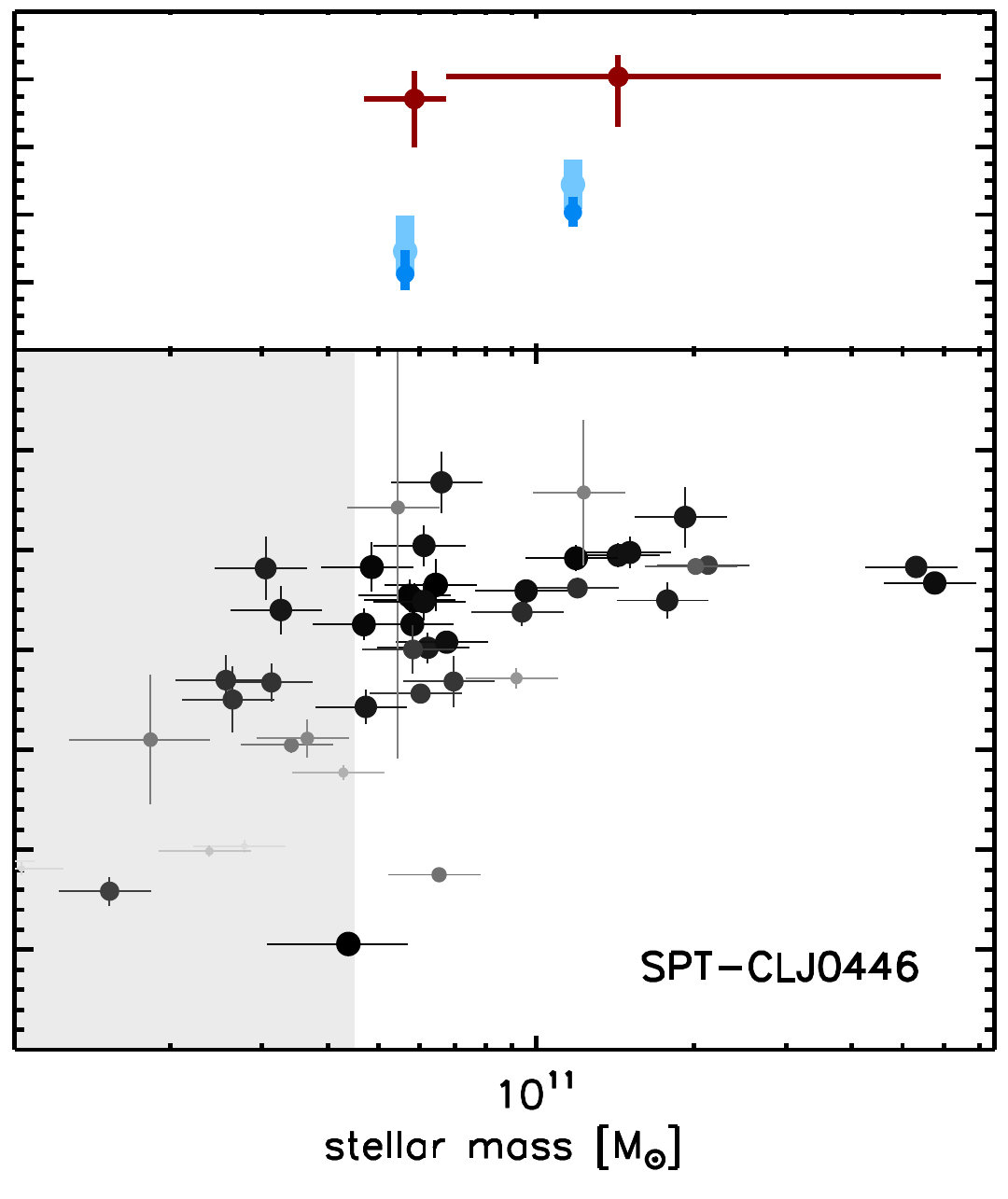}%
 \includegraphics[width=0.3\textwidth,viewport= 100 350 411 720, clip]{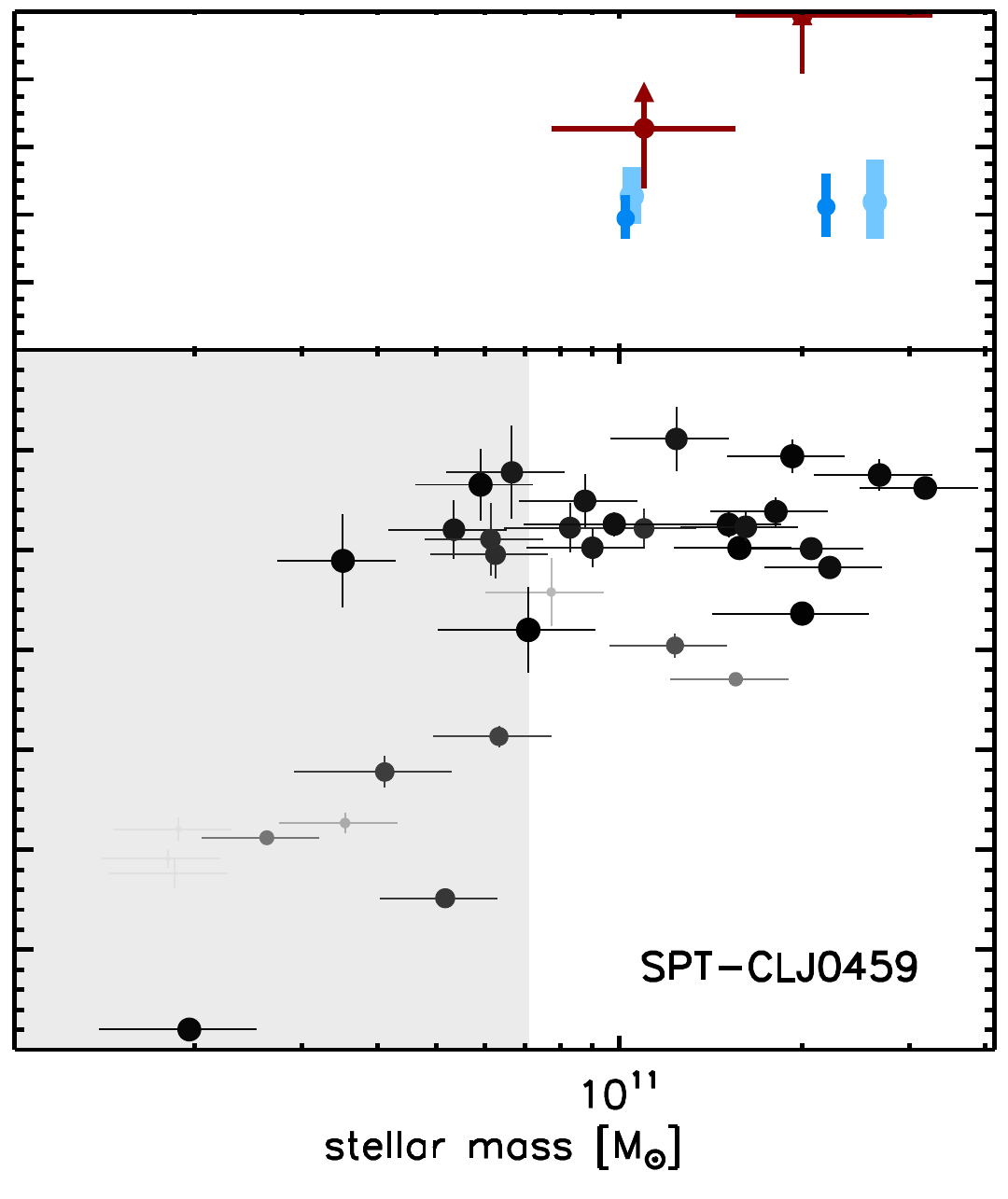}
\caption{For each cluster: {\it Bottom panel: } m814-m140 color
  vs.\ stellar mass of candidate cluster members within $r<0.7
  r_{500}$. Symbol size and color reflect the statistical background
  subtraction weight as in Fig.~\ref{fig:cmd}. The light-gray shading
  shows the stellar mass range below the mass completeness limit. {\it
    Top panel:} quiescent fraction within $r<0.7 r_{500}$ at
  masses above the individual mass completeness limit of the cluster
  (red symbols; error bars show the width of the stellar mass bin and
  the 1$\sigma$ binomial confidence intervals, see
  Sect.~\ref{sec:uvj}). Where applicable, red triangles show
  quiescent fractions assuming that all unclassified galaxies are
  quiescent (rather than all star-forming, see
  Sect.~\ref{sec:uvj}). Blue symbols show the quiescent fraction in the
  same stellar mass bins for photo-z and color-selected field samples
  as in Fig.~\ref{fig:cmd}, as indicated. All symbols are plotted at
  the median mass of the given sample.
\label{fig:colormass} }
\end{figure*}

 As mentioned above, there are differences between the color-magnitude
 diagrams of different clusters, and in particular their color
 distributions and red galaxy
 fractions. Figure~\ref{fig:redfracvsNgal/Mhalo} (bottom panel) shows
 the red galaxy fraction estimated to a common limit of
 $\sim$M$^*$+1.3 for all clusters\footnote{We stress that, in contrast
   with the quiescent fractions discussed later, the red fractions
   quoted here refer to flux-limited rather than mass complete
   samples.  }, highlighting in a more direct, quantitative way the
 more significant blue population in SPT-CLJ0607 and SPT-CLJ2040 that
 can already be seen in Fig.~\ref{fig:cmd}. While such differences in
 galaxy population properties can obviously be due to
 cluster-to-cluster variation, in principle they could also be
 artificially produced, at least to some extent, by differences in the
 local background of the different clusters. In
 Appendix~\ref{sec:redfracvsbackground}, we redetermine the red galaxy
 fraction adopting as control fields the random sub-fields described
 above for the estimation of the effect of small-scale cosmic
 variance. Fig.~\ref{fig:redfracvsNgal/Mhalo} shows that the red
 galaxy fraction is affected by at most $10\%$ even considering
 sub-fields with a galaxy density at the 10$^{th}$ and 90$^{th}$
 percentile level across the default control field. As further
 discussed in Appendix~\ref{sec:redfracvsbackground}, we do not have
 any significant evidence from the data studied here that differences
 in the local background density are the actual cause of differences
 in the estimated galaxy population properties among these clusters,
 although we have already noted that SPT-CLJ0607 could be affected by
 the presence of a background structure close to the line of sight
 (Sect.~\ref{sec:clustercenter}), and that the inferred background
 density for SPT-CLJ2040 tends to be marginally higher than the
 typical expectations from the control field
 (Sect.~\ref{sec:complcontsamples}). Nonetheless, we present in
 Fig.~\ref{fig:redfracvsNgal/Mhalo} our current best estimates of the
 possible systematics on the red galaxy fraction (which then translate
 to the quiescent galaxy fraction) coming from differences in the
 background densities of the cluster and adopted control
 fields. According to these estimates, the cluster that could be,
 potentially, more significantly affected is SPT-CLJ2040, for which
 our default red fraction measurement might in principle underestimate
 by up to $\sim$30\% the actual value. We note once again, though,
 that Appendix~\ref{sec:redfracvsbackground} presents these results
 for completeness, and we currently have no significant evidence to
 suggest that differences in the local cluster background densities
 may be the main source of differences in their observed galaxy
 population properties.

  Differences in galaxy population properties of clusters in this
  sample are discussed further in the following Sections.

    \begin{table*}
  \centering
  \caption{The quiescent fraction (corrected for area coverage and
    background contamination) estimated for each cluster in the
    $r<0.45r_{500}$ and $r<0.7r_{500}$ regions, for stellar masses
    above the individual cluster mass completeness limit as well as
    the common limit of log(M/M$_{\odot}$)=10.85. For comparison, the
    quiescent fraction estimated down to the same stellar mass limits
    on the corresponding photo-z selected control field samples is
    also listed. The errors refer to 1$\sigma$ binomial confidence
    intervals. Where applicable, quiescent fractions determined assuming
    that all unclassified galaxies in the sample are quiescent (rather
    than all being star-forming, see Sect.~\ref{sec:uvj}) are shown in
    parentheses.
\label{tab:passfrac}}
\vspace{0.2cm}

\begin{tabular}{l l l l l l l}
            \cline{2-7}
             & \multicolumn{6}{|c|}{Quiescent fraction (\%) }\\
            \cline{2-7}
            & \multicolumn{3}{|c|}{M$>$M$_\mathrm{compl,cluster}$} & \multicolumn{3}{|c|}{log(M/M$_{\odot}$)$>$10.85}\\
            \hline
            \multicolumn{1}{|c|}{cluster} &  \multicolumn{1}{|c|}{$r<0.45r_{500}$} &  \multicolumn{1}{|c|}{$r<0.7r_{500}$} &  \multicolumn{1}{|c|}{field}  &  \multicolumn{1}{|c|}{$r<0.45r_{500}$} &  \multicolumn{1}{|c|}{$r<0.7r_{500}$} &  \multicolumn{1}{|c|}{field}   \\
     \midrule
     SPT-CLJ0421-4845 &  $70^{+8}_{-11}$ ($79^{+6}_{-11}$)  & $71^{+6}_{-9}$~~~($76^{+6}_{-8}$) & $41^{+8}_{-7}$  &  $91^{+3}_{-16}$ ($100^{+0}_{-14}$) &  $80^{+6}_{-12}$ ($84^{+5}_{-11}$) &  $56^{+11}_{-12}$  \vspace{0.1cm}\\
     SPT-CLJ0607-4448 &  $58^{+11}_{-12}$ &   $62^{+9}_{-10}$ ($66^{+8}_{-11}$) &   $37^{+7}_{-6}$ &   $78^{+8}_{-20}$ &  $64^{+11}_{-16}$ &   $59^{+10}_{-12}$  \vspace{0.1cm}\\
     SPT-CLJ2040-4451 &  $66^{+7}_{-9}$ &   $73^{+5}_{-7}$ &   $36^{+8}_{-7}$ &   $81^{+6}_{-13}$ &   $88^{+4}_{-9}$ &   $53^{+11}_{-12}$  \vspace{0.1cm}\\
     SPT-CLJ0446-4606 &   $91^{+3}_{-11}$ &   $77^{+6}_{-10}$ &  $39^{+9}_{-8}$ &   $92^{+3}_{-17}$ &   $85^{+5}_{-16}$ &   $56^{+11}_{-12}$  \vspace{0.1cm}\\
     SPT-CLJ0459-4947 &  $90^{+3}_{-17}$ ($100^{+0}_{-15}$) &  $83^{+6}_{-12}$ ($88^{+4}_{-12}$) &  $49^{+8}_{-8}$ &  $90^{+3}_{-17}$ ($100^{+0}_{-15}$)  &  $83^{+6}_{-12}$ ($88^{+4}_{-12}$) &  $49^{+8}_{-8}$ \vspace{0.1cm}\\
            \hline
        \end{tabular}
    \end{table*}

\subsection{The quiescent population}
\label{sec:uvj}

Figure~\ref{fig:colormass} (bottom panel of each subfigure) shows the
stellar mass vs.\ m814-m140 color diagram for candidate members of the
five clusters, with the same symbol size and color coding as the
color-magnitude diagrams in Fig.~\ref{fig:cmd}. The massive cluster
galaxy population is dominated by red sources, which may be due to
older stellar populations of mostly quiescent galaxies, but also to
higher dust attenuation of more massive star-forming galaxies
\citep[e.g.,][]{garnbest2010,pannella2015}. 

To disentangle quiescent vs.\ dusty star-forming populations, we adopt
an approximate version of the restframe U-V vs.\ V-J (hereafrer UVJ)
color-color diagram \citep[e.g.,][]{labbe2005,williams2009}, which has
been routinely used over the last decade for studying star-forming
vs.\ quiescent populations up to high redshift.  In Fig.~\ref{fig:uvj}
we show the m814-m140 vs.\ m140-[3.6] colors of candidate cluster
members in all cluster fields. In the probed redshift range
$1.4\lesssim z \lesssim1.8$ the F814W, F140W and 3.6$\mu$m bands
probe, respectively, the rest-frame ranges $\sim2900-3400$~\AA~,
$\sim5000-6000$~\AA~, and $\sim13000-15000$~\AA~. In fact, these
passbands were explicitly selected to approximate the restframe UVJ
color diagram.

Green points in Fig.~\ref{fig:uvj} show all candidate members,
color-coded by their estimated stellar mass, with point size scaling
according to the statistical background subtraction weight determined
in Sect.~\ref{sec:statsub}. As discussed in Sect.~\ref{sec:statsub}, although
these weights do not directly translate into membership likelihood on
a galaxy-by-galaxy basis, the distribution of weights across the
m814-m140-[3.6] diagram is expected to be representative of the
cluster galaxy population.

All panels in Fig.~\ref{fig:uvj} refer to the $r < 0.7 r_{500}$
region. The (m140-selected) galaxy samples shown in the upper set of
panels are not mass complete -- the mass completeness limit is
indicated in each panel.  In the lower set of panels we only show
cluster candidate members above a common mass completeness limit of
log(M/M$_{\odot}$)$>$10.85, which is reached in the most distant
cluster. Therefore, although most of the star-forming population (by
number) falls below this mass threshold, the bottom panels show a more
proper comparison of the relevance of the quiescent and star-forming
components of massive galaxy populations across the different clusters
in our sample, and with respect to the field, which is discussed in
detail in Sections~\ref{sec:passfrac} and \ref{sec:conclusions}.

Fig.~\ref{fig:uvj} also shows the color-color criterion adopted to
separate quiescent from star-forming galaxies (thick gray line), which
was determined based on \citet{bc03} stellar population models as
described in \citet{strazzullo2016}. The thin dark-gray lines show how
the quiescent vs.\ star-forming separation would be displaced if the
cluster redshift were moved at the edges of the white+black error bar
range shown in the top panel of Fig.~\ref{fig:redshifts}.  For
reference, we show with light red (blue) squares field galaxies from
the GOODS-S sample at similar redshift as the cluster, with the same
magnitude (top panels) or mass (bottom panels) selection as candidate
cluster members, and classified as quiescent (star-forming,
respectively) based on their restframe UVJ colors. This confirms that
our adopted color-color criterion is analogous to the usual UVJ
classification in terms of selecting the same quiescent and
star-forming sub-samples, and also visually confirms the locations in
the color-color diagram of the quiescent and star-forming galaxy
populations in the cluster fields with respect to a reference (field)
population at the same redshift. A quiescent clump and a star-forming
sequence consistent with the expected colors are indeed clearly
visible in the color-color diagrams of all clusters.

\begin{figure}
\includegraphics[height=0.17\textwidth,viewport= 50 478 74 670, clip]{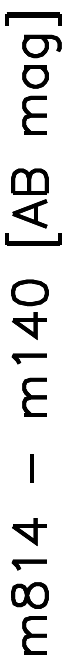}%
\includegraphics[height=0.17\textwidth,viewport= 74 406 269 720, clip]{fig_histopass_0421.pdf}%
\includegraphics[height=0.17\textwidth,viewport= 102 406 269 720, clip]{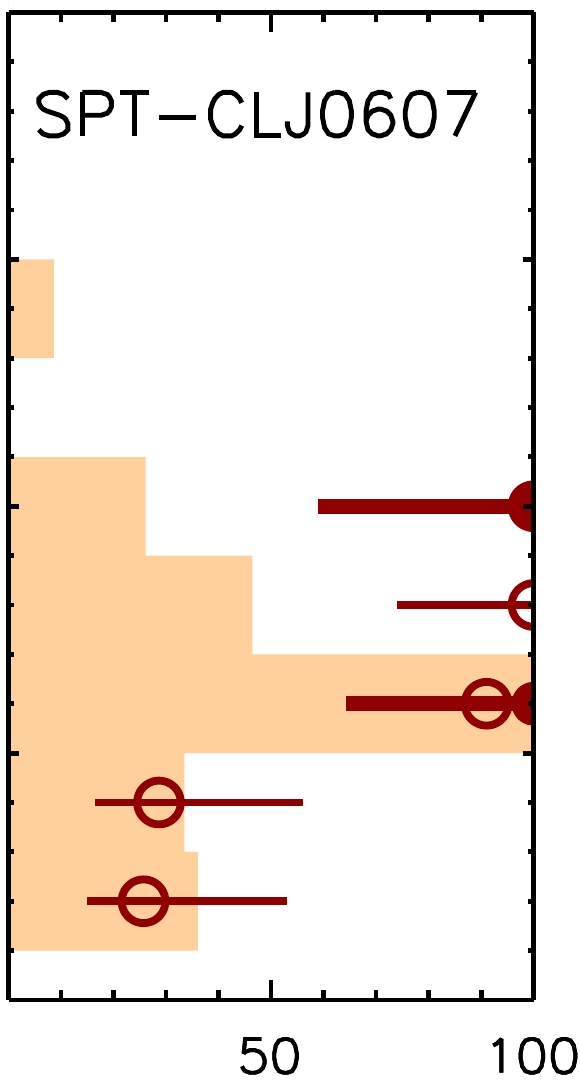}%
\includegraphics[height=0.17\textwidth,viewport= 102 406 269 720, clip]{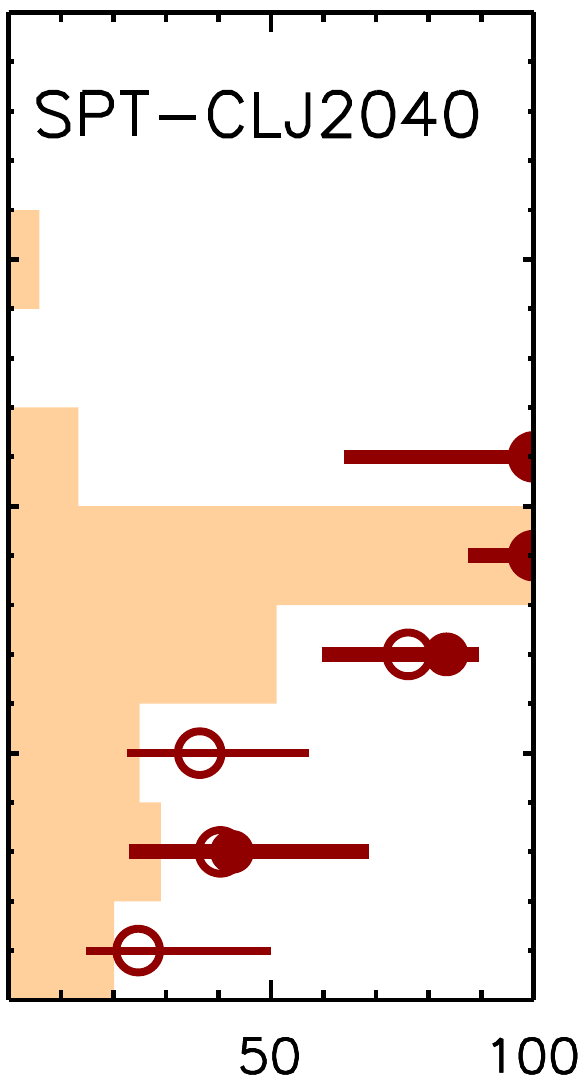}%
\includegraphics[height=0.17\textwidth,viewport= 102 406 269 720, clip]{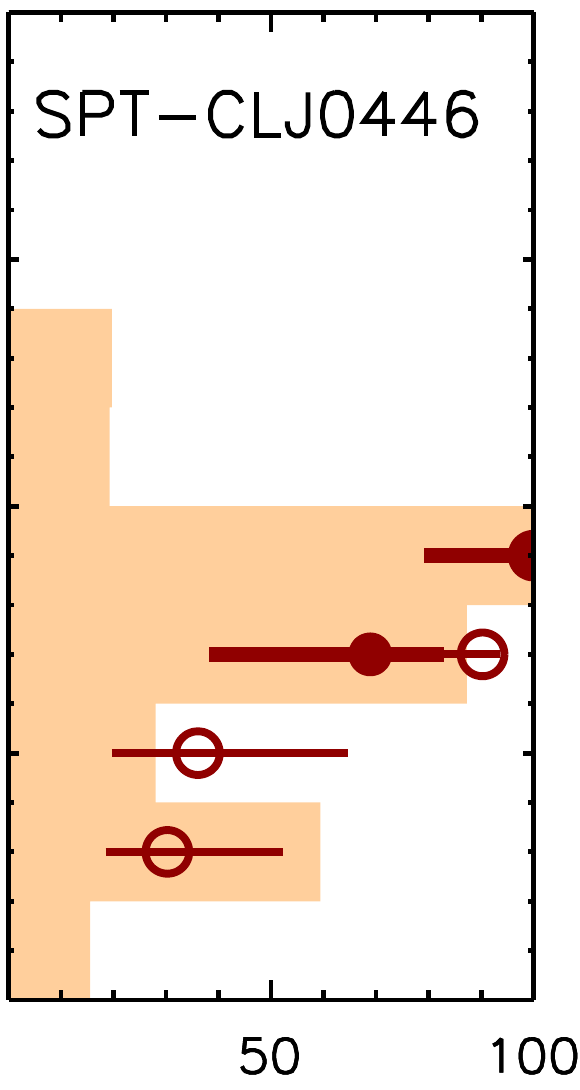}%
\includegraphics[height=0.17\textwidth,viewport= 102 406 269 720, clip]{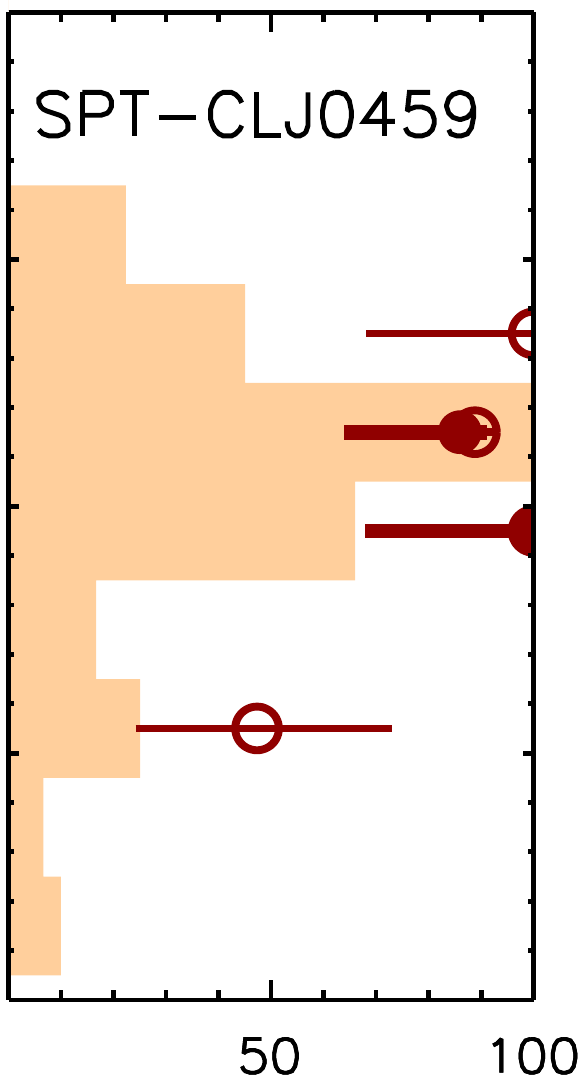}
\centering
\includegraphics[height=0.025\textwidth,viewport= 50 385 300 407, clip]{fig_histopass_0459.pdf}
\caption{Quiescent fraction vs. m814-m140 color for red
  (m814-m140$>$1.5) galaxies in all five clusters.  Galaxies bluer
  than m814-m140=1.5 are not shown as they are always classified as
  star-forming (thus 0\% quiescent fraction, see Fig.~\ref{fig:uvj}).
  Shaded histograms represent the color distribution of cluster
  galaxies as in Fig.~\ref{fig:cmd}, arbitrarily rescaled. Empty and
  solid symbols show, respectively, quiescent fractions for the
  m140-selected and for the log(M/M$_{\odot}$)$>10.85$ (common) mass
  complete samples.  Error bars show binomial confidence intervals
  (1$\sigma$). Color bins with $<2$ galaxies are not shown.
 \label{fig:histopass}}
\end{figure}

\begin{figure*}
 \includegraphics[width=0.383\textwidth,viewport= 30 376 459 716,
   clip]{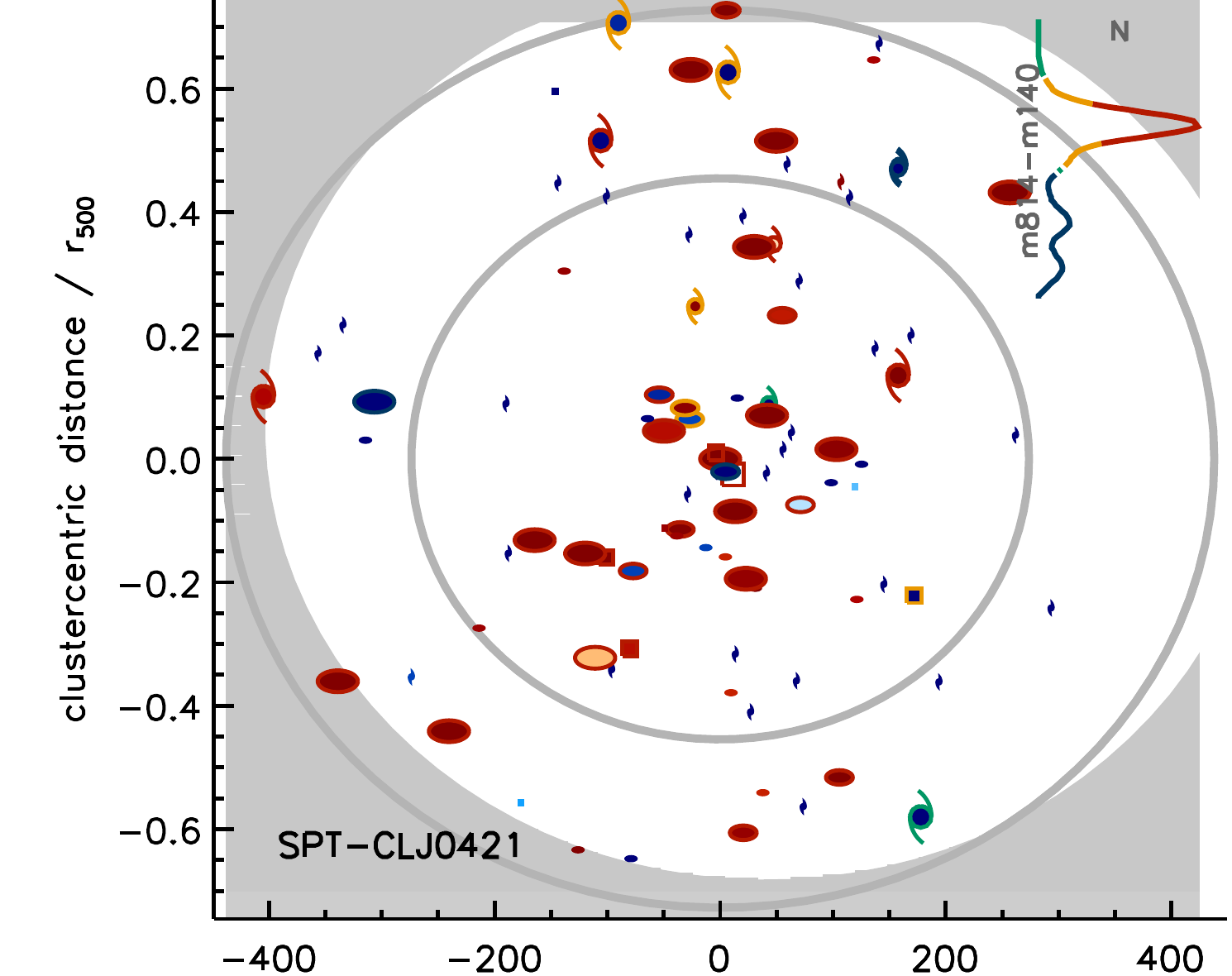}%
 \includegraphics[width=0.32\textwidth,viewport= 101 376 459 716 ,
   clip]{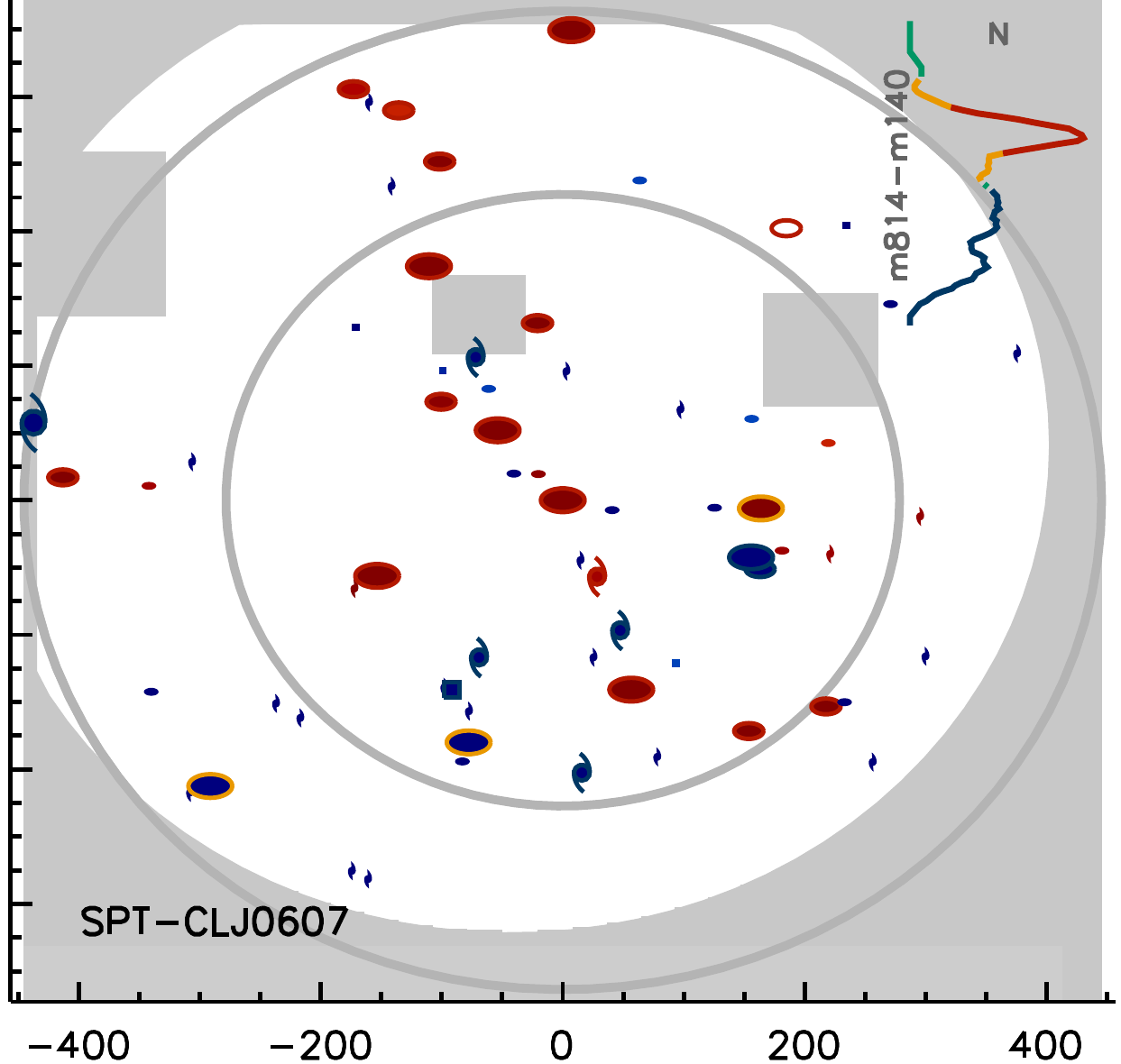}%
 \includegraphics[width=0.32\textwidth,viewport= 128 300 560 630,
   clip]{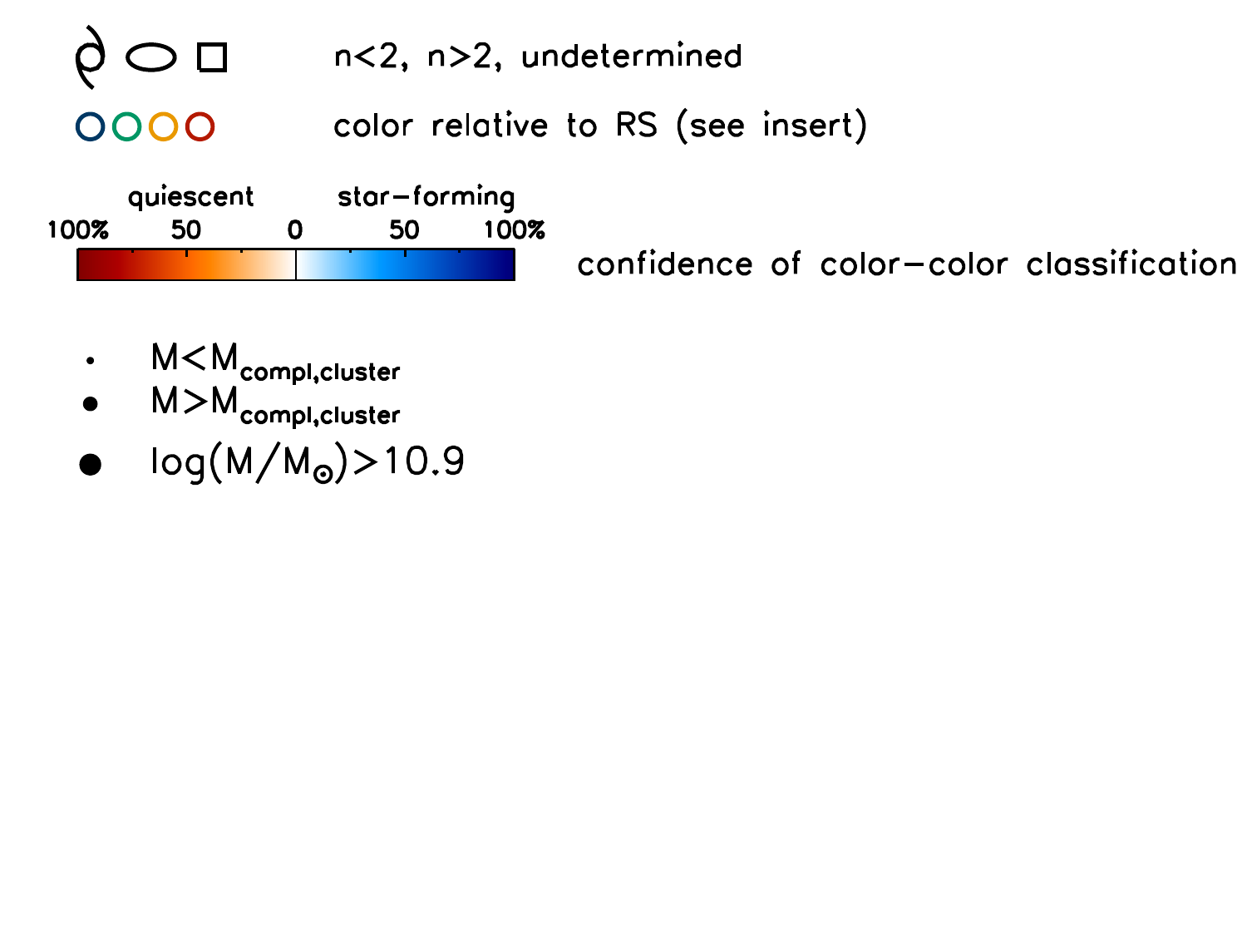}
\includegraphics[width=0.383\textwidth,viewport= 30 358 459 716,
   clip]{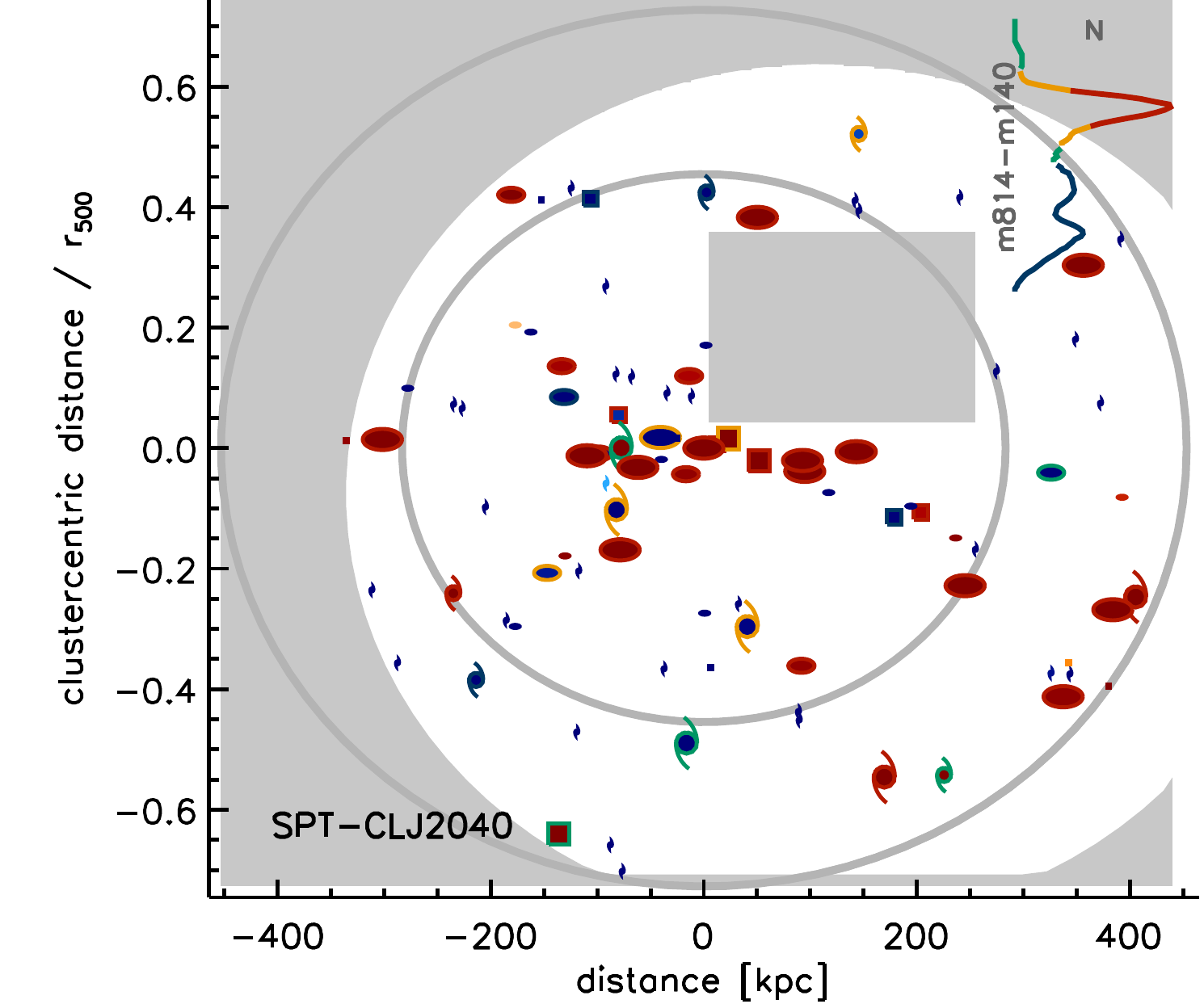}%
 \includegraphics[width=0.32\textwidth,viewport= 101 358 459 716,
   clip]{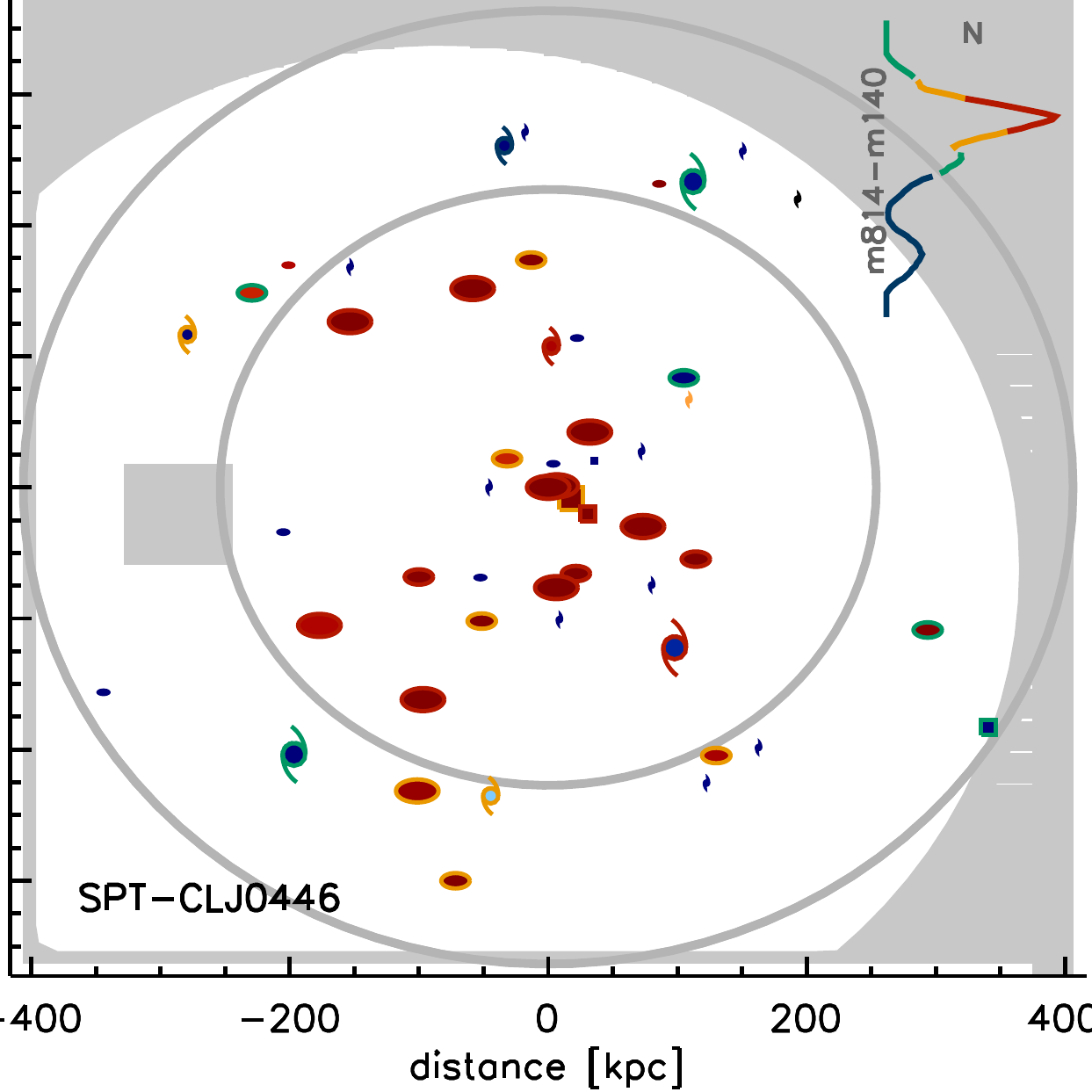}%
 \includegraphics[width=0.32\textwidth,viewport=101 358 459 716,
   clip]{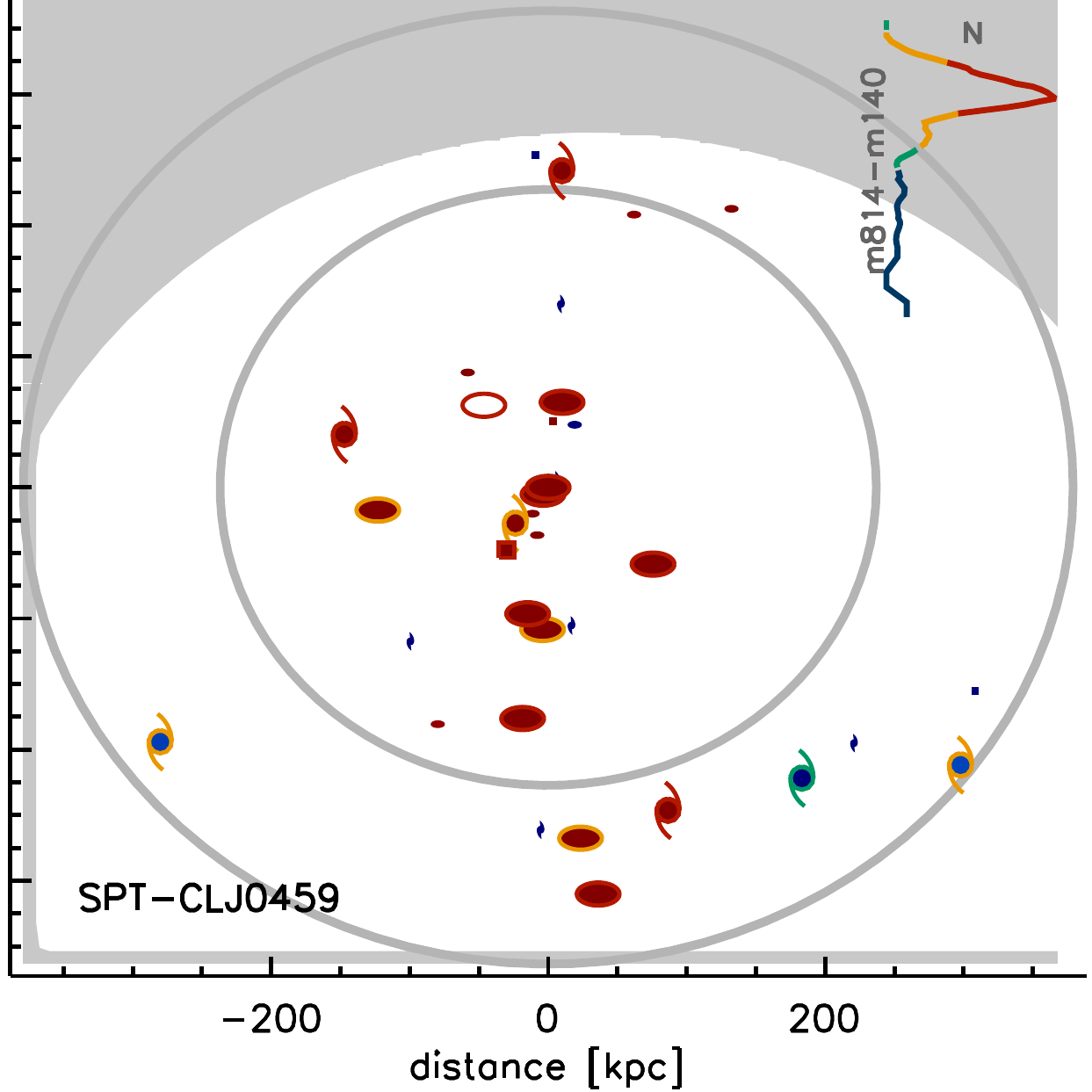}
\caption{Projected distribution of candidate cluster members. All
  candidate members are shown regardless of their statistical
  background subtraction weight. Each symbol has two colors.  The
  color of the symbol boundary scales with the m814-m140 color of each
  galaxy relative to the cluster RS color (see insert). The internal
  symbol color, in red or blue shade according to the galaxy
  classification as quiescent or starforming, scales with the
  reliability of this classification only accounting for the distance
  of the galaxy colors from the adopted selection dividing line and
  the relevant photometric errors. Symbol shape reflects the estimated
  Sersic index as indicated.  Masked areas and regions deemed to be
  outside of our uniform coverage are gray shaded. The two circles
  show the limiting radii of the adopted cluster regions ($r<0.45
  r_{500}$, $r<0.7 r_{500}$).
\label{fig:panoramix}}
\end{figure*}

\begin{figure}
 \includegraphics[width=0.49\textwidth,viewport=65 410 512 720,clip]{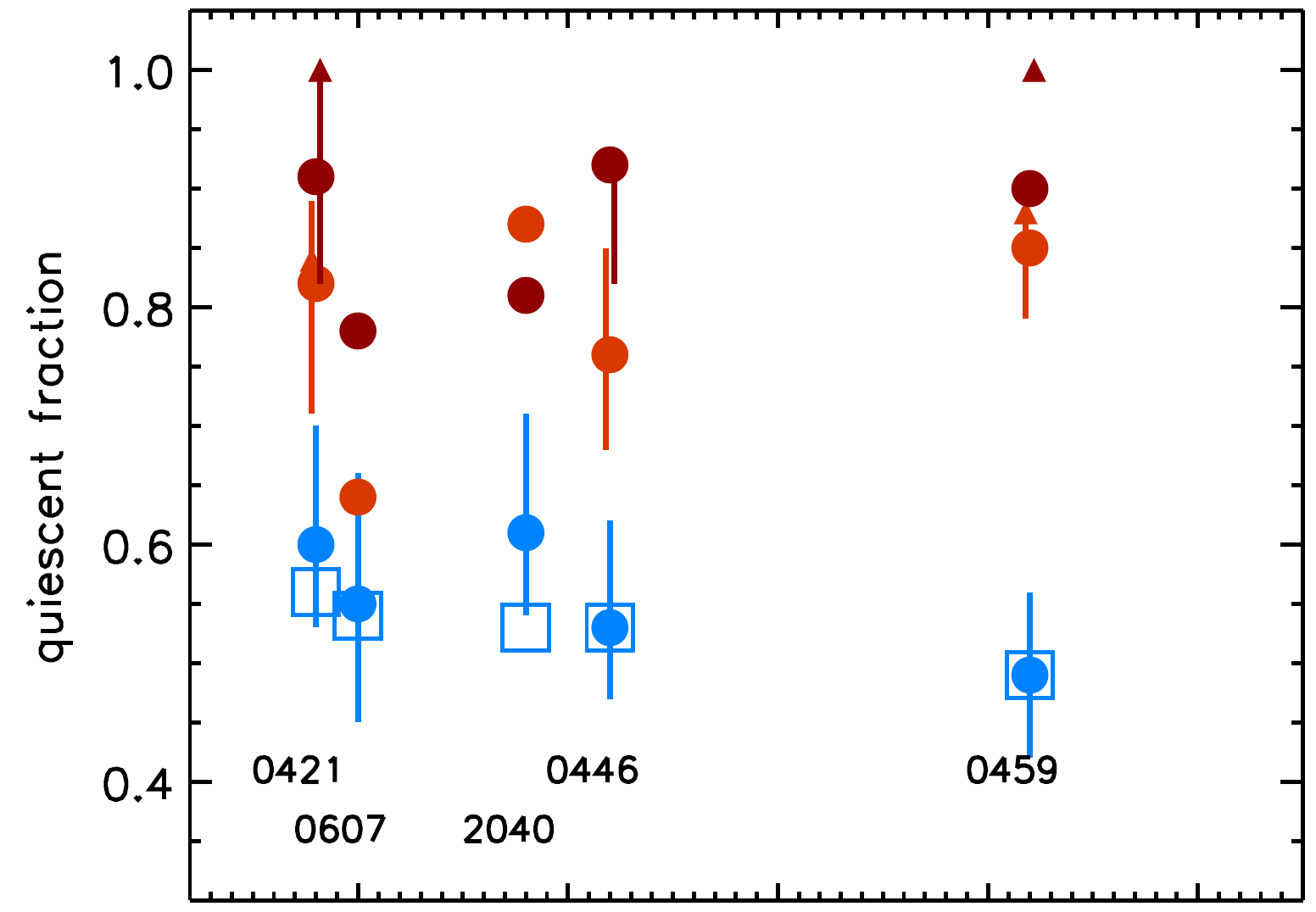}
 \includegraphics[width=0.49\textwidth,viewport=65 368 512 720,clip]{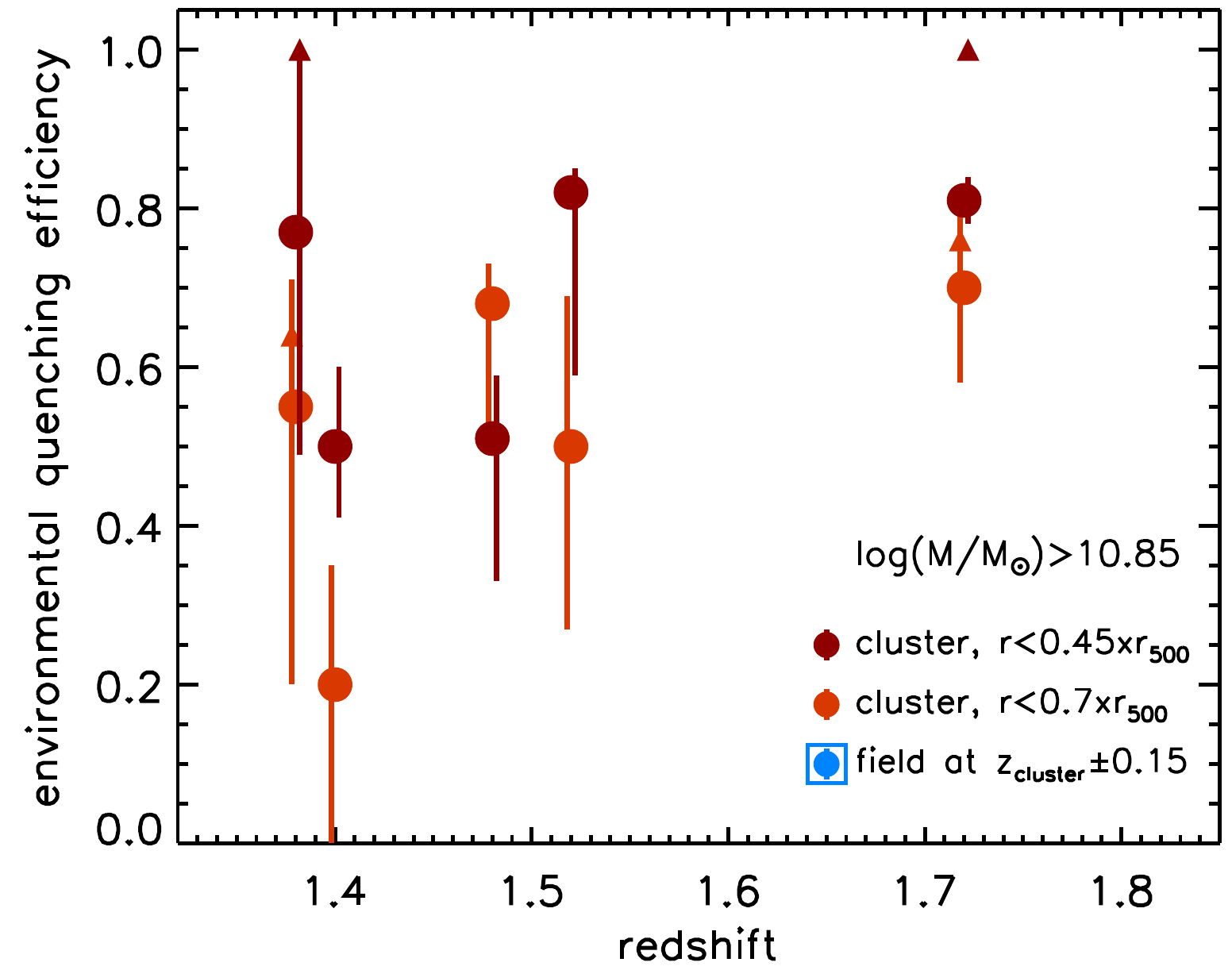}
\caption{{\it Top:} Quiescent fraction of cluster galaxies within
  $r<0.45 r_{500}$ (dark red) and $r<0.7 r_{500}$ (light red) above
  the common mass completeness limit log(M/M$_{\odot}$)$>$10.85. Error
  bars account for uncertainties in the quiescent vs. star-forming
  classification as described in Sect.~\ref{sec:passfrac}. Blue
  symbols show the quiescent fraction in corresponding photo-z
  selected control field samples (see Sect.~\ref{sec:passfrac}). Large
  empty blue squares show values from the COSMOS field (see text).
  {\it Bottom:} Environmental quenching efficiency as derived from
  cluster and field quiescent fractions in the top panel. Color coding
  reflects the top panel.  In both panels empty triangles show, where
  applicable, the quiescent fraction and derived quenching efficiency
  assuming that galaxies lacking a quiescent vs. star-forming
  classification are all quiescent (rather than all star-forming, see
  Sect.~\ref{sec:uvj}, Table~\ref{tab:passfrac}).
\label{fig:enveff}}
\end{figure}

In the following analysis, we classify as quiescent (star-forming)
sources those falling in the upper-left (bottom-right, respectively)
part of the diagram according to the color criterion shown in
Fig.~\ref{fig:uvj}.  Due to unreliable IRAC flux measurements for some
sources (Sects.~\ref{sec:spitzerobs}, \ref{sec:colselmembers}), the
m140-[3.6] color is not available for the full candidate member
samples. The small fraction of sources for which we miss both IRAC
fluxes (see Sect.~\ref{sec:colselmembers}) are not shown in this
figure. However, we still show the very few sources for which we only
have a 4.5$\mu$m flux measurement (empty circles), translating it to a
3.6$\mu$m magnitude using the average color of the IRAC sequence for
the cluster's red galaxy sample
(Sect.~\ref{sec:colselmembers}). Furthermore, even where a 3.6$\mu$m
flux measurement is not available, we classify as star-forming all
galaxies having a m814-m140 color bluer than the blue color-selection
cutoff. All sources that, after this procedure, still lack a quiescent
vs. star-forming color classification, are conservatively considered
as star-forming in the following. Depending on the cluster, these
amount to 0-6\% of the candidate member sample above the individual
cluster mass completeness limit over the $r < 0.7 r_{500}$ area. The
effect of these unclassified sources on the main results of this work
is shown in Table~\ref{tab:passfrac} and Figures~\ref{fig:colormass}
and \ref{fig:enveff}.

According to this color classification, the bulk of the RS population
in the probed cluster central regions is indeed made of quiescent
sources rather than very dusty star-forming galaxies, as shown more
specifically in Fig.~\ref{fig:histopass}. This is consistent with
other studies in massive clusters at similar redshifts
\citep[e.g.,][]{strazzullo2010b,newman2014,andreon2014,cooke2016}.

Figure~\ref{fig:panoramix} shows the projected distribution of
candidate members in the cluster fields, summarizing their estimated
properties as derived above, and highlighting the spatial
concentration of massive red and quiescent galaxies.

For each cluster, we then compute the quiescent galaxy fraction in
different stellar mass ranges (above the stellar mass completeness
limit) by adopting for each candidate cluster member the quiescent
vs.\ star-forming classification discussed above. More specifically,
from the statistical background subtraction and area coverage weights
of all (and quiescent, respectively) candidate members in a given
stellar mass range, we compute the background-corrected number of all
(and quiescent) cluster galaxies in that mass range, giving the
corresponding background-corrected quiescent fraction.

In Figure~\ref{fig:colormass} (top panels), these fractions are shown
for each cluster down to its individual mass completeness limit,
dividing the mass-complete candidate member sample in two mass bins
with approximately the same number of galaxies (stellar mass bins are
not the same for different clusters). The quiescent fraction of the
matched field control sample (photo-z or color-selected, analogous to
Fig.~\ref{fig:cmd}) is also shown.

Fig.~\ref{fig:colormass} thus clearly shows that the massive
population is in fact dominated by quiescent galaxies rather than very
dusty star-forming sources, and generally especially so at higher
stellar masses, as expected from previous work up to this redshift in
both clusters and average density fields.  Fig.~\ref{fig:colormass}
also shows that the quiescent fraction in the probed cluster region is
typically higher than in the field at the same redshift for all
clusters and in all stellar mass bins shown, although there are
variations across different clusters and different mass bins.

Table~\ref{tab:passfrac} summarizes the quiescent fraction
measurements described above, presenting the quiescent fraction down
to the common ($\log(M/M_{\odot})>10.85$) and individual stellar mass
completeness limits, for both the $r<0.45 r_{500}$ and $r<0.7 r_{500}$
regions. The quiescent fraction of field galaxies at the cluster
redshift to the same stellar mass limits is also listed. With the
(mild) exception of cluster SPT-CLJ2040 (whose central region is
affected by the masking of a significant area very close to the
cluster center, see Fig.~\ref{fig:sigmamaps}), the quiescent fraction
of massive galaxies within $r<0.45 r_{500}$ is higher than within
$r<0.7 r_{500}$, as observed in lower redshift clusters as well as at
least in most massive clusters at $z \sim 1.5$ \citep[e.g.,][ and more
  references in Sect.~\ref{sec:intro} also for lower-mass
  systems]{strazzullo2010b,newman2014,andreon2014,cooke2016}.

\subsection{The environmental quenching efficiency in the most massive distant cluster cores}
\label{sec:passfrac}

The comparison of the quiescent fraction in the cluster ($f_{q,cl}$)
and control field ($f_{q,fld}$) can be directly translated into the
so-called environmental quenching efficiency $\left(f_{q,cl} -
f_{q,fld}\right)/\left(1-f_{q,fld}\right)$, which is the fraction of
galaxies that would be star-forming in the field and that instead have
had their star formation suppressed by the cluster environment
\citep[e.g.,][]{vandenbosch2008,peng2012,wetzel2012,balogh2016,nantais2017}.
This conversion is shown in Fig.~\ref{fig:enveff}, where the top panel
shows the quiescent fraction for the five clusters and their matched
photo-z selected field samples, and the bottom panel shows the
corresponding environmental quenching efficiencies. Filled red circles
show the quiescent fractions and quenching efficiencies within $r <
0.45 r_{500}$ and $r < 0.7 r_{500}$ computed down to the common
stellar mass completeness limit of the full sample.  For reference, we
also show the corresponding measurements down to the individual mass
completeness limit of each cluster in Fig.~\ref{fig:enveffappendix}
(these cannot be properly compared across different clusters due to
the different stellar mass limits). In contrast with
Fig.~\ref{fig:colormass} and Table~\ref{tab:passfrac}, the error bars
reported in Fig.~\ref{fig:enveff} show the impact on the estimated
quiescent fraction of uncertainties in the source photometry and in
the definition of the quiescent color selection region.  The quoted
cluster quiescent fractions and related uncertainties in
Fig.~\ref{fig:enveff} correspond to the median and $16^{th}-84^{th}$
percentile range of 1000 realizations obtained by randomly shifting
the m814, m140, [3.6] photometry of each candidate member according to
a Gaussian with $\sigma$ given by the source photometric
uncertainties, and randomly offsetting the borders of the color
selection by $\pm0.1$~mag (see for comparison Fig.~\ref{fig:uvj}).
The quoted quiescent fractions and uncertainties for the corresponding
field samples are the median and $16^{th}-84^{th}$ percentile range of
1000 bootstraps on the photo-z and mass selected field
samples. Correspondingly, the quoted values and uncertainties for the
environmental quenching efficiency in the bottom panel show the median
and $16^{th}-84^{th}$ percentile range of environmental quenching
efficiencies obtained for each cluster from the different 1000
realizations for both cluster and field samples.  In the top panel of
Fig.~\ref{fig:enveff}, large empty blue squares show for comparison
the quiescent fraction in the $\sim$1.6~deg$^2$ COSMOS/UltraVISTA
field, estimated for log(M/M$_{\odot}$)$>$10.85 galaxies with photo-z
within $\pm$0.15 from the clusters' redshift, based on the
\citet{muzzin2013b} catalogs and the \citet{williams2009} UVJ
classification. Although, due to differences in the available data, we
cannot reproduce the analysis as described in this work on the COSMOS
field, the quiescent fractions estimated in the smaller GOODS-S field
are still representative of analogous measurements in the
significantly larger COSMOS survey.

\subsection{Is this sample really unbiased with respect to galaxy population properties?}
\label{sec:samplebias}

As mentioned in Section~\ref{sec:intro}, the SZE cluster selection is
approximately a halo mass selection with no a-priori dependence on
cluster galaxy properties.  However, given the high star formation
rates observed in some clusters in this redshift range, we need to
examine the possibility that mm-wave emission produced by high levels
of star formation may offset the SZE decrement, thus effectively
resulting in a biased cluster sample disfavoring systems with low
quiescent fractions. A general modeling of the effect of increased
star formation rates at high redshift on cluster SZE detection will be
presented elsewhere.  We focus here on the potential impact of mm-wave
emission from star formation on the SZE selection of the five clusters
studied here.  In particular, we seek to quantify the potential
selection bias that could impact our conclusions about quiescent
fractions and environmental quenching efficiencies for the broader,
massive cluster populations in this redshift range.

We start from the measured quiescent fractions within $r<0.45 r_{500}$,
and consider whether these five clusters (or more generally clusters
of similar mass and richness as those in this sample) would still be
in our sample if their quiescent fractions were lower than we
observe. We describe our modeling in full detail in
Appendix~\ref{sec:appendixa}, summarizing here the adopted approach,
assumptions and results.

For each cluster, we start from our mass complete sample of cluster
members within $r<0.45 r_{500}$ and their quiescent vs. star-forming
classification, and assume that all star-forming cluster galaxies form
stars at the same Main Sequence \citep[MS, e.g.,][]{elbaz2011} rate of
their field analogs (and that quiescent galaxies have a negligible
star formation rate, SFR). This gives an estimate of the total SFR of
cluster galaxies above mass completeness at $r<0.45 r_{500}$ (see
Appendix~\ref{sec:appendixa}). To account for the contribution of
galaxies below our mass completeness limit, we further assume that
(see Appendix~\ref{sec:appendixa1}): 1) the cluster galaxy stellar
mass function is to first order the same as in the field at the
cluster redshift; and 2) the quiescent fraction vs.\ stellar mass of
cluster galaxies can be modeled starting from our measured quiescent
fraction at high stellar masses and the quiescent fraction vs. stellar
mass observed in the field at the cluster redshift.

For each cluster in our sample, we thus obtain an estimate of the
total SFR within $r<0.45 r_{500}$. We finally estimate the SFR
contribution from cluster galaxies at $r>0.45 r_{500}$ by assuming an
NFW \citep{nfw1997} galaxy number density profile, and a quiescent
fraction vs.\ clustercentric radius profile determined based on the
measured quiescent fraction at $r<0.45 r_{500}$ and on the
corresponding field value at the cluster redshift (see
Appendix~\ref{sec:appendixa2}).

In practice, for the adopted assumptions and given a quiescent
fraction at $r<0.45 r_{500}$ above the mass completeness limit, our
modeling yields a SFR density profile of cluster galaxies (see
Appendix~\ref{sec:appendixa3}) that can be translated into flux
density maps at 95 and 150 GHz assuming an appropriate flux density to
SFR conversion (see Appendix~\ref{sec:appendixa4}).  If we consider
the actually measured quiescent fraction at $r<0.45 r_{500}$, such
modeling provides an estimate of the contamination of the observed SZE
signal from mm-wave emission of star-forming cluster galaxies. If
instead we consider a lower quiescent fraction at $r<0.45 r_{500}$,
such modeling yields an estimate of the additional contamination from
mm-wave emission that would be further reducing the SZE signal if the
star-forming galaxy fraction were higher than actually measured.

We estimate such additional contamination as a function of the $r<0.45
r_{500}$ quiescent fraction above mass completeness, for such
quiescent fraction values down to 10\% (see
Appendix~\ref{sec:appendixa3}). We then add the derived additional
flux density profile at 95 and 150 GHz to the observed 95 and 150 GHz
maps of the cluster, and estimate the S/N of the resulting SZE
detection (see
Appendix~\ref{sec:appendixa4}). Figure~\ref{fig:filteredSN} shows this
retrieved S/N for all five clusters as a function of the assumed
$r<0.45 r_{500}$ quiescent fraction above mass completeness, from the
actual measured value down to 10\%. We note that a quiescent fraction
of 10\% is significantly lower than the field value appropriate for
the cluster redshift and mass completeness limit. In
Fig.~\ref{fig:filteredSN}, solid lines show our S/N estimates for
$r<0.45 r_{500}$ quiescent fractions down to the relevant field level,
while dotted lines show the S/N for central quiescent fractions below
the field level. Although there is no evidence from this work that
quiescent fractions below the field level are common in these massive
cluster cores at the probed stellar masses, we show our S/N estimates
down to very low quiescent fractions below the field in the context of
the studies reporting a possible reversal of the star formation
vs. density relation in clusters at $z\gtrsim1.3$, as discussed in
Sect.~\ref{sec:intro}.

To show the approximate sensitivity of the results in
Fig.~\ref{fig:filteredSN} to the assumed conversion from SFR to 95 and
150~GHz flux density, the figure shows the S/N obtained with both of
the adopted MS SEDs \citep{bethermin2015,schreiber2018} used to
convert the SFR density profile into 95 and 150 GHz flux density
maps. On the other hand, the figure shows the most conservative result
(that is, producing lower S/N ratios) with respect to the modeling of
the quiescent fraction vs. stellar mass (see
Appendix~\ref{sec:appendixa4}).

In summary, Fig.~\ref{fig:filteredSN} shows that, as expected, the
lower the quiescent fraction the lower the S/N with which we retrieve
the cluster SZE detection. However, although this modeling clearly
relies on the several assumptions discussed above and in full detail
in Appendix~\ref{sec:appendixa}, it shows that at face value all five
clusters would have still been included in our S/N$>5$ sample even if
their central quiescent fractions were significantly lower than what
we observe, and  at least down to the field level (that is, even
if the environmental quenching efficiency were actually zero).  In
fact, according to Fig.~\ref{fig:filteredSN} most clusters in this
sample would still have been detected at S/N$>$5 even if 90\% of their
central massive galaxy population were still forming stars at the MS
level.

A full modeling of the impact of star formation in cluster galaxies on
cluster SZE detection for a range of cluster masses and as a function
of redshift, also considering in particular the possibility of a bias
on inferred galaxy population properties coming from cluster
selection, will be discussed in a future work. Nonetheless, from the
modeling presented here we conclude that this sample is not
significantly biased by cluster selection in terms of galaxy
population properties, and that thus the high quiescent fractions
observed in these clusters should be considered as representative of
the properties of galaxy clusters in this mass and redshift range.

\begin{figure}
 \includegraphics[width=0.49\textwidth,viewport=83 370 545 701, clip]{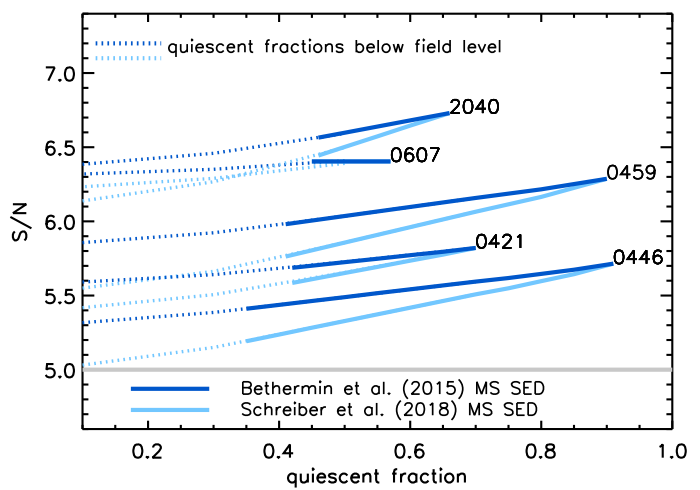}
\caption{Retrieved S/N of the cluster SZE detection obtained as a
  function of the $r<0.45 r_{500}$ quiescent fraction, from the
  observed value down to the field level (solid lines) at the cluster
  redshift and mass completeness limit (see
  Sect.~\ref{sec:samplebias}). Dotted lines show for completeness the
  retrieved S/N for central quiescent fractions down to 10\%.  For
  each cluster, the S/N is shown for both MS SEDs adopted to convert
  SFRs to 95, 150~GHz fluxes, as indicated. The gray line marks the
  S/N$>$5 limit of our original sample selection. According to the
  modeling in Sect.~\ref{sec:samplebias}, all five clusters would have
  been included in our S/N$>$5 sample even if environmental quenching
  were negligible.
\label{fig:filteredSN}}
\end{figure}

\section{Discussion and conclusions}
\label{sec:conclusions}

This work presents first results from a study of galaxy populations in
the five highest redshift ($1.35\lesssim z \lesssim 1.75$) massive
clusters that were selected from the 2500~deg$^2$ SPT-SZ survey via
their SZE signatures.  As has been shown in previous work
\citep[e.g.,][]{bleem2015,deHaan16}, the SPT-SZ selection produces a
roughly mass limited sample of clusters over the ($z\gtrsim0.2$)
redshift range where these systems exist.  Importantly, the selection
is based on the integrated pressure within the ICM, and therefore --
to first order-- is not impacted by the galaxy populations.  For the
specific clusters studied here, based on the analysis discussed in
Sect.~\ref{sec:samplebias}, even a quiescent fraction as low as that
measured in the field at these redshifts would not have impacted their
selection.

Furthermore, as discussed in Section~\ref{sec:intro}, our
$z\gtrsim1.4$, S/N$>$5 sample is expected to probe the
$M_{500}>2.74\times10^{14}M_\odot$, $z>1.4$ galaxy cluster population
with a completeness of $\sim70$\%.  Sample incompleteness is caused by
scatter in the SZE observable--mass relation, and this comes from both
the measurement noise in the extraction of the SZE observable from the
mm-wave maps (20\% for a S/N=5 cluster) and the intrinsic scatter in
the noise-free SZE observable at fixed mass \citep[empirically
  estimated to be $20\pm7$\%;][]{deHaan16}.  These two sources of
scatter combine to $\sim28$\%, which in combination with the mass
trend parameter $B_\mathrm{SZ}\sim1.6$ of the SZE observable mass
relation results in a $\sim$20\% mass scatter at fixed observable and
redshift. Interesting for our study is to understand the significance
of the contribution of contaminating flux from star formation to this
scatter.  With a large volume hydrodynamical simulation,
\citet{gupta2017b} estimate a scatter in SZE integrated Compton
$Y_{500}$ at fixed mass of $\sim$13.5\% at
M$_{500}=3\times10^{14}M_\odot$.  This scatter, due to variation in
cluster morphology and large scale structure projected onto the
cluster line of sight, is the dominant source of intrinsic scatter in
the SZE observable--mass relation. The contribution to the scatter due
to high frequency cluster radio galaxies is estimated to be marginal
\citep[$\sim$3\%;][]{gupta2017}.  When added in quadrature, these
components of the scatter correspond roughly to 24\% scatter in the
observable at fixed mass.  From Fig.~\ref{fig:filteredSN} we can
estimate the scatter due to contamination from star forming cluster
galaxies at these cluster masses and redshifts to be $<$5\% ($<$10\%
with the more conservative assumption on the mm-wave SED) even if
assuming zero environmental quenching in the cluster core
(Sect.~\ref{sec:samplebias}). The scatter would be larger (up to
$<15$\% in the worse case considered in Sect.~\ref{sec:samplebias}) if
a relevant part of the cluster population at these masses had
quiescent fractions below the field level, but as discussed in
Sect.~\ref{sec:samplebias} there is no evidence for this from our
current results.  At these levels, the contaminating flux from star
formation is expected to play a lesser role in comparison to the
dominant sources of scatter. Therefore, with our current estimates,
the expected 30\% incompleteness in our sample can be no more than
marginally correlated with the star formation properties of cluster
galaxies.  Following the discussions above, we conclude that the
galaxy population properties we report here should be representative
of the populations in massive clusters in this redshift range.

This study is based on deep imaging follow-up of the clusters in four
broad bands (F814W and F140W from {\it HST}, 3.6$\mu$m and 4.5$\mu$m
from {\it Spitzer}). Additional imaging has been acquired on part of
this sample (in particular with {\it HST}, within the {\it See Change}
program, and Cycle 24 GO-14677 program (PI
Schrabback) for weak lensing analyses), and will be used in future
investigations. However, the work presented here completely relies on
homogeneous, four band photometry across the full sample, and may be
considered as a field-test of the minimal requirements adopted in
designing our follow-up program to study cluster galaxy populations at
this redshift.

Clusters in this mass and redshift range are exceedingly rare. Thus,
although the current sample is very small for statistical purposes,
this work is still the first ``sample study'' of quiescent fractions
and environmental quenching efficiency in such massive, distant
clusters.

The cluster redshifts estimated from RS and IRAC colors
(Section \ref{sec:redshifts}) are consistent with the original
redshift selection ($z>1.4$) of this sample from the \citet{bleem2015}
catalog, and with the existing spectroscopic redshift determinations
\citep{bayliss2014,khullar2018}. Cluster SPT-CLJ0459 seems to be not
only the most distant cluster found in the SPT-SZ survey but likely
the most distant ICM-selected massive cluster found to date, and among
the very few extremely massive systems known near $z\sim2$,
irrespective of their selection \citep{andreon2009,newman2014,
  stanford2012,brodwin2016}.

\subsection{Derived galaxy population properties and comparison with previous work}

All clusters in this sample show a very strong galaxy concentration
near the position of the SPT detection (Fig.~\ref{fig:sigmamaps}),
with the exception of SPT-CLJ0607, which has a seemingly milder, less
concentrated galaxy overdensity at least in the probed magnitude
range. All clusters show an excess of red galaxies with respect to
similarly selected field galaxy samples at the same redshift, with
colors consistent with evolved stellar populations formed at $z>2$
(Section ~\ref{sec:redpops}).

According to the adopted $UVJ$-like color classification, the bulk of
the massive red galaxy population within $r < 0.7 r_{500}$ consists of
quiescent galaxies rather than very dusty star-forming sources.
Quiescent fractions above the common mass completeness limit
log(M/M$_{\odot}$)$>$10.85 are in the range $\sim 60-90$\% ($\gtrsim
80$\% within $r < 0.45 r_{500}$), compared to field levels of
$\sim50-60\%$ above the same stellar mass limit and across the probed
redshift range (Sections \ref{sec:uvj}, \ref{sec:passfrac}).  The
higher quiescent fractions in clusters relative to the field translate
into environmental quenching efficiencies of typically $\sim
0.50-0.70$ at $r < 0.7 r_{500}$ with the exception of SPT-CLJ0607, and
$\sim 0.5-0.9$ at $r < 0.45 r_{500}$.  This level of environmental
quenching efficiency is comparable to that observed in cluster cores
at $z\sim1$ \citep[for a similar stellar mass threshold,
  e.g.,][]{muzzin2012}, and already close to that observed in the
densest environments in the nearby Universe
\citep[e.g.,][]{peng2012,raichoor2014}. The observed variations
between clusters (see Figures~\ref{fig:colormass} and
\ref{fig:enveff}, and Table~\ref{tab:passfrac}), particularly with
reference to SPT-CL0607, may suggest a possibly non negligible range
of quiescent fractions and  environmental quenching
efficiencies even among clusters of similar halo mass and 
redshift \citep[see also related findings from ][as also discussed
  below]{nantais2017}. However, given the small cluster sample size
and the uncertainties on the environmental quenching efficiency of
individual clusters, it is not possible to draw firm conclusions.

SPT-CLJ0459, in spite of being the most distant cluster in the sample,
still shows a quiescent fraction of $\gtrsim 0.8$ in the probed mass
range and cluster region, close to the striking value observed in
JKCS~041 at z=1.8 \citep{newman2014,andreon2014}. As already discussed
above, cluster selection bias - that could be considered in the case
of JKCS~041, which was indeed selected as an overdensity of red
sources - is not relevant for this sample.  Finally, as described in
Sect.~\ref{sec:uvj}, the quoted quiescent fractions are formally lower
limits in the sense that galaxies redder than the color selection blue
cutoff without a reliable IRAC flux measurement cannot be classified
as quiescent or star-forming based on our criteria, and are
conservatively considered as star-forming. In the case of SPT-CLJ0459,
the fraction of candidate members without a quiescent vs. star-forming
classification and above the mass completeness limit is 5\% at $r <
0.7 r_{500}$, and 10\% at $r < 0.45 r_{500}$.
Figures~\ref{fig:colormass}, \ref{fig:enveff} and
Table~\ref{tab:passfrac} also show for reference the impact of
unclassified galaxies on the measured quiescent fractions, when
assuming they are all quiescent rather than all star-forming.

Although high quiescent fractions in cluster central regions are the
obvious expectation at $z\lesssim 1$, several studies (see
Sect.~\ref{sec:intro}) have reported high fractions of star forming
galaxies even in cluster cores at $z \gtrsim 1.4$.  Some mid-IR and
far-IR based studies of IR-selected cluster samples
\citep{brodwin2013,alberts2016} have shown a star-forming fraction for
$z\gtrsim1.4$ clusters at clustercentric radii $<0.5$~Mpc consistent
with or even higher than field levels.  Although our probed
$r<0.7r_{500}$ areas are comparable to $r<0.5$~Mpc (our corresponding
clustercentric radius is on average 430~kpc, ranging from 390 to 460
kpc), given the high mass of our clusters compared to the typically
lower masses ($\approx10^{14}$M$_{\odot}$) of clusters in the
IR-selected samples, the area we probe corresponds to a more central
region with respect to the cluster virial radius.  In addition, the
selection of the galaxy samples in our work differs from that in the
mentioned studies. Nevertheless, the results of our analysis do not
generally lend support to the view that $z\gtrsim1.4$ corresponds to
an era before significant quenching in cluster cores, at least for
clusters as massive as those studied here.

It is also conceivable -- perhaps even expected -- that galaxy
population properties depend on cluster mass and assembly history
\citep[see e.g., discussion in][and references therein]{nantais2017},
with more massive clusters typically hosting more evolved populations
\citep[see e.g.,][for a study of clusters and groups at
  $z\sim1$]{balogh2016}.  Indeed, JKCS~041 at z=1.8 \citep[as
  mentioned above,][]{andreon2009,andreon2014,newman2014} and
XMMU~J2235-2557 at z=1.39
\citep{mullis2005,rosati2009,strazzullo2010b} are two examples of very
massive clusters bracketing the redshift range probed here, and both
show strongly suppressed star formation in their core. On the other
hand, the far-IR based work of \citet{santos2015} indicates
significant star formation activity in the inner $r<250$~kpc core of
the very massive cluster XDCP~J0044.0-2033 at z=1.58
\citep{santos2011,tozzi2015}, with a star formation rate approaching
2000~M$_{\odot}$/yr just accounting for three {\it Herschel}-detected
ULIRGs associated with massive cluster members.  Thus, it is not clear
that what we are seeing is a simple cluster mass dependence of the
quiescent fraction.

In the recent work by \citet{nantais2017}, the evolution of the
environmental quenching efficiency at $z\sim 0.9$ to 1.6 was
investigated with a sample of RS selected \citep[SpARCS,
][]{wilson2009} clusters. They measure quiescent fractions within a
clustercentric radius of 1~Mpc which are very close to the field
level, resulting in an environmental quenching efficiency consistent
with zero, although the authors note that there is considerable
dispersion in environmental quenching efficiencies of different
clusters in their $z\sim1.6$ sample (three clusters), larger than in
any of their lower-redshift samples. For reference, tentative masses
from velocity dispersions of the $z\sim1.6$ clusters in
\citet{nantais2017} are estimated to be of order $0.4 - 2.4 \times
10^{14}$M$_{\odot}$ \citep{lidman2012}. The very low quenching
efficiency they measure at $z\sim1.6$ makes for a significant drop
from the $\sim$0.6 and 0.7 values measured at $z\sim$1.3 and 0.9,
respectively, from which \citet{nantais2017} conclude that
environmental quenching in clusters is a relatively subdominant
process earlier than $z\sim 1.5$, and then rapidly rises, increasing
its relevance up to $z\sim1$.  The clustercentric radius of our probed
region is significantly smaller than 1~Mpc ($\sim r_{200}$ for our
clusters), and we clearly expect quiescent fractions to decrease at
increasing clustercentric distance. This complicates any quantitative
comparison of our current results with
\citet{nantais2017}. Nonetheless, we must conclude that environmental
quenching efficiency seems to be significantly larger than zero in the
cluster central regions, at least in the approximately halo mass
selected $z\gtrsim 1.4$ sample we have studied here.

\subsection{Future directions}

Obvious future directions for this study involve at least two
aspects. First, obtaining an accurate measurement of the star
formation rate in the cluster core with tracers not biased by dust
attenuation remains critical. Our current data do not allow us to
estimate star formation rates -- especially for red sources -- with any
reasonable accuracy. Although the classification of quiescent and
star-forming galaxies based on the adopted color criterion seems to be
well-behaved with respect to separation of the quiescent and
star-forming sequences, and to match well with expectations based on
equivalent galaxy samples in the control field, the ultimate
confirmation that these environments are already so efficiently
quenched remains with direct dust-unbiased star formation rate
measurements, especially given the mid-/far-IR based results mentioned
above. The cluster sky locations and needed sensitivity place limits
on the possible choice of instruments for such observations to JWST,
ALMA, or SKA pathfinders.

Second, an extension of this analysis out to the virial radius and
even beyond would allow us to probe the relevance and timescales of
environmental effects in suppressing star formation as galaxies are
accreted by the clusters. We stress again that the accurate knowledge
of cluster masses, and thus of their virial radii, allows us to
consider meaningful apertures for comparing quenching efficiencies
across different clusters. This is obviously crucial when studying any
property that exhibits a radial dependence.

Ultimately, larger well-defined cluster samples at redshifts well
beyond $z\sim1$ and up to $z\gtrsim2$ over a range of halo masses
still remain a critical missing piece in defining a quantitative
picture of early-time environmental effects on galaxy evolution in the
first clusters. New and upcoming surveys, notably SPTpol
\citep{austermann12}, SPT-3G \citep{benson14}, Advanced ACTPOL
\citep{thornton16} and eROSITA \citep{merloni12}, will contribute to
shaping our view in the near future, in preparation for the next
generation of distant cluster surveys.

\begin{acknowledgements}
  We thank Emiliano Merlin for helpful suggestions on the use of {\tt
    T\_PHOT}, and Corentin Schreiber for discussions and
  clarifications on the MS SED model adopted in
  Sect.~\ref{sec:samplebias}. We thank the anonymous referee for a
  constructive report that improved the presentation of this
  work. M.P., N.G., R.C.  and V.S. acknowledge support from the German
  Space Agency (DLR) through {\it Verbundforschung} project
  ID~50OR1603. We acknowledge the support by the DFG Cluster of
  Excellence ``Origin and Structure of the Universe'', the
  Ludwig-Maximilians-Universit\"at (LMU-Munich), and the Transregio
  program TR33 ``The Dark Universe''. A.S. is supported by the ERC-StG
  "ClustersXCosmo", grant agreement 71676.  B.B. is supported by the
  Fermi Research Alliance LLC under contract no. De-AC02-07CH11359
  with the U.S. Department of Energy. T.S. acknowledges support from
  the German Federal Ministry of Economics and Technology (BMWi)
  provided through DLR under projects 50 OR 1407, 50 OR 1610, and 50
  OR 1803. D.R. is supported by a NASA Postdoctoral Program Senior
  Fellowship at the NASA Ames Research Center, administered by the
  Universities Space Research Association under contract with
  NASA. C.L.R. acknowledges support from Australian Research Council’s
  Discovery Projects scheme (DP150103208). The South Pole Telescope is
  supported by the National Science Foundation through grant
  PLR-1248097.  Based on observations made with the NASA/ESA Hubble
  Space Telescope under program GO-14252, obtained from the Data
  Archive at the Space Telescope Science Institute, which is operated
  by the Association of Universities for Research in Astronomy, Inc.,
  under NASA contract NAS 5-26555.  Based on observations made with
  the Spitzer Space Telescope (program ID~12030) which is operated by
  the Jet Propulsion Laboratory, California Institute of Technology
  under NASA contract.

\end{acknowledgements}

\begin{appendix}

    \section{Effect of the background estimation on the red galaxy fraction in relation to different red fractions across clusters in this sample}
\label{sec:redfracvsbackground}

\begin{figure}[]
\begin{center}
 \includegraphics[width=0.45\textwidth,viewport= 75 377 512 721,
   clip]{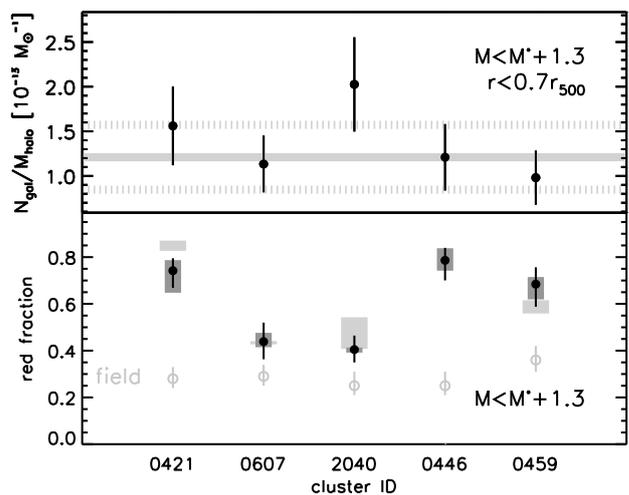}
\end{center}
\caption{ {\it Top panel -- } Black points show for all clusters the
  estimated total number of background-subtracted cluster members at
  m140$<$M$^{*}$+1.3 and within $r<0.7 r_{500}$ divided by the total
  projected halo mass within the same region
  (N$_{gal}$/M$_{halo}$). The solid gray line shows the median value
  for this cluster sample.  The dashed lines show an estimate of the
  intrinsic scatter of N$_{gal}$/M$_{halo}$. See text for details.
  {\it Bottom panel -- } Estimate of the effect of differences in the
  background density along the line of sight of the clusters on the
  estimated red fraction of cluster galaxies. Black points show the
  red fraction for all clusters as estimated adopting the default
  control field for the statistical residual background subtraction
  (Sect.~\ref{sec:statsub}). Dark gray bars show the effect on the
  estimated red fraction of adopting control sub-fields with densities
  in the 10$^{th}$ to 90$^{th}$ percentile range across the default
  control field. Light gray bars show the effect of considering the
  background-subtracted cluster member samples contaminated or
  incomplete according to the face-value N$_{gal}$/M$_{halo}$ of each
  cluster with respect to the median N$_{gal}$/M$_{halo}$ of the
  sample. See text for details. For comparison, light gray empty
  circles show the red fraction in field galaxies in the same m140
  magnitude range, and with a photo--z within $\pm0.15$ of the cluster
  redshift.
\label{fig:redfracvsNgal/Mhalo} }
\end{figure}

As mentioned in Sect.~\ref{sec:redpops}, different clusters in this
sample exhibit different color distributions in their color-magnitude
diagram, resulting in a range of red galaxy fractions, with
SPT-CLJ0607 and SPT-CLJ2040 showing a more significant blue
population. Figure \ref{fig:redfracvsNgal/Mhalo} (bottom panel) shows
the red fraction of all clusters down to a same limit of
m140$<$M$^*$+1.3, making for a more proper comparison than what can be
done on Fig.~\ref{fig:cmd}, that reaches different depths ranging from
$\sim$M$^{*}$+2 to $\sim$M$^{*}$+1.3 depending on the cluster redshift
(see Sect.~\ref{sec:redpops}). We note that, in contrast with the
quiescent fractions discussed in Sections~\ref{sec:uvj},
\ref{sec:passfrac}, \ref{sec:samplebias}, \ref{sec:conclusions}, the
red fractions discussed here are based on flux limited, not mass
complete galaxy samples. For the purpose of
Fig.~\ref{fig:redfracvsNgal/Mhalo}, we define as ``red'' those
galaxies with a m814-m140 color redder than 0.4~mag below the RS model
with $z_f =3$ (see Fig.~\ref{fig:cmd}).

Black symbols with error bars show the red fraction estimated based on
the statistical background subtraction weights calculated in
Sect.~\ref{sec:statsub} on the full control field. Similarly to what
can be inferred from Fig.~\ref{fig:cmd}, clusters SPT-CLJ0421,
SPT-CLJ0446, and SPT-CLJ0459 have a higher red fraction ($\sim70\%$ at
m140$<$M$^*$+1.3) than SPT-CLJ0607 and SPT-CLJ2040 ($<50\%$). We then
redetermine the statistical background subtraction weights using 100
sub-fields at random positions in the control field of the same area
as the probed cluster region (see Sect.~\ref{sec:complcontsamples}),
estimating each time the corresponding red galaxy fraction. As
expected, the red fraction depends on the assumed control field, so
that for each given cluster we obtain a higher red fraction when using
a higher density\footnote{ By density we mean here the galaxy density
  in the control field after applying the magnitude and color
  selection criteria as described in Sect.~\ref{sec:statsub}.} control
field. The dark gray bars in Fig.~\ref{fig:redfracvsNgal/Mhalo}
(bottom) show the variation of the estimated red fraction when
assuming control fields with densities spanning from the 10$^{th}$ to
the 90$^{th}$ percentile of the density distribution across the 100
random sub-fields. As the figure shows cosmic variance on small
scales, as can be probed with this approach, results in a relatively
minor impact (10\% at most) on the estimated red fraction. On the
other hand, we do not probe with this approach cosmic variance on
large scales.

The top panel of Fig.~\ref{fig:redfracvsNgal/Mhalo} shows the number
of background-subtracted cluster galaxies at m140$<$M$^*$+1.3 and
$r<0.7 r_{500}$ as estimated with our default background subtraction
weights (Sect.~\ref{sec:statsub}) divided by the projected halo mass
at $r<0.7 r_{500}$ \citep[calculated based on the cluster mass
  M$_{500}$ and redshift assuming a concentration according
  to][]{duffy2008}. This number of galaxies per halo mass (hereafter,
N$_{gal}$/M$_{halo}$) is similar across our sample, as can be expected
given the small cluster mass range. Most values of
N$_{gal}$/M$_{halo}$ are indeed consistent within the
uncertainties\footnote{We note that the N$_{gal}$/M$_{halo}$
  uncertainties in Fig.~\ref{fig:redfracvsNgal/Mhalo} are, if
  anything, somewhat underestimated, as they only account for Poisson
  error on the estimated total number of cluster galaxies, and for the
  error on M$_{500}$.}. The solid gray line shows the median
N$_{gal}$/M$_{halo}$ across the sample, with the dashed gray lines
showing an intrinsic scatter on N$_{gal}$/M$_{halo}$ based on the
estimates of \citet{hennig2017}. Based on
Fig.~\ref{fig:redfracvsNgal/Mhalo} all clusters are thus consistent
with having similar N$_{gal}$/M$_{halo}$ according to our
expectations. This suggests that there is no evidence for large
variations in the local background (corresponding to the redshift
range resulting from our color selections) of the individual clusters,
because such large variations would affect the estimated N$_{gal}$
(and thus N$_{gal}$/M$_{halo}$) if using the same control field for
all clusters. The largest deviation of N$_{gal}$/M$_{halo}$ with
respect to the median of the sample occurs for SPT-CLJ2040. Although
given the uncertainties we cannot take this as actual evidence, we
consider here the possibility that the higher face-value
N$_{gal}$/M$_{halo}$ of SPT-CLJ2040 results from an higher local
background density (than in our control field) which is not accounted
for by our statistical background subtraction. That is, we assume that
the excess of SPT-CLJ2040's N$_{gal}$/M$_{halo}$ with respect to the
median value is due to unaccounted interlopers contaminating the
number of background-subtracted cluster members. In this assumption,
we can estimate the number of such unaccounted interlopers assuming
that SPT-CLJ2040 has intrinsically the median N$_{gal}$/M$_{halo}$ of
the sample. This would result in 29 of the estimated 71
background-subtracted cluster members within $r<0.7 r_{500}$ being
actually interlopers. Given our initial color and magnitude selection
(Sect.~\ref{sec:samplesel}), we can reasonably assume (see
e.g. Figs.~\ref{fig:samplesel}, \ref{fig:redshifts}) that the bulk of
the background contaminating the color-selected candidate member
sample is roughly at $1.2\lesssim z \lesssim 2$. Therefore, we can
estimate the fraction of galaxies in the control field at these
redshifts and within our magnitude limit that would appear as ``red''
with the m814-m140 color threshold applied for SPT-CLJ2040. This is
about 25\% over the full $1.2\lesssim z \lesssim 2$ redshift range,
ranging from $\sim$20\% to $\sim$40\% across the range when calculated
in $\Delta z = 0.2$ redshift bins. We thus consider a minimum and
maximum ``red fraction'' of the unaccounted interlopers of 20\% and
40\%, respectively. We can thus estimate a ``corrected'' red fraction
for SPT-CLJ2040 as:

   ``corrected red fraction'' =
     $\frac{N_\textrm{red,bkgsub} - [0.2,0.4]
       N_\textrm{interlopers}}{N_\textrm{total,bkgsub}-N_\textrm{interlopers}}$

  where $N_\textrm{red,bkgsub}$ and $N_\textrm{total,bkgsub}$ are our
  default estimates of the number of red and all background-subtracted
  candidate members (with standard background subtraction from
  Sect.~\ref{sec:statsub}), respectively, while
  $N_\textrm{interlopers}$ is the estimated number of interlopers
  contaminating $N_\textrm{total,bkgsub}$ in the assumptions described
  above. The adopted 0.2-0.4 ``red fraction'' range of the unacconted
  interlopers corresponds to the minimum and maximum values estimated
  above, and results in a ``corrected red fraction'' ranging from 41\%
  to 54\%, as shown by the light gray band for SPT-CLJ2040 in the
  bottom panel of Fig.~\ref{fig:redfracvsNgal/Mhalo}.

   We estimate in an analogous way the ``corrected red fractions''
   shown by light gray bands in Fig.~\ref{fig:redfracvsNgal/Mhalo} for
   all other clusters. The deviation from the median
   N$_{gal}$/M$_{halo}$ for the other clusters is more marginal than
   for SPT-CLJ2040 (all being consistent at $<1\sigma$ with the median
   value, even not accounting for the intrinsic scatter). Therefore,
   as mentioned above, there is no evidence for significant
   cluster-to-cluster background variations. We show nonetheless for
   completeness the ``corrected red fractions'' as a reference. We
   note that assuming a local background density significantly lower
   (rather than higher) than our default control field (e.g. in
   particular for SPT-CLJ0459) has the effect of reducing (rather than
   increasing) the estimated red fraction (due to the observed color
   distribution of field galaxies in the magnitude and redshift range
   considered).

    \section{Environmental quenching efficiency down to the individual mass completeness limit of each cluster}
\label{sec:appendixenveffplot}

\begin{figure}
 \includegraphics[width=0.49\textwidth,viewport=65 410 512 720,clip]{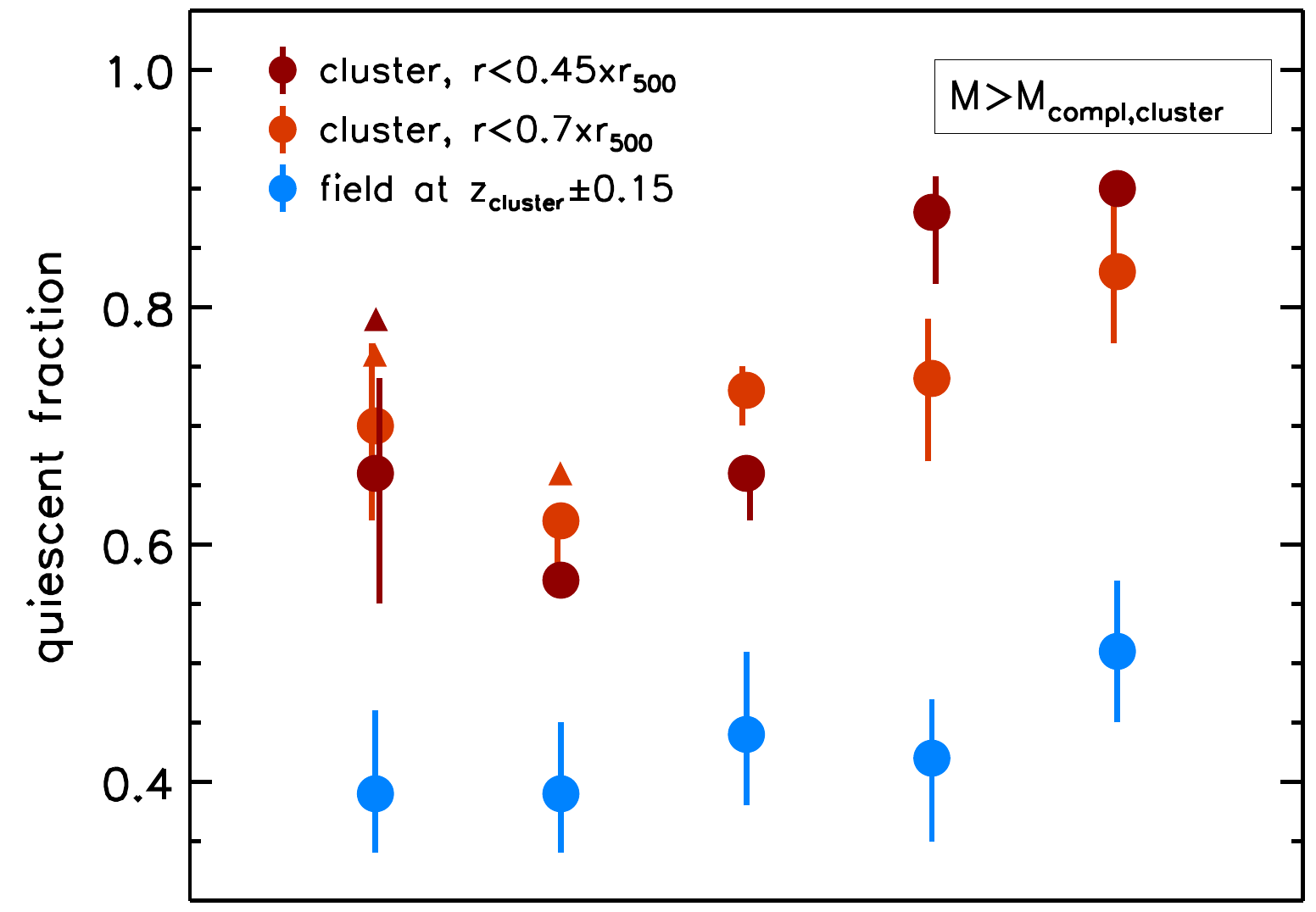}
 \includegraphics[width=0.49\textwidth,viewport=65 368 512 720,clip]{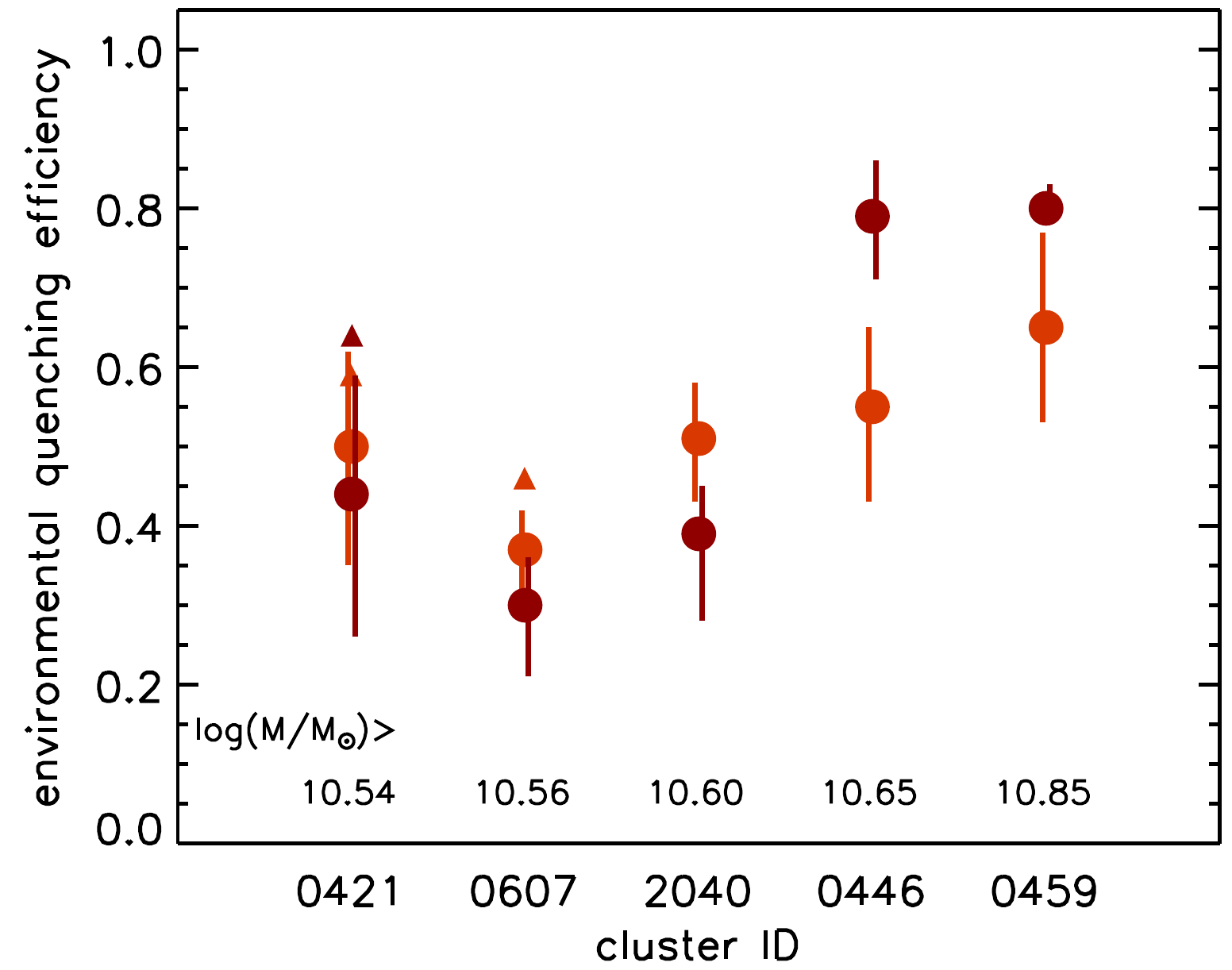}
\caption{ {\it Top:} Quiescent fraction of cluster galaxies within
  $r<0.45 r_{500}$ (dark red) and $r<0.7 r_{500}$ (light red) above
  the mass completeness limit of each cluster as reported in the
  bottom panel. Error bars account for uncertainties in the
  quiescent vs. star-forming classification as described in
  Sect.~\ref{sec:passfrac}. Blue symbols show the quiescent fraction in
  corresponding photo-z selected control field samples (see
  Sect.~\ref{sec:passfrac}).  {\it Bottom:} Environmental quenching
  efficiency as derived from cluster and field quiescent fractions in
  the top panel. Color coding reflects the top panel.  In both panels
  empty triangles show, where applicable, the quiescent fraction and
  derived quenching efficiency assuming that galaxies lacking a
  quiescent vs. star-forming classification are all quiescent (rather than
  all star-forming, see Sect.~\ref{sec:uvj}, Table~\ref{tab:passfrac}).
\label{fig:enveffappendix}}
\end{figure}

Figure~\ref{fig:enveffappendix} shows the quiescent fraction for the
five clusters and their matched photo-z selected field samples (top
panel) and the corresponding environmental quenching efficiencies
(bottom panel) down to the individual stellar mass completeness limit
of each cluster. Because the individual mass completeness limits
differ, in contrast with the results shown in Fig.~\ref{fig:enveff}
those in Fig.~\ref{fig:enveffappendix} cannot be properly compared
across different clusters.

\section{Modeling of mm-wave emission from star-forming cluster galaxies and its effect on SZE detection and completeness of this cluster sample}
\label{sec:appendixa}

\begin{figure*}
  \centering \includegraphics[height=0.22\textheight,viewport=55 355
    463 716,clip]{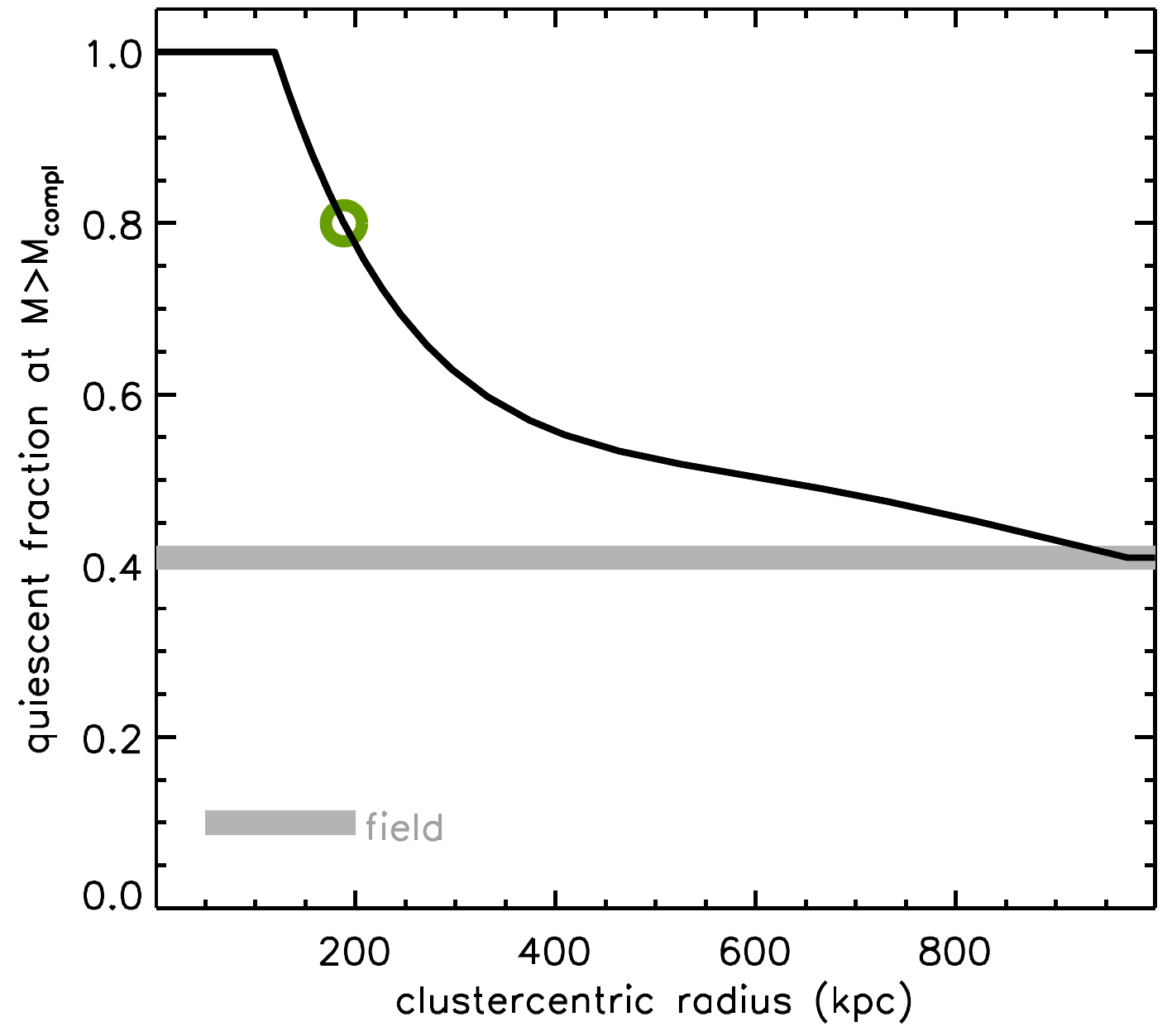}%
  \includegraphics[height=0.22\textheight,viewport=106 355 463
    716,clip]{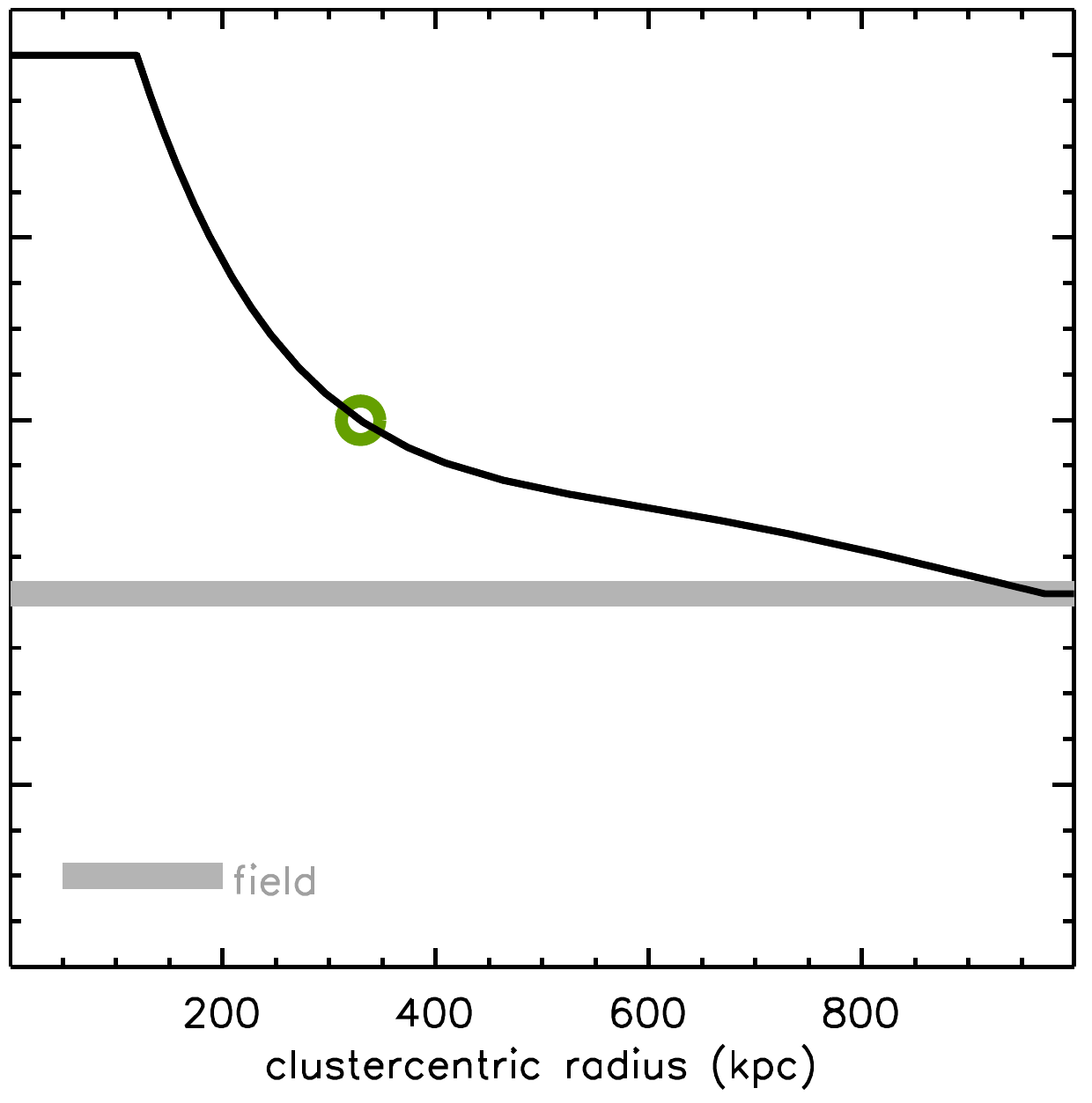}%
  \includegraphics[height=0.22\textheight,viewport=106 355 463
    716,clip]{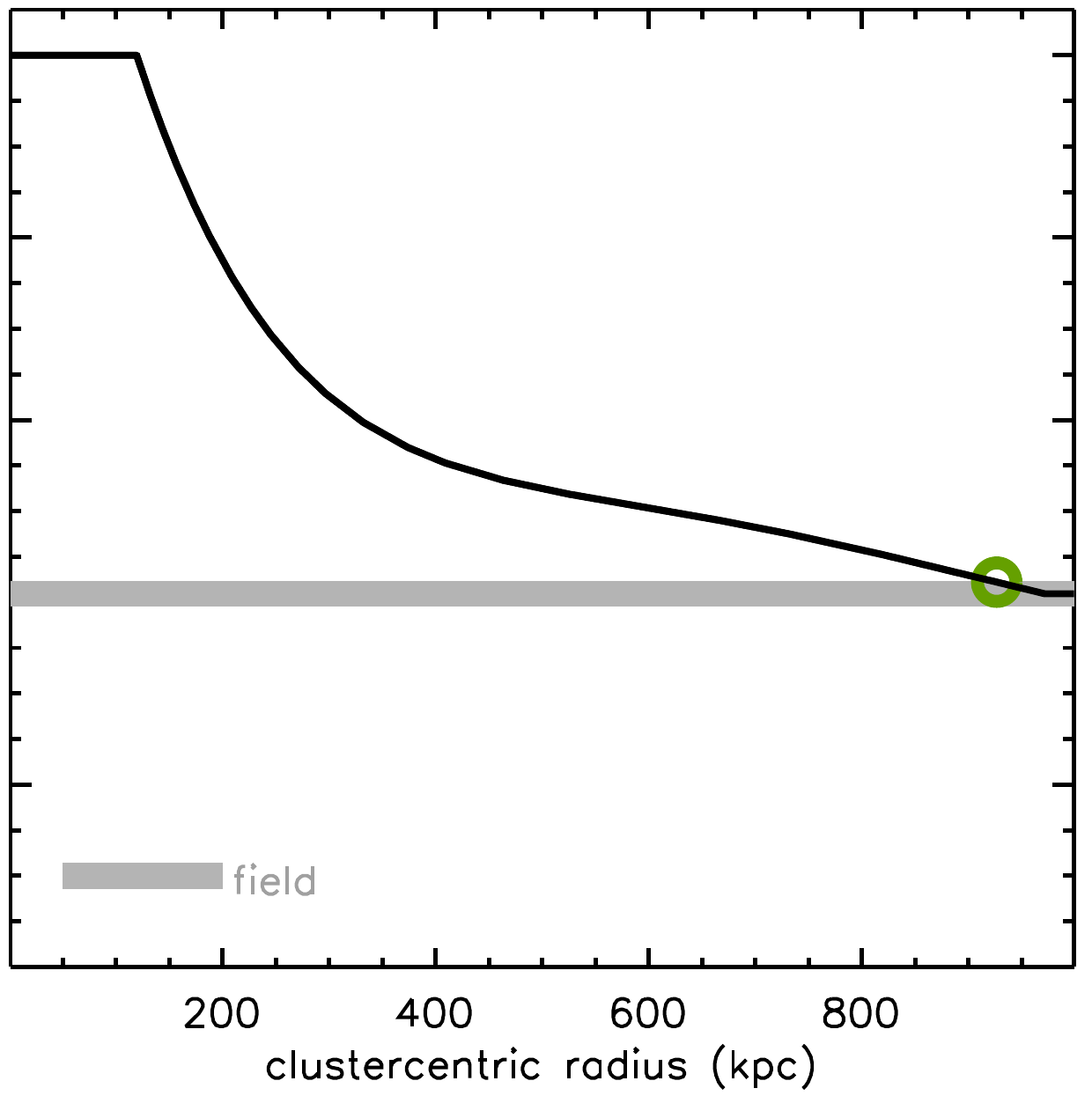}
  \includegraphics[height=0.195\textheight,viewport=55 396 463
    716,clip]{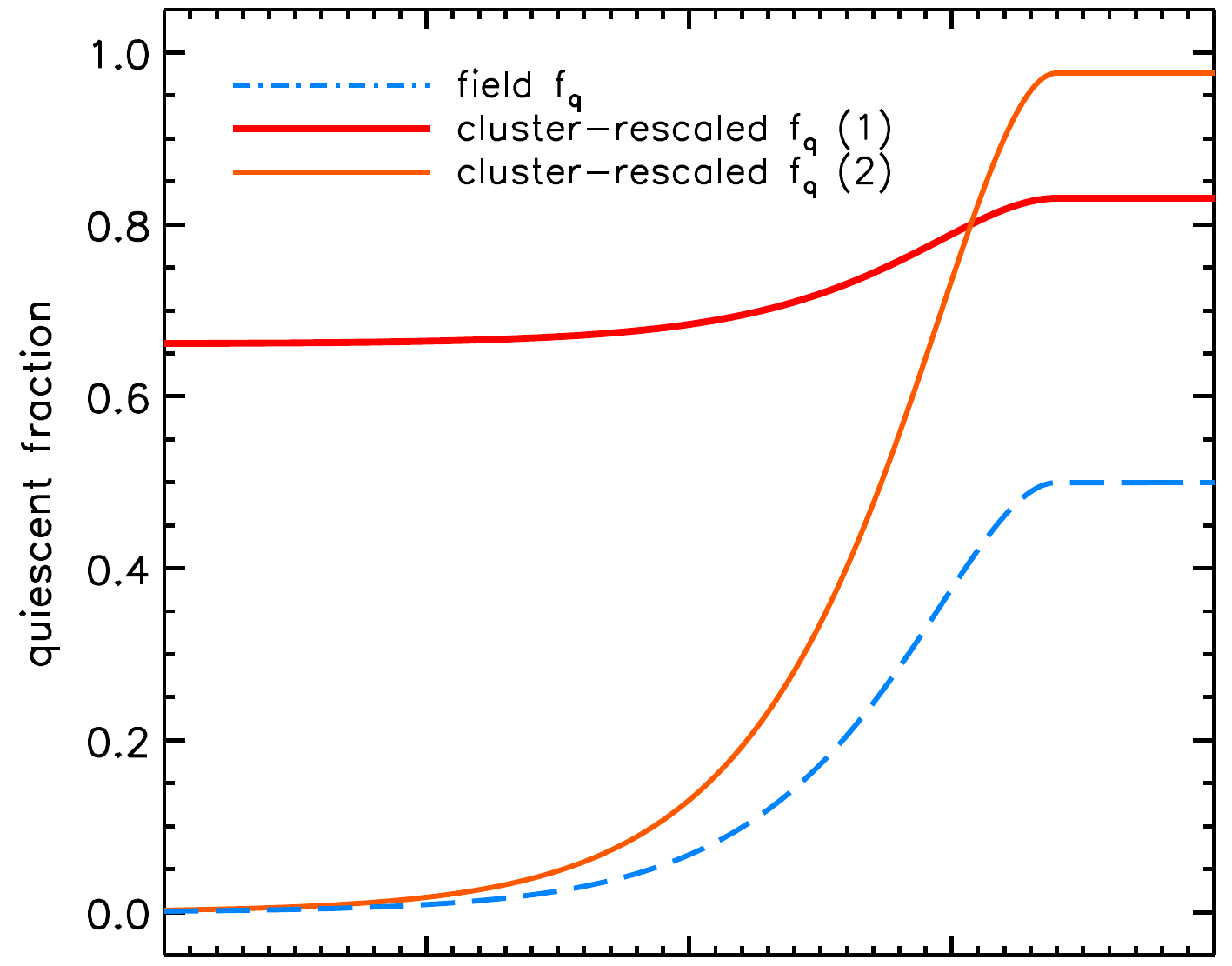}%
  \includegraphics[height=0.195\textheight,viewport=106 396 463
    716,clip]{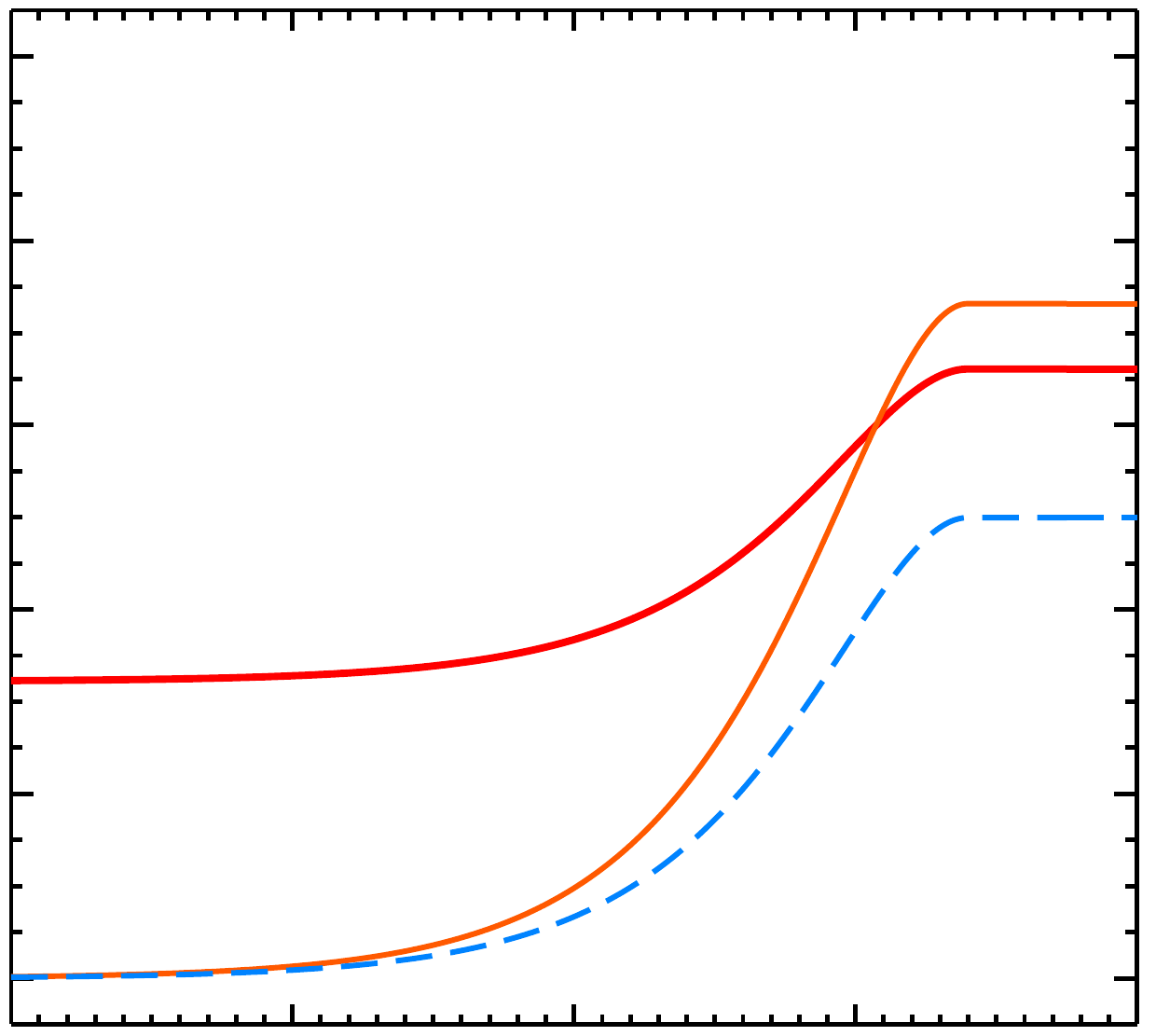}%
  \includegraphics[height=0.195\textheight,viewport=106 396 463
    716,clip]{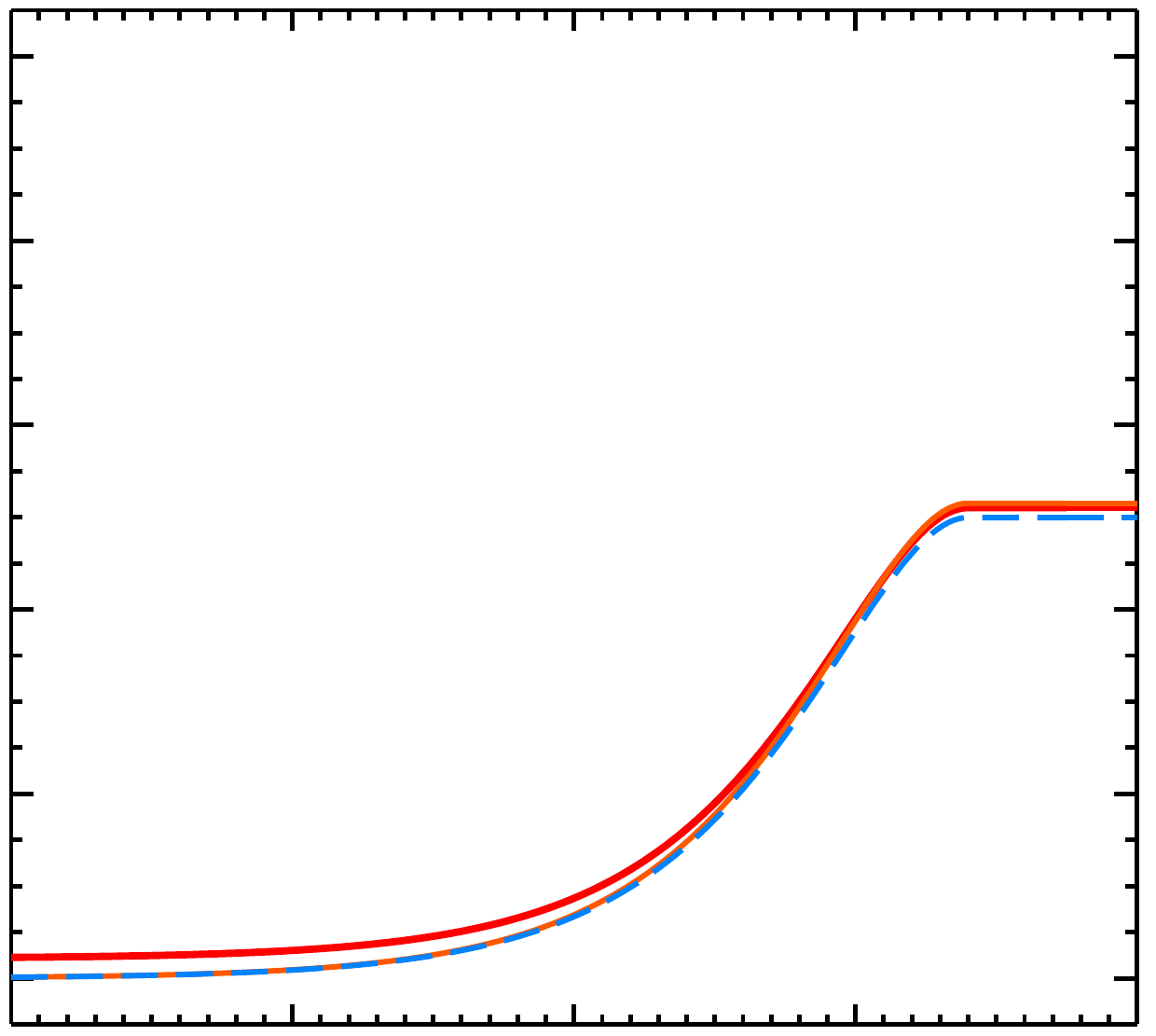}
  \includegraphics[height=0.22\textheight,viewport=66 355 463
    716,clip]{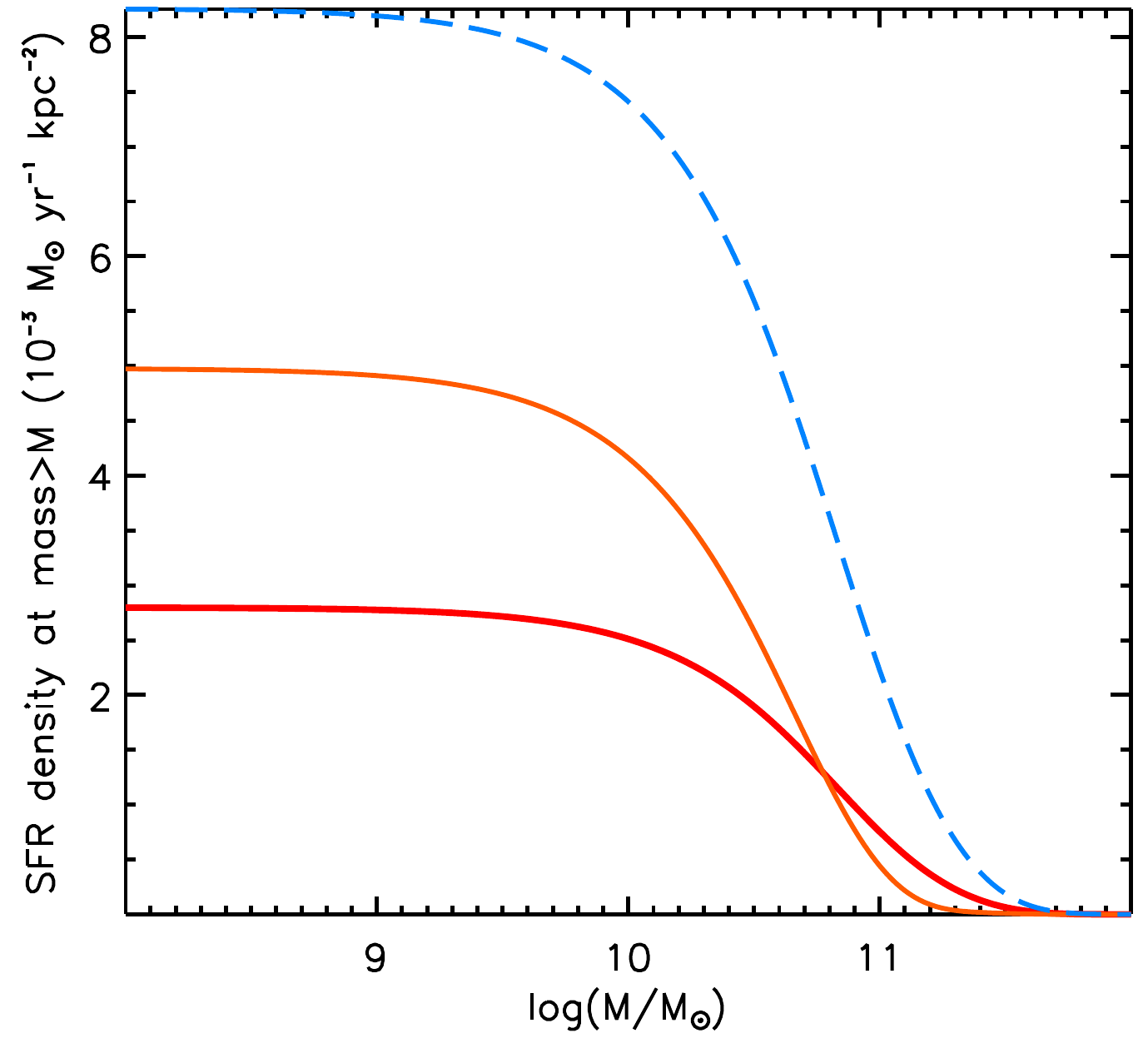}%
  \includegraphics[height=0.22\textheight,viewport= 106 355 463
    716,clip]{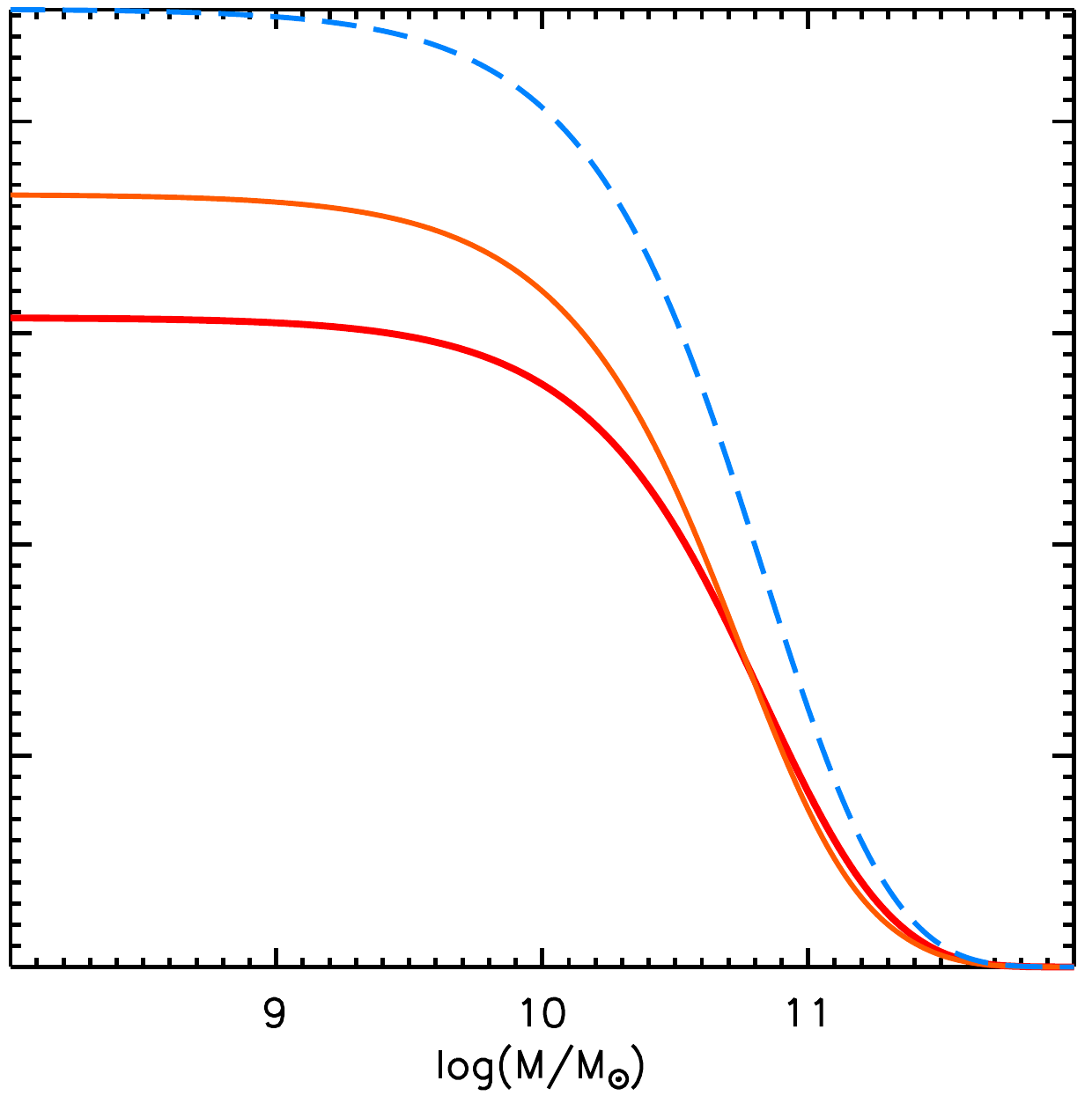}%
  \includegraphics[height=0.22\textheight,viewport= 106 355 463
    716,clip]{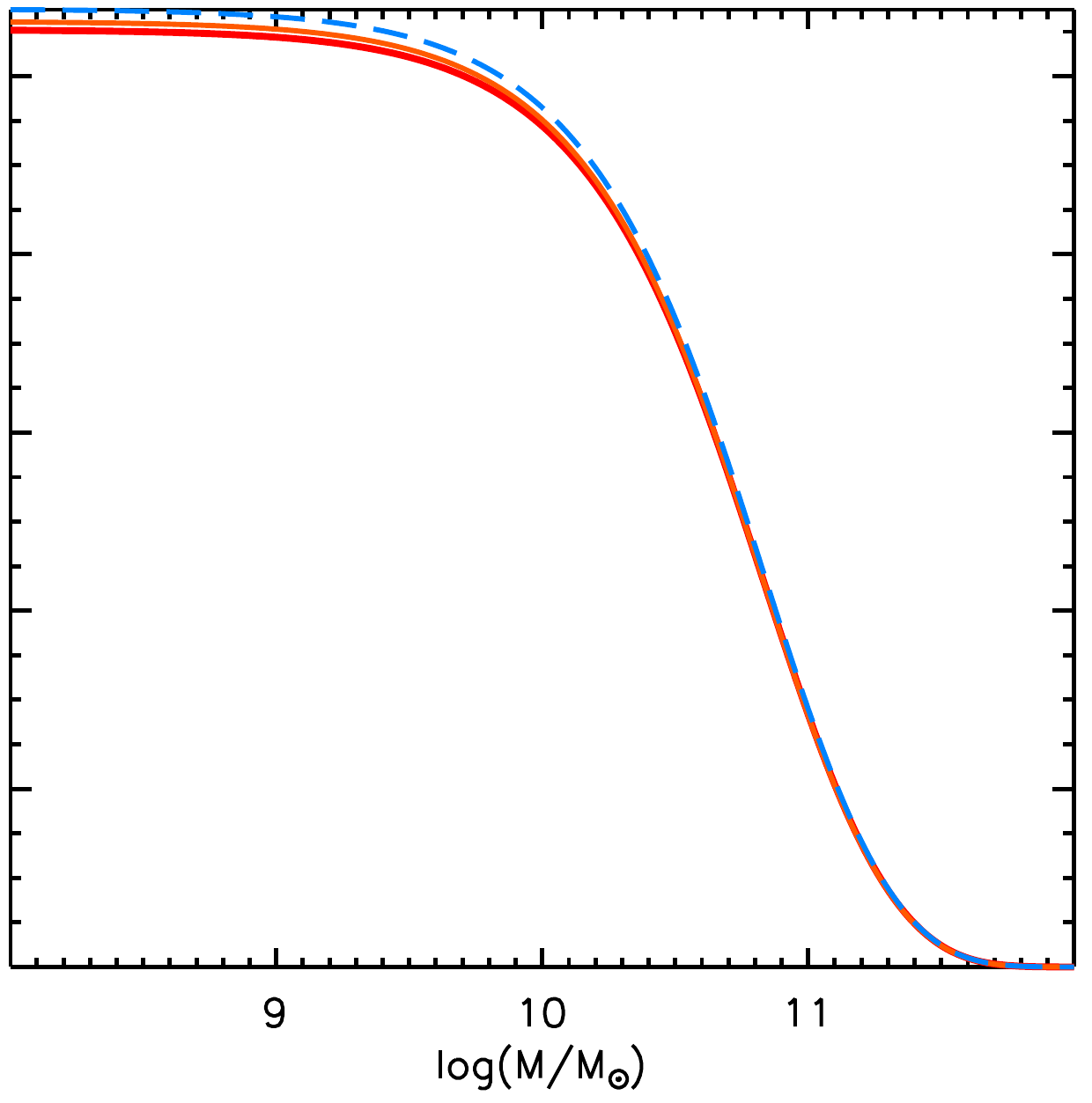}

\caption{Example illustration of the modeling of the quiescent
  fraction vs.\ stellar mass. {\it Top panels: } The assumed radial
  profile of the quiescent fraction above mass completeness. In the
  three panels a green circle highlights a given clustercentric
  distance, and thus quiescent fraction value. Based on these values,
  the corresponding {\it middle panels} show the quiescent fraction
  vs.\ stellar mass as estimated with both approaches described in
  Sect.~\ref{sec:samplebias} (red, orange lines).  The blue line shows
  the quiescent fraction vs.\ mass in the field at the cluster
  redshift. The corresponding {\it bottom panels} show the overall
  contribution of galaxies above a given stellar masses to the total
  SFR density at the given clustercentric distance assuming the
  quiescent fraction vs.\ mass from the middle panels (see
  Sect.~\ref{sec:samplebias}).
\label{fig:passfracmod}}
\end{figure*}

We describe here in detail the modeling outlined in
Sect.~\ref{sec:samplebias}, devised to investigate a possible bias of
this cluster sample against clusters with higher star-forming galaxy
fractions.

For each cluster, from the background subtracted and area weighted
mass complete sample of cluster members we first estimate the total
SFR in the $r<0.45 r_{500}$ region above the mass completeness limit,
by assuming that all galaxies classified as star forming with our
color criterion (Sect.~\ref{sec:uvj}) are forming stars at the Main
Sequence \citep[MS, e.g.,][]{elbaz2011} rate. We adopt the MS modeling
of \citet{schreiber2015}, which includes bending at high stellar
masses; adopting a straight MS modeling as from
e.g., \citet{sargent2014} would produce here only marginal
differences. No contribution to the SFR is considered from galaxies
classified as quiescent.

\subsection{Contribution of cluster galaxies below the mass completeness limit}
\label{sec:appendixa1}

We then estimate the star formation rate contribution from cluster
galaxies below the mass completeness limit and down to
$M=10^{8}$M$_{\odot}$ by assuming:

1) that the shape of the galaxy stellar mass function in clusters is
to first order the same as in the field at the same redshift
\citep[but see e.g.,][]{vanderburg2013}. We adopt the
\citet{muzzin2013c} stellar mass functions.

2) the \citet{schreiber2015} MS SFR for all star-forming galaxies
depending only on their stellar mass. We neglect here (as above for
the total SFR estimate of the mass complete sample) modeling the
intrinsic scatter of the MS as we are averaging over the full
star-forming galaxy population to obtain total star formation rates
for the whole cluster galaxy sample. On the other hand, we also
neglect to model non-MS populations: all galaxies classified as
quiescent are assumed to have a negligible SFR, and we neglect the
minority population ($\sim 2\%$ by number) of starbursts, estimated to
contribute $<10-15\%$ of the star formation rate density (in the
field, at this redshift) \citep[e.g.,][]{rodighiero2011,sargent2012}.
In this respect, we note though that the possible relevance to the
mm-wave emission contamination of the higher SFRs of the small
fraction of starburst galaxies is reduced by their different SED:
according to \citet[e.g.,]{bethermin2012,bethermin2015} SEDs, the
150~GHz (95~GHz) flux of a starburst galaxy with a SFR six times
greater than the MS SFR, is within a factor 2-3 ($\lesssim3-4$) of the
flux of a MS galaxy of the same stellar mass.

3) that the quiescent fraction at stellar masses below our mass
completeness limit can be estimated from our measured quiescent
fraction with the following approach. We start from the
\citet{muzzin2013c} estimate of the quiescent fraction as a function
of stellar mass in the field at the cluster redshift. Since above the
mass completeness limit we observe higher quiescent fractions than in
the field, we model the quiescent fraction vs.\ stellar mass for
cluster galaxies in two ways, both sketched in
Figure~\ref{fig:passfracmod}. The first approach assumes that mass and
environmental quenching are separable, and following \citet{peng2010} we
estimate the quiescent fraction vs.\ stellar mass as $f_{q} =
\epsilon_{m} + \epsilon_{\rho} - \epsilon_{m} \times \epsilon_{\rho}$,
with $\epsilon_{m}$ and $\epsilon_{\rho}$ the mass and environmental
quenching efficiencies, respectively. As an estimate of
$\epsilon_{m}$, we take the field quiescent fraction of \citet[][
  which indeed drops to very low values at low masses at the clusters'
  redshift, so that it can be interpreted to first approximation as a
  mass quenching efficiency if considering the ``reference'' mass
  where no mass quenching occurs below
  $10^{9}$M$_{\odot}$]{muzzin2013c}. As an estimate of
$\epsilon_{\rho}$ we adopt the environmental quenching efficiency
measured for each given cluster in Sect.~\ref{sec:passfrac}
\citep[again, this assumes that mass and environmental quenching are
  completely separable and thus that the environmental quenching
  efficiency is independent of stellar mass, but see e.g.,][for
  contrasting results on this assumption at this
  redshift]{kawinwanichakij2017}. This gives us the adopted quiescent
fraction vs.\ mass from the first approach, labeled as (1) in
Fig.~\ref{fig:passfracmod}.

\begin{figure}
 \includegraphics[width=0.49\textwidth,viewport=80 371 540 701, clip]{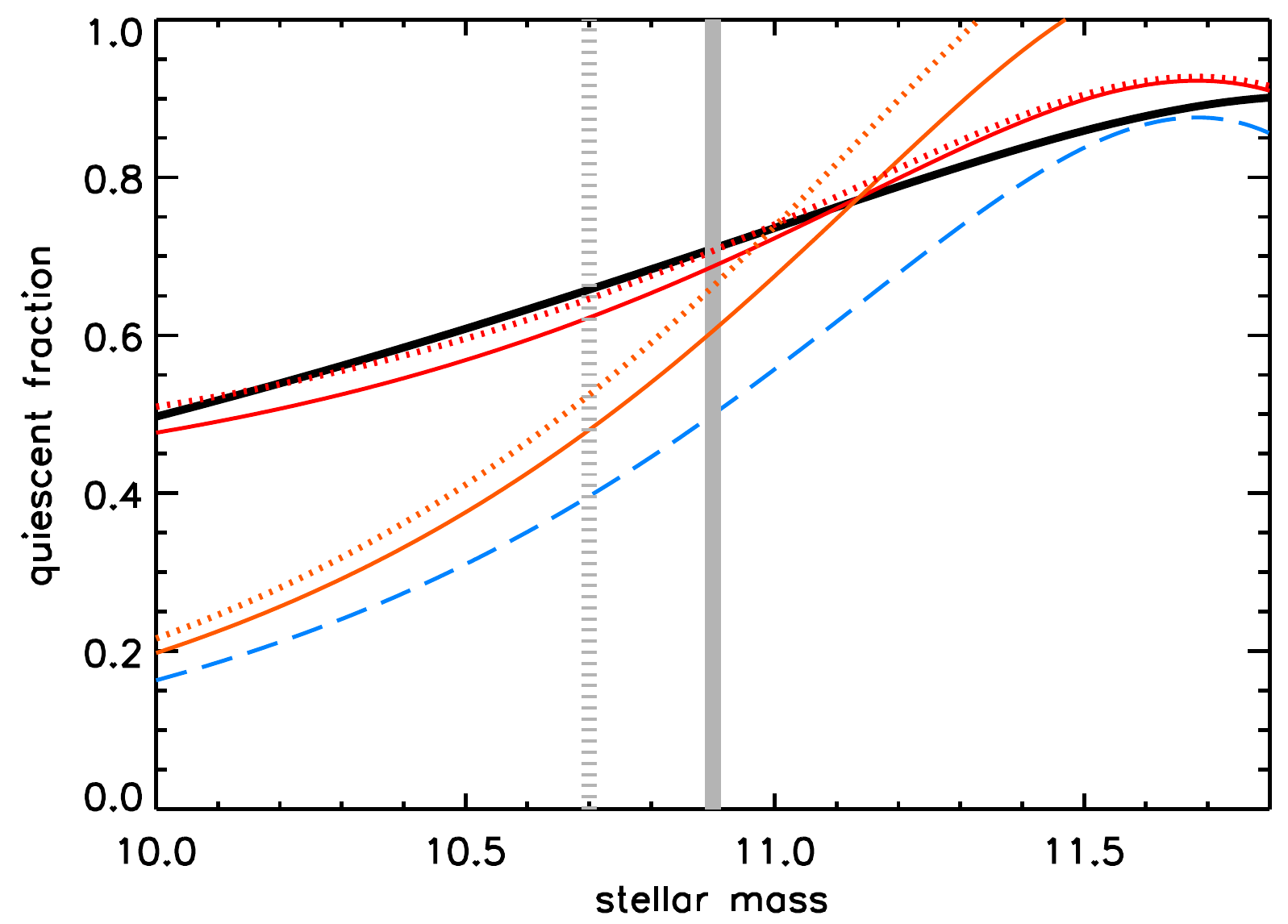}
\caption{Test of the quiescent fraction vs. stellar mass modeling
  discussed in Sect.~\ref{sec:samplebias}. The black line shows the
  observed quiescent fraction vs. stellar mass in $z\sim1$ clusters
  from \citet{vanderburg2013}. The red and orange solid (dotted) lines
  (color coding as in Fig.~\ref{fig:passfracmod}) show the estimated
  quiescent fraction vs. mass using the two approaches shown in
  Fig.~\ref{fig:passfracmod}, starting from the observed quiescent
  fraction in a mass complete sample with log(M/M$_{\odot}$)$>10.9$
  (10.7, respectively, as shown by vertical gray lines). Blue dashed
  line shows the field quiescent fractions vs. mass at the same
  redshift.
\label{fig:passfracvsmasstest}}
\end{figure}

The second approach just rescales the field quiescent fraction
vs.\ stellar mass by multiplying it by the ratio of the measured
quiescent fraction in the cluster (from Sect.~\ref{sec:passfrac}) and
in the field \citep[from][]{muzzin2013c} above the mass completeness
limit of the given cluster. This second approach produces by
definition the same quiescent fraction above mass completeness, but
considerably lower quiescent fractions at lower stellar masses,
quickly approaching the field levels rather than reaching the quiescent
fraction plateau determined by the assumed environmental quenching
efficiency in the first approach. This second approach is labeled (2)
in Fig.~\ref{fig:passfracmod}.  By definition, both approaches
reproduce the assumed field quiescent fraction vs.\ stellar mass when
the quiescent fraction in the mass complete sample of cluster galaxies
is taken to be the same as in the field.

We believe that these two approaches reasonably bracket the plausible
range of quiescent fraction vs.\ stellar
mass. Figure~\ref{fig:passfracvsmasstest} presents a test of both
approaches against the quiescent fraction vs.\ stellar mass observed in
$z\sim1$ clusters from \citet{vanderburg2013}. We simulate our
modeling by considering from these data the quiescent fraction in two
mass complete samples with log(M/M$_{\odot}$)$>$10.7,10.9, and
estimating the quiescent fraction at lower masses with the two
approaches described above. As Fig.~\ref{fig:passfracvsmasstest}
shows, the first approach (1) based on the separable environmental and
mass quenching efficiencies reproduces a quiescent fraction vs. mass
in much better agreement with the observed trend than the second
approach. We thus mostly focus in the following analysis on the
quiescent fraction vs. stellar mass from the first aproach. We also
stress that, for the purpose of this Section, this is a conservative
choice as explained below.

\begin{figure*}
  \centering
 \includegraphics[height=0.166\textheight,viewport=76 408 538 698,clip]{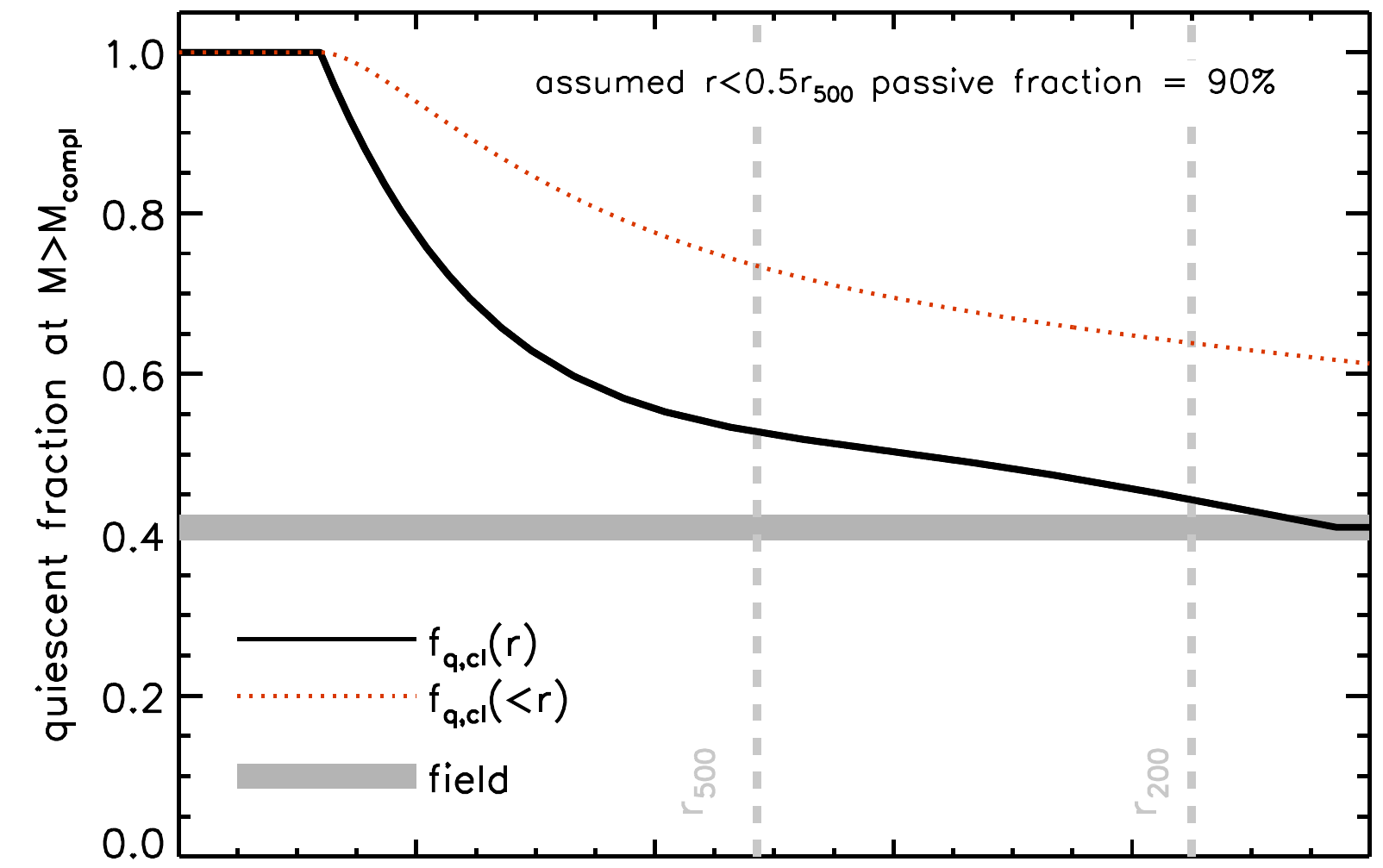}%
 \includegraphics[height=0.166\textheight,viewport=133 408 538 698,clip]{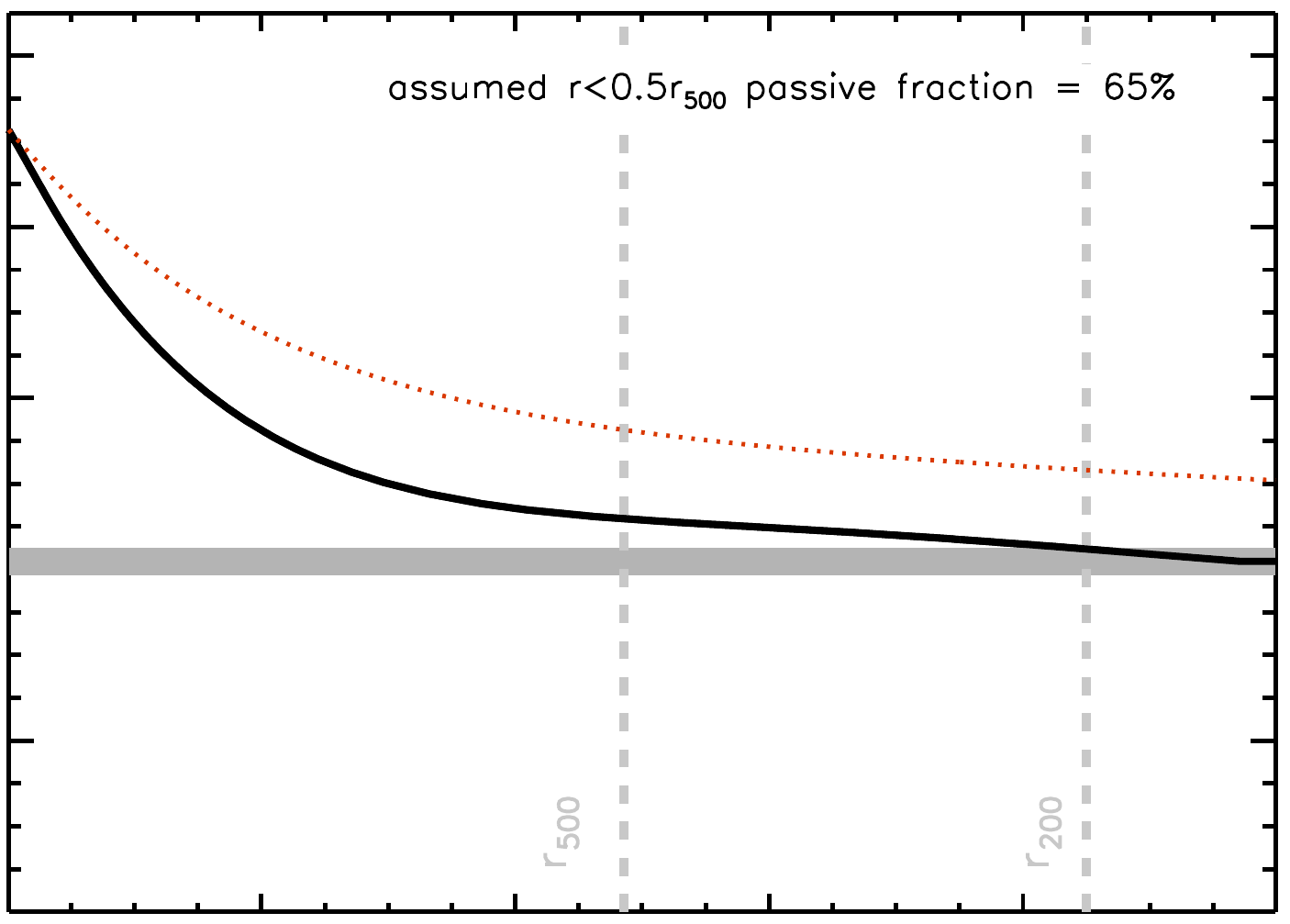}%
 \includegraphics[height=0.166\textheight,viewport=133 408 538 698,clip]{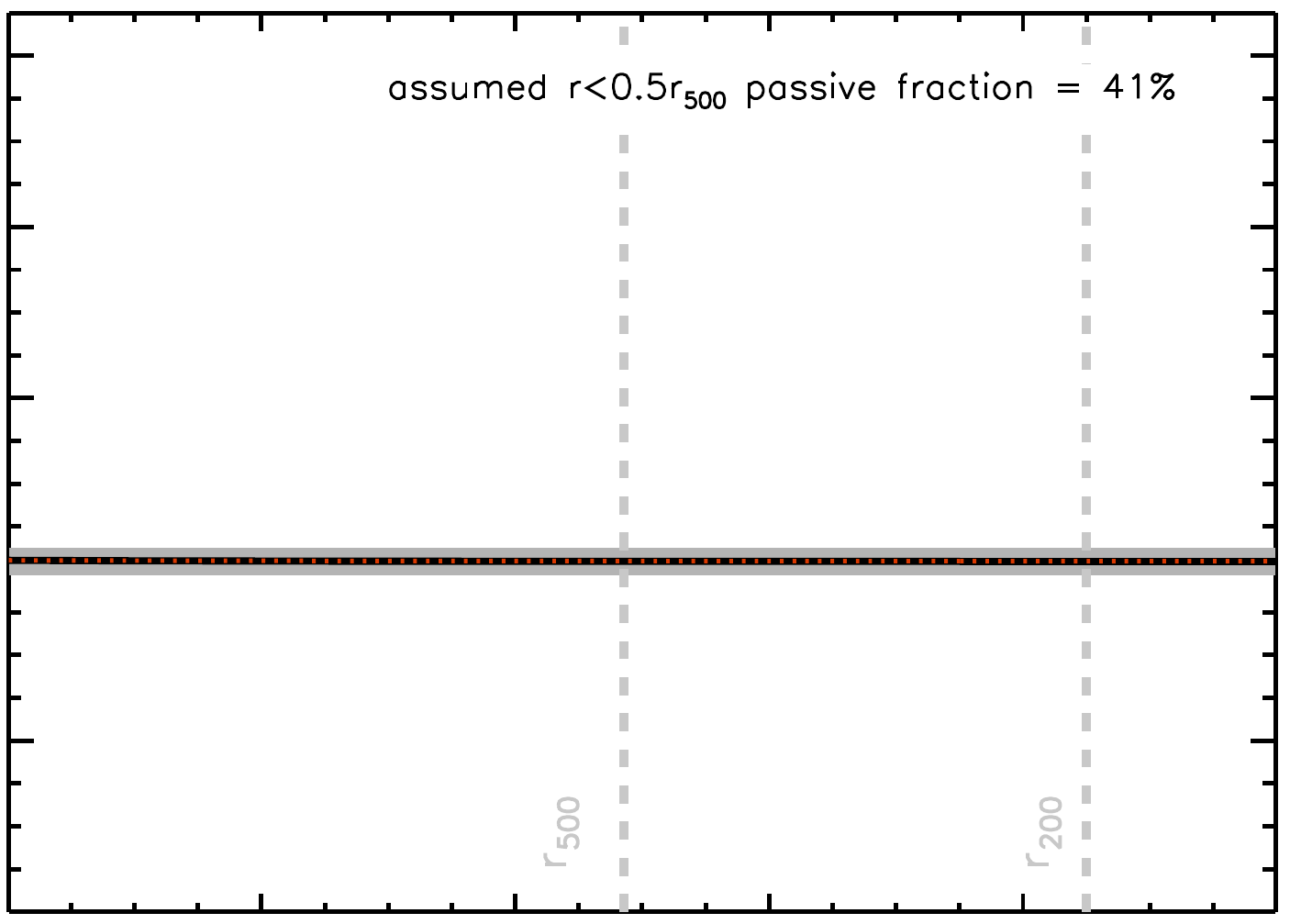}
 \includegraphics[height=0.166\textheight,viewport=76 408 538 698,clip]{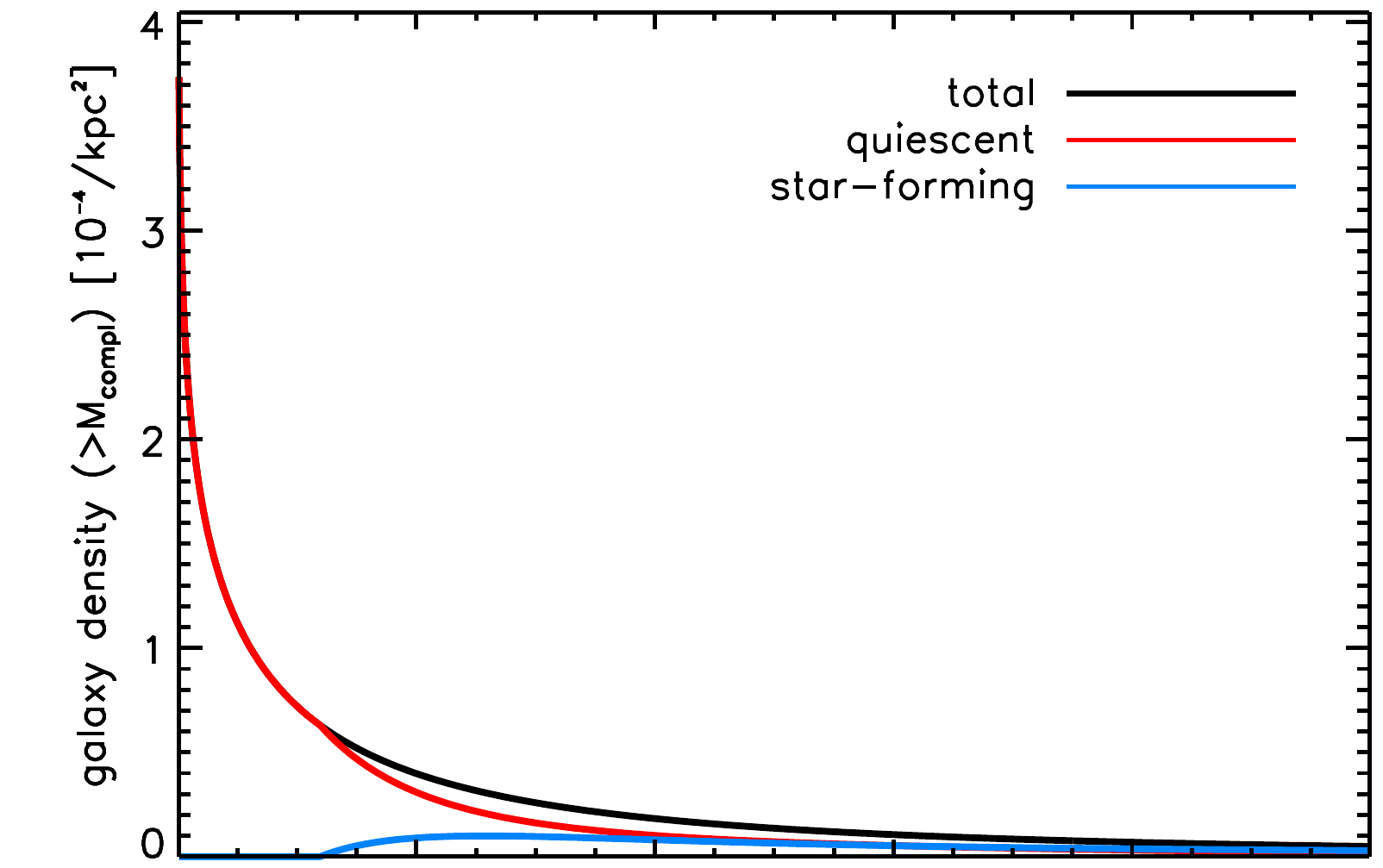}%
 \includegraphics[height=0.166\textheight,viewport=133 408 538 698,clip]{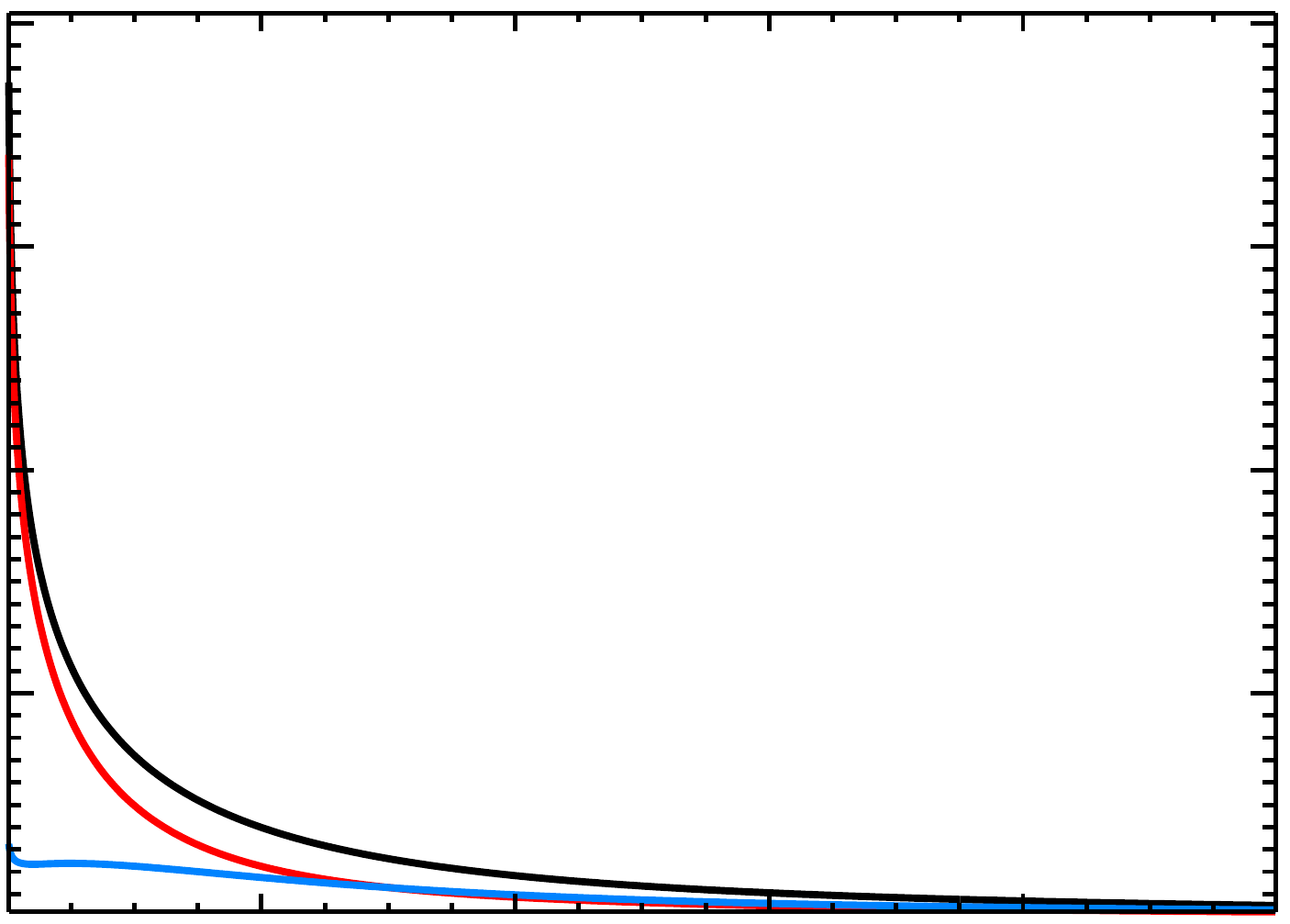}%
 \includegraphics[height=0.166\textheight,viewport=133 408 538 698,clip]{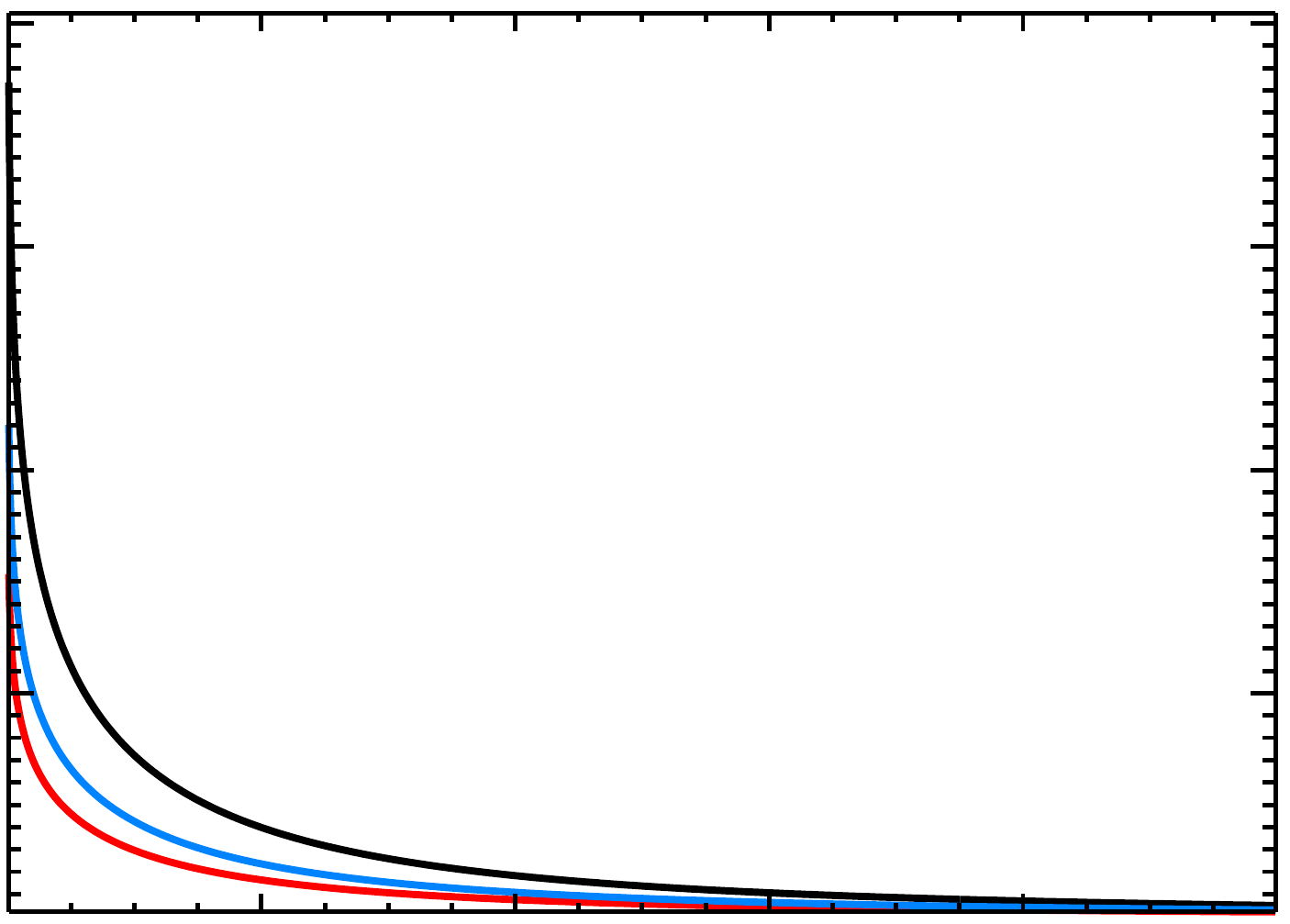}
 \includegraphics[height=0.189\textheight,viewport=76 369 538 698,clip]{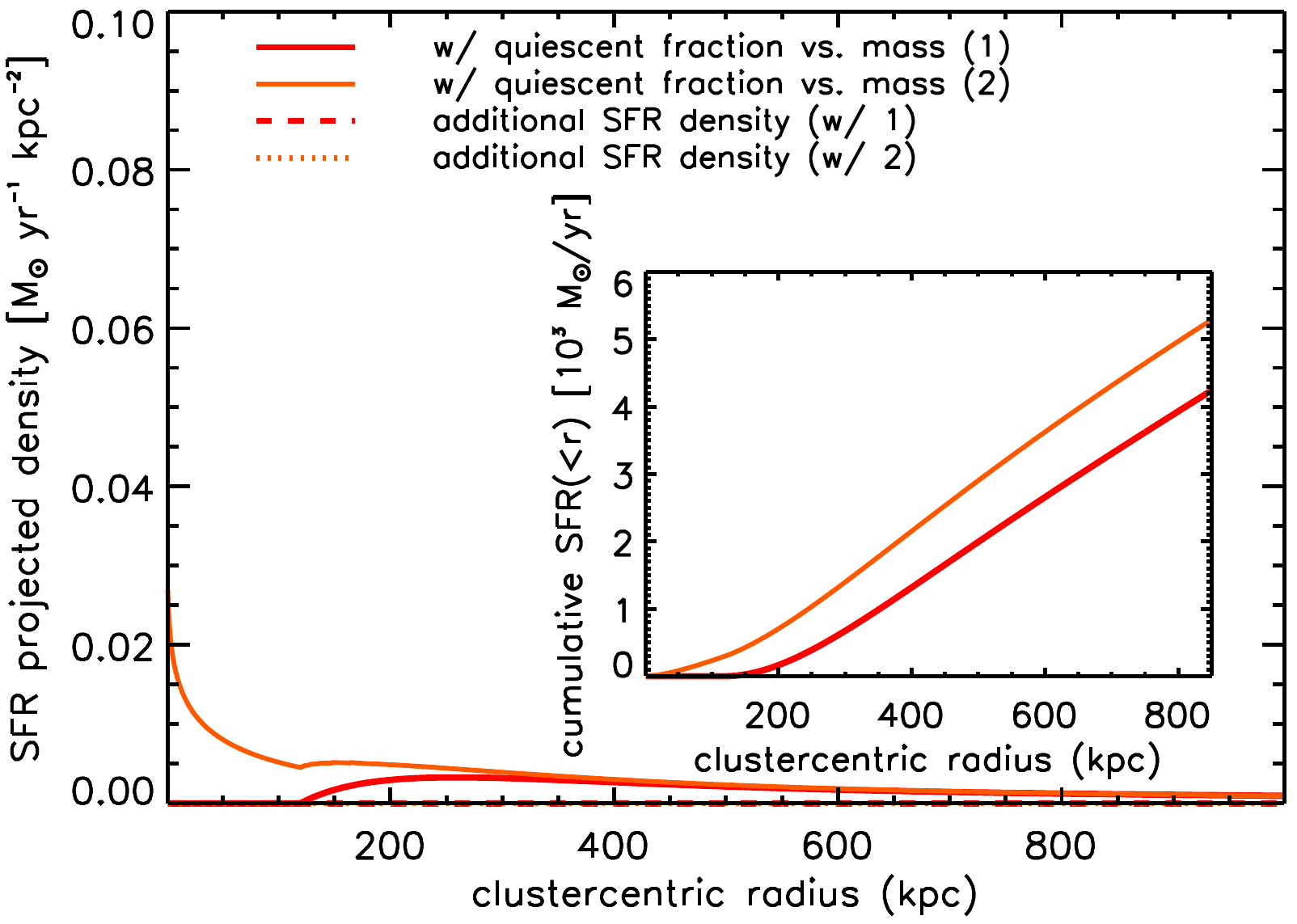}%
 \includegraphics[height=0.189\textheight,viewport=133 369 538 698,clip]{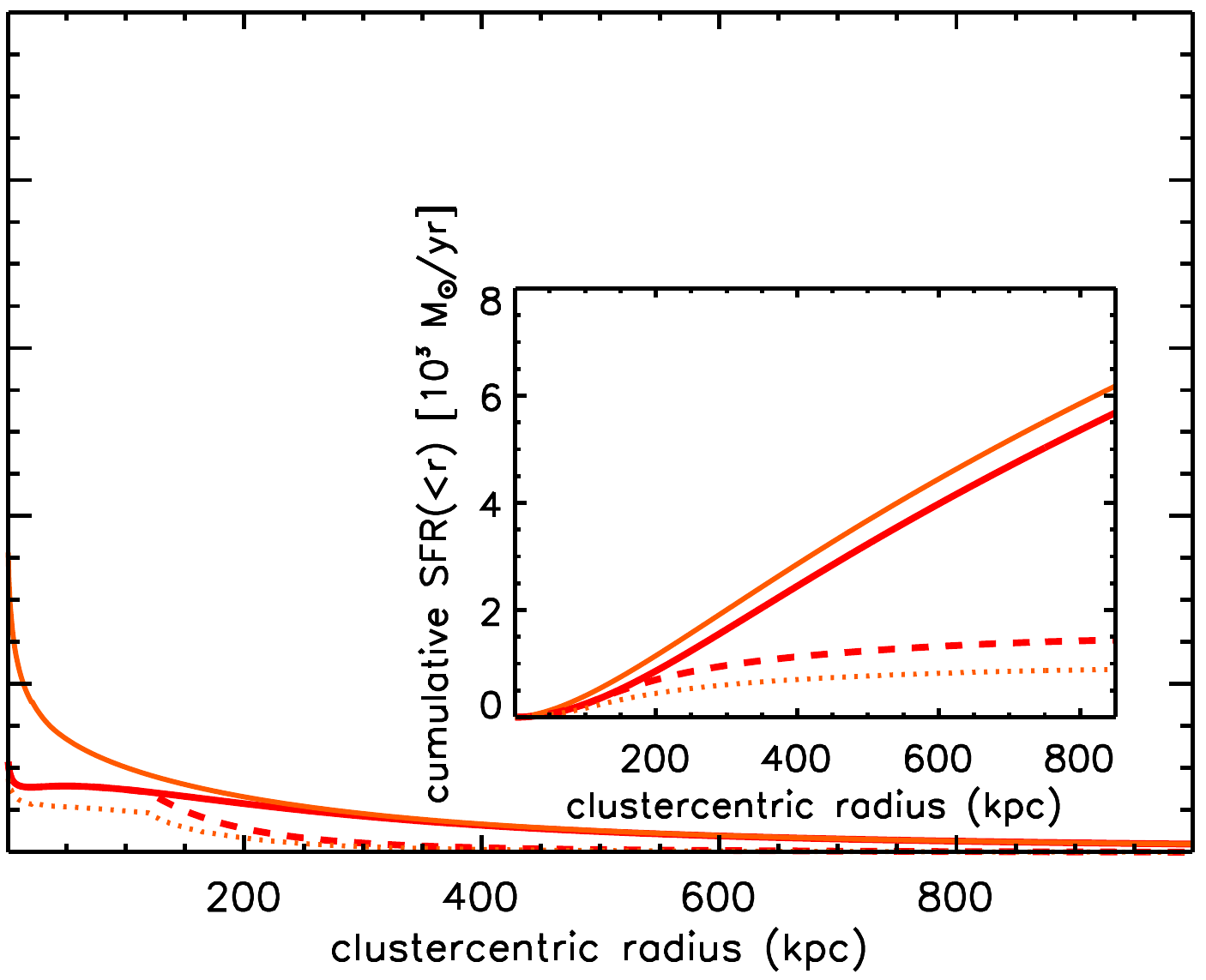}%
 \includegraphics[height=0.189\textheight,viewport=133 369 538 698,clip]{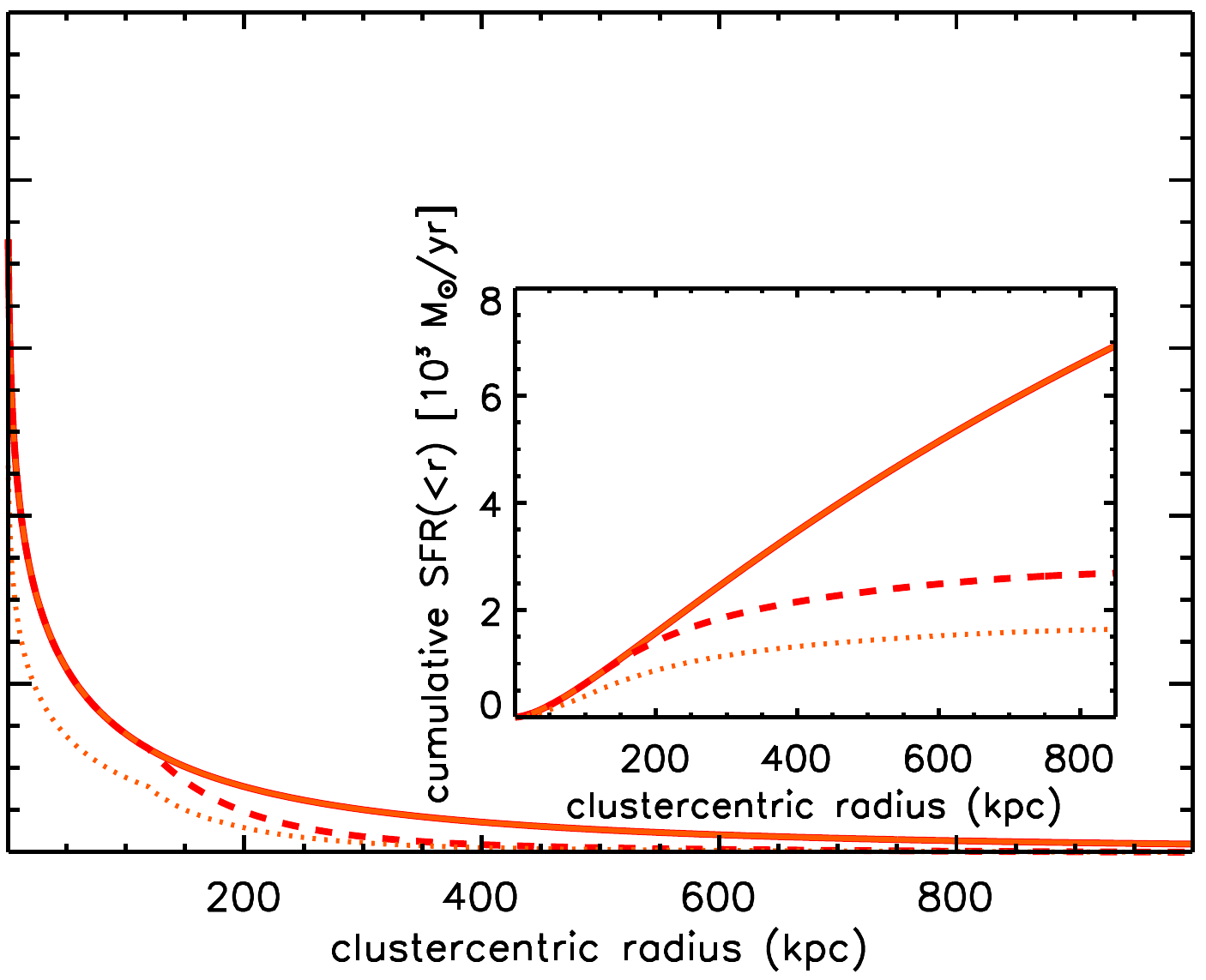}
  \caption{Example illustration of the modeling of the SFR density profiles
    described in Sect.~\ref{sec:samplebias}, see text for full
    details. Left, middle and right panel columns correspond to three
    different values assumed for the $r<r_{500}$ quiescent fraction
    above mass completeness, as indicated. {\it Top panels} show the
    radial profile of quiescent fraction above mass completeness
    (black, the field value is shown in gray). The dotted red line
    shows the quiescent fraction of the full mass complete sample out to
    the given clustercentric radius.  {\it Middle panels} show the
    projected galaxy number density profile above mass completeness
    (black), and the derived profiles of star-forming and quiescent
    galaxies (blue and red) given the quiescent fraction profile in
    the corresponding top panel. {\it Bottom panels} show the
    corresponding projected SFR density profiles, derived with the two
    models of quiescent fraction vs.\ stellar mass as indicated (solid
    lines). Dashed and dotted lines show the \emph{additional} SFR density
    profile over the prediction for the actually observed quiescent
    fraction (see text). Inserts show the corresponding cumulative SFR
    as a function of clustercentric radius.
\label{fig:SFprofiles}}
\end{figure*}

For each cluster in our sample, we thus obtain an approximate estimate
of the total SFR within the $r<0.45 r_{500}$ region by adding up the
estimated MS SFRs of all (background subtraction and area coverage
weighted) candidate members classified as star forming, and the
estimated SFR of cluster galaxies below the mass completeness limit
described above.

\subsection{Contribution of cluster galaxies at $r>0.45 r_{500}$}
\label{sec:appendixa2}

We estimate the total SFR density profile of the cluster beyond $r=0.45
r_{500}$ by assuming a NFW \citep{nfw1997} galaxy number density
profile (normalized to the $r<0.45 r_{500}$ region\footnote{The galaxy
  number density profile is obviously normalized by the total galaxy
  number at $r<0.45 r_{500}$ estimated from the (weighted) candidate
  members above mass completeness, but we check that the total stellar
  mass and SFR above the mass completeness limit predicted by our
  modeling at $r<0.45 r_{500}$ are indeed in agreement (on average
  within $\sim 10-15\%$, or $<20-25\%$ at worst) with those estimated
  from the adopted stellar masses and SFRs of the individual
  (weighted) candidate members.}) with concentration according to
\citet{duffy2008}, and a quiescent fraction profile determined as
follows.

We start from a red fraction profile determined from the stacked total
and red galaxy projected number density profiles of $0<z<1$ clusters
from \citet{hennig2017}. This red fraction profile is then distorted
to force a match with the central cluster quiescent fraction above
mass completeness at $r < 0.45 r_{500}$, and with the corresponding
field quiescent fraction at $r=2 r_{500}$
(Figure~\ref{fig:SFprofiles}, top panels). The assumed total (NFW)
projected galaxy number density profile, and the described quiescent
fraction profile above mass completeness, together yield a projected
number density profile of star-forming cluster galaxies
(Figure~\ref{fig:SFprofiles}, middle panels). In the assumption that
the galaxy stellar mass function and MS SFR do not depend on
clustercentric radius, and that the quiescent fraction vs.\ stellar
mass can be determined as described above at each clustercentric
distance given the adopted quiescent fraction profile above mass
completeness, we thus derive for each cluster a SFR projected density
profile (Figure~\ref{fig:SFprofiles}, bottom panels).

\subsection{Modeling of the total cluster SFR vs.\ quiescent fraction}
\label{sec:appendixa3}

To carry out a first-order investigation of how much the cluster SZE
detection would be affected by a quiescent fraction lower than what
observed, we then reduce in the modeling above the quiescent fraction
above mass completeness in the $r<0.45 r_{500}$ region, to
progressively lower values down to a quiescent fraction of 10\%, and
compute the SFR density profile of the cluster for the given $r<0.45
r_{500}$ quiescent fraction.  As discussed in
Sect.~\ref{sec:samplebias}, the smallest adopted value of 10\% for the
central quiescent fraction is well below the field value at the
cluster redshift for the considered stellar mass range.

Figures~\ref{fig:passfracmod} and \ref{fig:SFprofiles} show an actual
example of the overall derivation based on the cluster
SPT-CLJ0459. The top panels of Fig.~\ref{fig:passfracmod} show the
adopted radial profile of the quiescent fraction above mass
completeness for the measured value of the $r<r_{500}$ quiescent
fraction (black line). The green circles highlight three radii -- and
thus three values of the quiescent fraction above mass completeness --
for which the middle and bottom panels present, respectively, the
modeling of the quiescent fraction vs.\ stellar mass with both
approaches described above, and the corresponding inferred
contribution of galaxies of different stellar mass to the total SFR
density at the given clustercentric distance.
Fig.~\ref{fig:SFprofiles} shows instead the modeling of the SFR
density profile for three different values assumed for the $r<0.45
r_{500}$ quiescent fraction above mass completeness: the actually
observed one (left panels), the field quiescent fraction (right), and
an intermediate value (middle). For each of these, the top panels show
the radial profile of the quiescent fraction above mass completeness
derived as described above (black line). For comparison, the red
dotted line shows the predicted quiescent fraction of the whole
cluster galaxy population above mass completeness out to a given
clustercentric distance. The middle panels show, for the assumed total
projected galaxy number density profile above mass completeness
(black), the star-forming galaxy projected density profile (blue, also
at M$>$M$_{compl}$) obtained for the quiescent fraction profile in the
corresponding top panel. Finally, the bottom panels show the
corresponding predicted SFR projected density profile obtained with
the described modeling for both estimated of the quiescent fraction vs.
stellar mass (solid lines, color coding as in
Fig.~\ref{fig:passfracmod}). We note that in fact we consider that the
actual SZE measurement of the cluster is indeed affected by some
amount of contamination from star formation in cluster galaxies,
corresponding to the star forming galaxy fraction actually observed.
Decreasing the $r<0.45 r_{500}$ quiescent fraction increases the
contamination of the cluster SZE signal from mm-wave emission due to
star formation, and we estimate this additional contamination (dashed
and dotted lines, labeled as ``additional SFR density profile'' in the
bottom panels of Fig.~\ref{fig:SFprofiles}) as the difference between
the SFR density profiles estimated with the assumed quiescent fraction
and with the actually measured one.

\subsection{Modeling of the S/N of the SZE detection  vs.\ quiescent fraction}
\label{sec:appendixa4}

We translate this additional SFR projected density profile into flux
density maps at 95 and 150 GHz assuming the MS SEDs from either
\citet{bethermin2015} or \citet{schreiber2018}. We then add these
additional flux density maps to the observed 95 and 150 GHz maps of
the cluster and reanalyze the resulting maps with the same filtering
procedure used to detect clusters in the SPT-SZ survey
\citep{bleem2015}, estimating a S/N for the resulting SZE
detection. Figure~\ref{fig:filteredSN} shows this retrieved S/N as a
function of the assumed $r<0.45 r_{500}$ quiescent fraction above mass
completeness down to a value of 10\%.  The figure shows the S/N
obtained with both the adopted MS SEDs, and for the main adopted
approach (1) for estimating the quiescent fraction vs. stellar mass
below the mass completeness limit. The other approach (2) resulting in
comparatively higher star-forming fractions below the mass
completeness limit (Fig.~\ref{fig:passfracmod}), produces even higher
predicted S/N due to the smaller difference in SFR between the
observed and reduced quiescent fraction cases. Further discussion of
Fig.~\ref{fig:filteredSN} and derived conclusions are presented in
Section~\ref{sec:samplebias}.

\end{appendix}



\end{document}